\pdfoutput=1                                      

\RequirePackage{fix-cm}                            

\documentclass[]{svjour3}

\usepackage{amsmath}                    

\usepackage{bm}                                                    

\usepackage{exscale,relsize}   

\usepackage{graphicx}                               

\usepackage{mathptmx}                             

\usepackage{stmaryrd}                             

\bmdefine{\nhat}{n}
\bmdefine{\ehat}{e}

\bmdefine{\nuhat}{\nu}
\bmdefine{\tauhat}{\tau}

\bmdefine{\bfa}{a}
\bmdefine{\bff}{f}
\bmdefine{\bfl}{l}
\bmdefine{\bfp}{p}
\bmdefine{\bfr}{r}
\bmdefine{\bfs}{s}
\bmdefine{\bft}{t}
\bmdefine{\bfv}{v}
\bmdefine{\bfw}{w}

\bmdefine{\bfB}{B}
\bmdefine{\bfD}{D}
\bmdefine{\bfE}{E}
\bmdefine{\bfF}{F}
\bmdefine{\bfG}{G}
\bmdefine{\bfH}{H}
\bmdefine{\bfK}{K}
\bmdefine{\bfM}{M}
\bmdefine{\bmP}{P}

\bmdefine{\bfchi}{\chi}
\bmdefine{\bfeps}{\varepsilon}
\bmdefine{\bfphi}{\varphi}
\bmdefine{\bfpsi}{\psi}
\bmdefine{\bfsigma}{\sigma}

\bmdefine{\bfzero}{0}

\newcommand*{\calF}{\mathcal{F}}
\newcommand*{\calV}{\mathcal{V}}

\newcommand*{\calFv}{\calF_{\text{v}}}
\newcommand*{\calFs}{\calF_{\text{s}}}

\newcommand*{\eps}{\epsilon}

\newcommand*{\epsa}{\Delta\eps}
\newcommand*{\epara}{\eps_{\scriptscriptstyle\parallel}}
\newcommand*{\eperp}{\eps_{\scriptscriptstyle\perp}}

\newcommand*{\chipara}{\chi_{\scriptscriptstyle\parallel}}
\newcommand*{\chiperp}{\chi_{\scriptscriptstyle\perp}}

\newcommand*{\bfI}{\mathbf{I}}
\newcommand*{\bfL}{\mathbf{L}}
\newcommand*{\bfP}{\mathbf{P}}
\newcommand*{\bfQ}{\mathbf{Q}}
\newcommand*{\bfR}{\mathbf{R}}
\newcommand*{\bfT}{\mathbf{T}}
\newcommand*{\bfW}{\mathbf{W}}

\newcommand*{\rhof}{\rho_\text{f}}
\newcommand*{\rhom}{\rho_\text{m}}

\newcommand*{\sigmaf}{\sigma_\text{f}}

\newcommand*{\Tc}{\bfT_{\text{c}}}
\newcommand*{\Te}{\bfT_{\text{e}}}
\newcommand*{\Ts}{\bfT_{\text{s}}}

\newcommand*{\TEL}{\bfT_{\text{E-L}}}
\newcommand*{\TM}{\bfT_{\text{M}}}
\newcommand*{\TP}{\bfT_{\!\bmP}}

\newcommand*{\We}{W_\text{e}}
\newcommand*{\Ws}{W_\text{s}}
\newcommand*{\WE}{W_\text{E}}
\newcommand*{\WH}{W_\text{H}}

\newcommand*{\tildeWs}{\widetilde{W}_\text{s}}
\newcommand*{\Ttilde}{\widetilde{T}}

\newcommand*{\Pf}{\bmP_\text{f}}

\newcommand*{\bfchie}{\bfchi_\text{e}}
\newcommand*{\bfchim}{\bfchi_\text{m}}

\newcommand*{\eb}{e_\text{b}}
\newcommand*{\es}{e_\text{s}}

\newcommand*{\Arg}{\text{arg}}

\newcommand*{\nutwo}{\nuhat_2}
\newcommand*{\tautwo}{\tauhat_2}

\newcommand*{\nablaS}{\nabla_{\!\text{S}}}
\newcommand*{\xS}{x_{\text{S}}}

\newcommand*{\Eext}{\bfE_{\text{ext}}}

\newcommand*{\dWdn}{\frac{\partial W}{\partial\nhat}}
\newcommand*{\dWdgn}{\frac{\partial W}{\partial\nabla\nhat}}

\newcommand*{\dWedn}{\frac{\partial\We}{\partial\nhat}}
\newcommand*{\dWedgn}{\frac{\partial\We}{\partial\nabla\nhat}}

\newcommand*{\dWEdgn}{\frac{\partial\WE}{\partial\nabla\nhat}}

\newcommand*{\dWdx}{\frac{\partial W}{\partial x}}

\newcommand*{\dndx}{\frac{\partial\nhat}{\,\partial x_1}}
\newcommand*{\dndxx}{\frac{\partial\nhat}{\,\partial x_2}}

\newcommand*{\epstensor}{\mathlarger{\bfeps}}

\newcommand*{\Eps}{\mathlarger{\eps}}

\DeclareMathOperator{\Div}{div}

\DeclareMathOperator{\myskew}{skew}
\DeclareMathOperator{\Skew}{Skew}

\newcommand*{\divS}{\Div_{\text{S}}}

\DeclareMathOperator{\axial}{axial}
\DeclareMathOperator{\curl}{curl}
\DeclareMathOperator{\interior}{int}
\DeclareMathOperator{\tr}{tr}

\begin{document}

\title{Forces and Variational Compatibility for Equilibrium \\
       Liquid Crystal Director Models with Coupled Electric Fields}


\titlerunning{Forces in Liquid Crystal Director Models with Coupled
  Electric Fields}

\author{Eugene C. Gartland, Jr.}


\institute{E. C. Gartland, Jr. \at
  Department of Mathematical Sciences,
  Kent State University,
  Kent, Ohio 44242, USA \\
  Tel.: +1-330-672-9112,
  Fax:+1-330-672-2209,
  \email{gartland@math.kent.edu}}

\date{\today}

\maketitle

\begin{abstract}
  Expressions are obtained for force and couple densities and stress
  tensors in macroscopic models for nematic liquid crystals subjected
  to electric fields.  The coupling between the liquid crystal
  orientational properties and the electric field is taken into
  account via a free energy of Oseen-Frank type expressed as an
  integral functional of the director field and the electric potential
  field.  The variational model here also allows for a gravitational
  field and a magnetic field, and the differences among these three
  common types of force fields (gravitational, magnetic, and electric)
  are discussed.  Also included in the free energy is a surface
  anchoring potential, and its effect on boundary traction and couple
  stress is explored.  The electric field is assumed to arise from
  electrodes held at constant potential.  Flexoelectric effects are
  included, and as a consequence, the material is no longer a linear
  dielectric medium.  It is shown that the equilibrium solutions of
  the Euler-Lagrange equations satisfy appropriate expressions of
  force balance and torque balance, obtained from a virtual-work
  principle, and a connection is made between this and ideas related
  to Noether's Theorem.  The development here builds from theories of
  Ericksen related to the notion of ``variational compatibility,'' and
  this connection is made.  Comparisons are also made to the extensive
  literature in physics and continuum mechanics on electromagnetic
  field/matter interaction.
\end{abstract}

\section{Introduction}

\label{sec:Introduction}

Our interest is in macroscopic continuum models for the orientational
properties of materials in a liquid crystal phase, a complex partially
ordered fluid phase exhibited by certain materials in certain
parameter ranges.  Such models are used at the scales of typical
devices and experiments involving these kinds of materials.  Our
primary objective is to obtain expressions for the various stress
tensors, couple stress tensors, and boundary tractions in situations
in which the material is subjected to an electric field.  These
tensors are required to describe conservation laws in the hydrodynamic
theory, and they are sometimes needed to model experiments.  Electric
fields have been used for decades to control liquid crystals, and
treating these fields as analogous to the better-understood magnetic
fields often produces adequate approximations.  There are, however,
aspects to the coupling between electric fields and liquid-crystal
orientational properties that have lacked complete understanding,
especially with respect to forces and stresses, and especially with
respect to situations in which flexoelectric polarization needs to be
taken into account.

The history of the macroscopic continuum theory of liquid crystals has
its origins in the works of Oseen \cite{oseen:33}, Zocher
\cite{zocher:33}, and Frank \cite{frank:58}.  It was brought into the
realm of modern continuum mechanics by Ericksen, who together with
Leslie developed the hydrodynamic theory (the Ericksen-Leslie
equations).  The review articles \cite{ericksen:76} and
\cite{leslie:79} contain numerous references.  The standard modern
reference in the physics literature is \cite{degennes:prost:93}; while
the standard references in the applied mathematics and continuum
mechanics literature are \cite{sonnet:virga:12,stewart:04,virga:94}.
At this level of modeling, the orientational state of the material is
characterized by a unit-length vector field $\nhat$ (referred to as
the ``director field''), which represents the average orientation of
the distinguished axis of the anisometric molecules in a fluid element
at a point.  Central to the modeling of both equilibria and dynamics
is an appropriate expression for the free energy of the system, which
serves as a work function (stored-energy function) for isothermal,
reversible processes.  In the models of interest to us, the material
is also treated as incompressible.

Liquid crystals are very responsive to external stimuli, such as
magnetic fields and electric fields, and this has been one of the keys
to their usefulness in technological applications.  Ericksen
deliberately ignored this aspect in his first papers, as he focused on
other features of liquid-crystal systems, such as their stress
tensors, which are not necessarily symmetric---these are examples of
what are known as ``polar materials''
\cite[Sec.\,98]{truesdell:noll:04}.  Ericksen finally sought to
introduce external fields into his development in \cite{ericksen:62},
where he presented sufficient conditions for the director field that
solved the equilibrium Euler-Lagrange equations associated with the
free energy to be consistent with the hydrostatic limit of the
Ericksen-Leslie equations in the presence of external force fields, an
issue which he termed ``variational compatibility.''  We discuss the
details of these ideas in Sec.\,\ref{sec:Variational_compatibility}.
They work well for systems involving gravitational and/or magnetic
fields, and this is the accepted approach in those settings.  The
additional complications associated with electric fields, however,
have been acknowledged for a long time---quoting Leslie
\cite[Sec.\,B.4]{leslie:79}: ``Although there are rather obvious
dielectric analogies for electric fields, we do not discuss this
further on account of other complications that can occur in such
cases.''  Similar cautionary notes can be found in
\cite[Sec.\,II.A.1]{ericksen:76} and
\cite[Secs.\,3.5,\,4.2.4]{stewart:04}, while the discussion of the
hydrostatics of nematics in \cite[Sec.\,3.5.1]{degennes:prost:93}
explicitly avoids electric fields and flexoelectric effects.

Three types of force fields of external origin are commonly discussed
in the context of liquid crystals: gravitational fields, magnetic
fields, and electric fields.  Gravitational fields seldom have enough
influence to warrant inclusion in a model, although there are
exceptions (see \cite[Sec.\,II.A.1]{ericksen:76}).  Magnetic fields
are commonly used in experiments, sometimes in combination with
electric fields.  Electric fields are very common in devices, as well
as in experiments.  Gravitational fields are true external fields,
i.e., independent of the state of the material at a point
\cite[Sec.\,9.01]{guggenheim:67}.  Magnetic fields are influenced by
liquid crystal materials, which are anisotropic with magnetic
susceptibilities that depend on the orientational state of the
material at a point.  For the parameter values of typical
liquid-crystal materials, however, this influence is negligible---we
elaborate below in Sec.\,\ref{sec:free-energy-density}.  Thus a
magnetic field in a liquid crystal can be treated as an external
field.  An electric field is influenced by the state of the
liquid-crystal material in a similar fashion; however the coupling is
much stronger and cannot be ignored---this also is quantified below.
In the development here, we consider all three types of force fields,
and we will see how they contribute differently to the body forces,
stresses, body couples, couple stresses, and tractions.

The literature on electromagnetic fields in matter is vast and
contains more than its share of controversies.  In the physics
literature, one can find decades-long debates about the appropriate
expressions for electromagnetic momentum and stress in dielectrics, a
prominent example being the ``Abraham-Minkowski controversy''---for a
contemporary view, see \cite{barnett:10,barnett:loudon:10,bocker:16}
and references contained therein.  In the continuum-mechanics
literature, there has been extensive development in the areas of
electromagnetics in deformable and fluent matter, tracing back to
Toupin \cite{toupin:56} and continuing to this day.  This work has
been motivated by a variety of applications, including
piezoelectricity, photoelasticity, electro/magneto-active polymers,
electrogels, electro-rheological fluids, and the like.  The models
take a variety of forms, due to several factors: different application
domains, no unique choice of basic electromagnetic variables, no
unique definition of electromagnetic stress and body force, and the
need to postulate expressions for such quantities.  More recent papers
and books have illustrated the sameness of several different-looking
models---see for example
\cite{bustamante:dorfmann:ogden:09a,bustamante:dorfmann:ogden:09b,liu:13}
and \cite[Ch.\,4]{hutter:vandeven:ursescu:06}.  We shall make contact
with this literature at points in our development.

Our setting here is simpler in several ways (isothermal,
incompressible, equilibrium, non-optical fields); yet it is more
complicated in others (inhomogeneous, anisotropic, nonlinear
dielectric medium).  A major advantage that we enjoy is that we are
able to work from an expression for the free energy that is of widely
accepted form, from which we can derive the quantities of interest,
thereby avoiding the postulational approach and associated
controversies.  Our approach is to employ a virtual work principle as
used by Ericksen in his original work \cite[Sec.\,IV]{ericksen:61}, as
recounted in \cite[Sec.\,2.4.2]{stewart:04} and
\cite[Secs.\,127--8]{truesdell:noll:04}, except that here we will take
into account the coupling between the director field and the electric
field.  The approach is similar in spirit to that found in
\cite[Sec.\,15]{landau:lifshitz:pitaevskii:93},
\cite[Sec.\,2.21]{stratton:41}, and \cite[Sec.\,10]{toupin:56}.

The paper is organized as follows.  In Section~\ref{sec:ModelProblem}
we introduce the variational model, including all fields and terms
that contribute to the free-energy density and surface anchoring
energy.  Section~\ref{sec:Variational_compatibility} concerns the
relationship between the variational equilibrium equations and the
hydrostatic limit of the Ericksen-Leslie flow equations.  There the
issue of ``variational compatibility'' and Ericksen's ``compatibility
potential'' are introduced.  Expressions for the balance of forces and
balance of torques that characterize hydrostatic equilibrium are
established in Sec.\,\ref{sec:VirtualWorkPrinciple}.  Also introduced
in that section is a virtual-work principle, from which expressions
are derived for the stress tensor, couple stress tensor, and boundary
conditions for our model.  Section~\ref{sec:Interpretation} contains
an attempt to interpret our results and to put them in context with
related results in the literature.  The final
Sec.\,\ref{sec:Conclusions} summarizes our main findings.

\section{Model Problem}

\label{sec:ModelProblem}

\begin{figure}
  \centering
  \includegraphics[width=.75\linewidth]{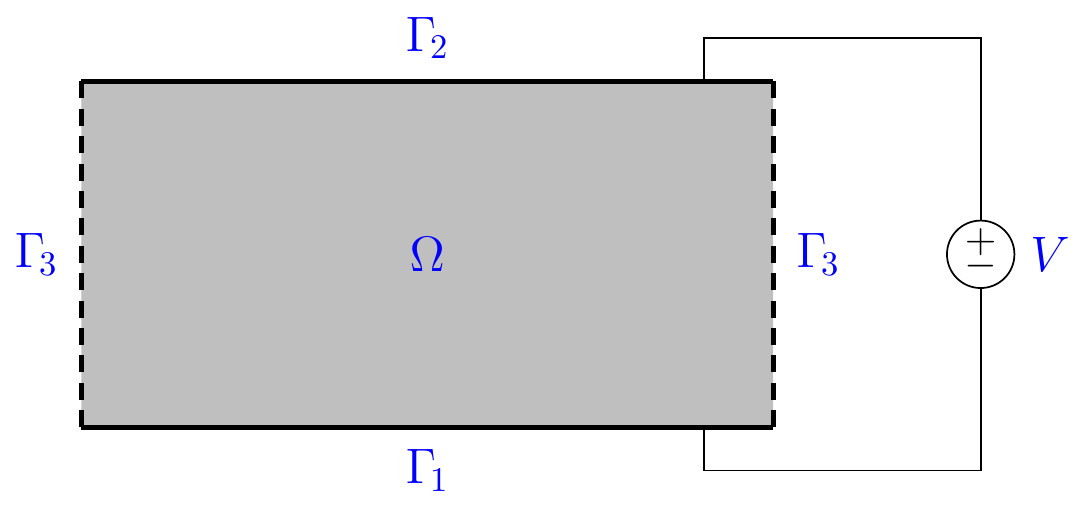}
  \caption{Model problem domain (two-dimensional depiction of actual
    three-dimensional, hexahedral domain).  Boundary conditions: on
    $\Gamma_1$ (strong anchoring, Dirichlet boundary condition
    $\nhat=\nhat_0$ on $\nhat$, $U=-V/2$), on $\Gamma_2$ (weak
    anchoring, natural boundary condition \eqref{eqn:naturalBC} on
    $\nhat$, $U=V/2$), on $\Gamma_3$ (periodic boundary conditions on
    both $\nhat$ and $U$ on opposing sides of $\Gamma_3$).}
  \label{fig:domain}
\end{figure}

We are interested in the modeling of liquid-crystal systems as found
in typical experiments with such materials: a thin-film liquid-crystal
cell sandwiched between a combination of glass substrates, alignment
layers, optical components, and conducting layers that are attached to
a variable voltage source.  The typical setup is, in essence, that of
a parallel-plate capacitor with a complex, changeable
dielectric---changes in the voltage cause changes in the orientational
state of the liquid crystal, which in turn cause changes in the
effective capacitance.  In general, the lateral dimensions of the cell
are much larger than the cell gap, which allows one to model the
behavior in the interior ignoring the influence of fringe fields and
the extended nature of the electric field.  Thus for a model problem
domain, we can take an interior subdomain $\Omega$, as shown in
Fig.\,\ref{fig:domain}---while this depiction is two dimensional, the
actual domain that we envision is three dimensional, hexahedral.

Confining substrates can be treated in various ways in order to coerce
(or at least encourage) the neighboring molecules to orient in desired
ways.  If the control exercised in this fashion is sufficiently
effective, then one can assume a ``strong anchoring condition''
(Dirichlet boundary condition) on the director; otherwise, one will
use a ``weak anchoring condition'' (a natural boundary condition
arising from a surface anchoring energy)---this is discussed in
Sec.\,\ref{sec:surface_anchoring_energy} below.  The boundary
tractions and couple stresses that we study here behave differently
with respect to strong versus weak boundary conditions, and so to
account for this, we incorporate both, imposing the following boundary
conditions on $\nhat$ in our model problem: on $\Gamma_1$ (strong
anchoring, $\nhat=\nhat_0$), on $\Gamma_2$ (weak anchoring,
\eqref{eqn:naturalBC} below), on $\Gamma_3$ (periodic)---it is common
(though not necessary) for $\nhat_0$ to be constant.  The electric
potential $U$ is constant on $\Gamma_1$ and also on $\Gamma_2$ and
must have a difference of $V$ between them, which is imposed by
requiring $U=-V/2$ on $\Gamma_1$ and $U=V/2$ on $\Gamma_2$---the
potential is relative to a value of zero at infinity.  As is the case
with $\nhat$, the potential $U$ satisfies periodic boundary conditions
on opposing sides of $\Gamma_3$.

While the model problem geometry and boundary conditions are somewhat
idealized, they embody several features that are of interest to us,
and they allow for several types of solutions that are encountered in
practice.  These include one-dimensional solutions
($\nhat = \nhat(x_3)$, uniform in $x_1$, $x_2$), two-dimensional
solutions ($\nhat = \nhat(x_1,x_3)$, periodic in $x_1$, uniform in
$x_2$, ``stripes''), and three-dimensional solutions
($\nhat = \nhat(x_1,x_2,x_3)$, periodic in $x_1$, periodic in $x_2$,
``cells'').  Here the coordinates are with respect to a frame with
$\ehat_1$ and $\ehat_2$ in the lateral, in-plane directions, and
$\ehat_3$ bridging the cell gap.  In reality, the periodicities of
such solutions would not be known a-priori but would instead be
determined in a self-consistent way with the equilibrium fields.  We
ignore this aspect, which would only add unnecessary complexity to our
efforts here.

The theories from which we work are variational in nature.  We
consider a model for a material in a nematic liquid crystal phase
(chiral or achiral) that can be expressed as an integral functional of
the director field $\nhat$ and the electric potential field $U$:
\begin{equation}\label{eqn:calF}
  \calF[\nhat,U] = \!
  \int_{\Omega} W(x,\nhat,\!\nabla\nhat,U,\!\nabla U) \, \text{d}V + \!
  \int_{\Gamma_2} \Ws(x,\nhat) \, \text{d}S .
\end{equation}
Here $\Omega$ is the region occupied by the material (as depicted in
Fig.\,\ref{fig:domain}), $W$ is the free-energy density (per unit
volume), $\Gamma_2$ is the weak-anchoring portion of the boundary of
$\Omega$, and $\Ws$ is the surface anchoring energy (per unit area).
Stationary points of $\calF$ (subject to the unit-length constraint on
$\nhat$ and appropriate boundary conditions) determine equilibrium
states of the system.  The indicated dependencies of $W$ and $\Ws$ are
sufficiently general to encompass the variety of effects of interest
to us, which we now elaborate.

\subsection{Free-energy density}

\label{sec:free-energy-density}

The free-energy densities in which we are interested may embody all or
some of the influences of distortional elasticity, an electric field,
flexoelectricity, a prescribed distribution of free electric charge, a
magnetic field, and gravity.  The relevant relations from
electrostatics (macroscopic Maxwell equations, Maxwell-Faraday theory)
are
\begin{equation*}
  \bfD = \eps_0 \bfE + \bmP , \quad
  \Div \bfD = \rhof , \quad
  \bfE = - \nabla U ,
\end{equation*}
where $\bfD$ is the electric displacement, $\eps_0$ the vacuum
permittivity, $\bfE$ the electric field, $\bmP$ the polarization, and
$\rhof$ the density of free charge \cite[Ch.\,III]{stratton:41}.  The
individual contributions to $W$ that we envision can be grouped as
follows, with dependencies indicated:
\begin{equation}\label{eqn:Wgeneral}
  W(x,\nhat,\!\nabla\nhat,U,\!\nabla U) =
  \We(\nhat,\!\nabla\nhat) + \WE(\nhat,\!\nabla\nhat,\!\nabla U) +
  \rhof(x) \, U + \WH(\nhat) + \rhom g h(x) .
\end{equation}
Here $\We$ denotes the distortional elasticity of the director field,
$\WE$ and $\WH$ stand for the free-energy contributions associated
with the electric and magnetic fields, $\rhom$ is the constant mass
density, $g$ is the gravitational constant, and the function $h$ gives
the height of the point $x$ above some reference elevation---$gh$ is
the gravitational potential per unit mass for an incompressible
material.

Through intermolecular forces, the liquid crystal director field is
encouraged to assume certain simple arrangements: uniform parallel
alignment (in the nematic phase), helical structures (in the
cholesteric or chiral nematic phase).  Other influences (such as
boundary conditions and external fields) can cause $\nhat$ to deviate
from these preferred orientations, and the distortional elasticity
$\We$ gives the energetic penalty for this.  In macroscopic models,
the most commonly used expression for $\We$ is the Oseen-Frank model
and is given by
\begin{equation*}
  2 \, \We = K_1 (\Div\nhat)^2 + K_2 (\nhat\cdot\curl\nhat+q_0)^2 +
  K_3 |\nhat\times\curl\nhat|^2 +
  K_{24} \bigl[ \tr (\nabla\nhat)^2 - (\Div\nhat)^2 \bigr] ,
\end{equation*}
with $K_1$, $K_2$, $K_3$, and $K_{24}$ the elastic constants and $q_0$
the spontaneous twist parameter (the wave number associated with the
cholesteric pitch) \cite[Sec.\,2.2]{stewart:04},
\cite[Sec.\,3.2]{virga:94}.  These parameters depend on various
factors: the type of material, the temperature, the concentration of
chiral dopant, etc.  Different symmetries are associated with the
presence or absence of $q_0$: the case $q_0=0$ corresponds to the
``nematic energy,'' which satisfies
\begin{equation*}
  \We(\bfQ\nhat,\bfQ\nabla\nhat\bfQ^T\!) =
  \We(\nhat,\nabla\nhat) , \quad \forall \, \bfQ \in O(\calV) ,
\end{equation*}
while $q_0\not=0$ gives the ``cholesteric energy,'' which satisfies
\begin{equation}\label{eqn:SO3_invar}
  \We(\bfR\nhat,\bfR\nabla\nhat\bfR^T\!) =
  \We(\nhat,\nabla\nhat) , \quad \forall \, \bfR \in SO(\calV) .
\end{equation}
Here $O(\calV)$ and $SO(\calV)$ denote the orthogonal group and the
special orthogonal group on the underlying vector space $\calV$\!.  We
note that while the $SO(\calV)$ invariance holds for the full
expression $\We$ above, it is generally the case that one takes
$K_{24}=0$ in the cholesteric energy---the reasons for this are given
in \cite[Sec.\,2.2.2]{stewart:04} and \cite[Sec.\,3.2
Remark~2]{virga:94}.  The simplest prototype $\We$ is given by the
``equal elastic constants model'' ($q_0=0$, $K_1=K_2=K_3=K_{24}= K$):
\begin{equation*}
  \We = \frac12 K | \nabla\nhat |^2 .
\end{equation*}
The precise form of $\We$ does not matter in our development.  All
that matters is that $\We$ depends only on $\nhat$ and $\nabla\nhat$
and that it is properly frame indifferent (i.e., satisfies
\eqref{eqn:SO3_invar} above).  Distortion of the director field can
lead to the creation and transmission of forces and torques
\cite[Secs.\,3.1.5,\,3.5.2]{degennes:prost:93}, and one of our main
objectives here is to determine expressions for the stress tensor and
couple stress tensor related to this in a more general context than
has been considered in the past.  In the hydrodynamic setting,
director distortion can induce fluid motion, and fluid motion can
induce reorientation of the director field---here, however, we
restrict our attention to statics.

The most common situation for liquid-crystal devices and experiments
involving electric fields is for the fields to be produced by
electrodes held at constant potential by some voltage source (or
battery), for which the contribution to the free energy is usually
found written in the form
\begin{equation}\label{eqn:WEgeneral}
  \WE = - \int_{\bfzero}^{\bfE} \!\! \bfD \cdot \text{d} \bfE .
\end{equation}
See for example \cite[Sec.\,3.3.1]{degennes:prost:93} or
\cite[Sec.\,6.2]{lagerwall:99} or \cite[Sec.\,2.3]{stewart:04}.  This
expression gives the electrostatic free-energy density at a point
expressed in terms of the intensity of the electric field as the field
is built up incrementally, reversibly from $\bfE = \bfzero$ to its
final value.  The equivalent relation is given in differential form in
\cite[Eqn.\,(10.9)]{landau:lifshitz:pitaevskii:93}.  To determine
$\WE$ in a particular context, one must prescribe the dependence of
$\bfD$ on $\bfE$.  Here we assume that the polarization $\bmP$
consists of an induced polarization that is linear in $\bfE$ plus a
flexoelectric polarization (which is caused by the distortion of the
director field itself):
\begin{equation*}
  \bmP = \eps_0 \bfchie \bfE + \Pf ~~ \Rightarrow ~~
  \bfD = \epstensor \bfE + \Pf , ~~ \epstensor = \eps_0 ( \bfI + \bfchie ) .
\end{equation*}
Here $\bfchie$ is the electric susceptibility tensor and $\epstensor$ the
dielectric permittivity tensor, which for a uniaxial nematic liquid
crystal has the ``transversely isotropic'' form
\begin{equation}\label{eqn:eps-tensor}
  \epstensor = \eps_0 \bigl[ \eperp \bfI + \epsa ( \nhat\otimes\nhat ) \bigr] ,
  \quad \epsa := \epara - \eperp ,
\end{equation}
with $\epara$ and $\eperp$ the relative permittivities parallel to
$\nhat$ and perpendicular to $\nhat$.  The tensor field $\epstensor$ is
real, symmetric, and positive definite, and the medium is
inhomogeneous and anisotropic in general.  Linearity of the induced
polarization is a valid assumption for the most common situations,
though it is acknowledged that it holds more properly in the frequency
domain \cite[Sec.\,I.4]{jackson:75}.  The flexoelectric polarization
is taken in the form
\begin{equation}\label{eqn:Pf}
  \Pf = \es (\Div\nhat) \nhat + \eb \, \nhat\times\curl\nhat ,
\end{equation}
with $\es$ and $\eb$ the ``splay'' and ``bend'' flexoelectric
coefficients \cite[Sec.\,4.1]{lagerwall:99}.

With this model for $\bfD(\bfE)$, one can compute $\WE$ (most readily
by expressing all fields in terms of an orthonormal basis for $\epstensor$
at the point):
\begin{equation}\label{eqn:WE}
  \WE = - \int_{\bfzero}^{\bfE} \! ( \epstensor \bfE + \Pf ) \cdot \text{d} \bfE =
  - \frac12 \epstensor \bfE \cdot \bfE - \Pf \cdot \bfE =
  - \frac12 \epstensor \nabla U \cdot \nabla U + \Pf \cdot \nabla U .
\end{equation}
See for example \cite[Sec.\,II.C]{mottram:newton:14}.  If
flexoelectric terms are ignored (taking $\Pf = \bfzero$), this
simplifies to the more familiar form
\begin{equation*}
  \WE = - \frac12 \bfD \cdot \bfE , \quad \bfD = \epstensor \bfE .
\end{equation*}
We note that the inclusion of flexoelectric terms formally takes us
outside the realm of a ``linear dielectric medium'' (since
$\bfD=\epstensor\bfE+\Pf$ is an affine function of $\bfE$ and not the
result of a linear transformation acting on $\bfE$).  We also note
that the presence of $\Pf$ creates a coupling between the electric
field and $\nabla\nhat$, in addition to the coupling between $\bfE$
and $\nhat$ already present in the simplified expression above.

The potential associated with the magnetic field can be taken in the
form
\begin{equation}\label{eqn:WH}
  \WH = - \frac12 \mu_0 \Delta\chi (\bfH\cdot\nhat)^2 , \quad
  \Delta\chi := \chipara - \chiperp ,
\end{equation}
with $\mu_0$ the free-space permeability, $\chipara$ and $\chiperp$
the magnetic susceptibilities parallel to $\nhat$ and perpendicular to
$\nhat$, and $\bfH$ the magnetic field (which can be assumed to be
constant, uniform throughout the material)
\cite[Sec.\,2.3]{stewart:04}, \cite[Sec.\,4.1]{virga:94}.  In reality,
$\bfH$ is affected by the medium, but at a negligible level for
typical liquid crystal materials, as we have noted in
Sec.\,\ref{sec:Introduction}.  This can be seen from the relevant
formulas of magnetostatics for a uniaxial nematic:
\begin{equation}\label{eqn:BM}
  \begin{gathered}
    \bfB = \mu_0 ( \bfM + \bfH ) , \quad
    \bfM = \bfchim \bfH , \quad
    \bfchim = \chiperp \bfI + \Delta \chi ( \nhat \otimes \nhat ) , \\
    \WH = - \frac12 \bfB \cdot \bfH = - \frac12 \mu_0 \bigl[
    ( 1+ \chiperp ) H^2 + \Delta \chi ( \bfH \cdot \nhat )^2 \bigr] .
  \end{gathered}
\end{equation}
Here $\bfB$ is the magnetic induction, $\bfM$ the magnetization, and
$\bfchim$ the magnetic susceptibility tensor
\cite[Ch.\,IV]{stratton:41}.  In terms of the scalar magnetic
potential $U_\text{m}$ (with $\bfH = - \nabla U_\text{m}$), the
Maxwell equation for the magnetic induction ($\Div\bfB=0$) takes the
form
\begin{equation*}
  \Delta U_\text{m} + \frac{\Delta\chi}{1 + \chiperp}
  \Div \bigl[ ( \nabla U_\text{m} \cdot \nhat ) \nhat \bigr] = 0 ,
\end{equation*}
and the term involving $\nhat$ is negligible (at our level of
modeling) by virtue of the fact that
$\Delta \chi / ( 1 + \chiperp ) = O\bigl(10^{-6}\bigr)$ for typical
liquid crystals---see \cite[Sec.\,3.2.1]{degennes:prost:93} or
\cite[Appendix D]{stewart:04}.  The analogous equation for the
electric potential $U$ (from $\Div\bfD=\rhof$) is
\begin{equation*}
  \Delta U + \frac{\epsa}{~\eperp}
  \Div \bigl[ ( \nabla U \cdot \nhat ) \nhat \bigr] =
  \frac1{~\eps_0\eperp} \bigl( \Div\Pf - \rhof \bigr) ,
\end{equation*}
and the coupling term satisfies $\epsa / \eperp = O(1)$---again see
\cite[Appendix D]{stewart:04}.  For this reason, the influence of the
medium on the electric field cannot be neglected---the electric field
influences the director field and vice versa, and the two fields must
be determined in a coupled, self-consistent way
(cf.~\cite[Sec.\,3.5]{stewart:04}).  The term
$-\frac12 \mu_0 ( 1 + \chiperp ) H^2$ in the expression for
$-\frac12 \bfB \cdot \bfH$ above would just add a constant to $W$ and
is generally discarded.  We note that while $\nabla\bfH$ is negligibly
small in typical liquid crystal settings (the justification for
discarding the term $-\frac12 \mu_0 ( 1 + \chiperp ) H^2$), if
circumstances were such that it needed to be taken into account, then
it would contribute to the body force---see
\cite[Sec.\,3.1.4]{sonnet:virga:12} or \cite[Sec.\,2.4.2]{stewart:04}.

The term $\rhof U$ is not common but is included here for the purpose
of illustration.  For the systems we are interested in modeling, the
only free-charge distributions typically encountered are very small
concentrations of ionic impurities, which are usually dealt with in
one of two ways.  If DC electric fields are employed, then the mobile
ions will cluster into very thin screening layers adjacent to the
electrodes of opposite polarity.  This can be modeled using
Poisson-Boltzmann theory, though simpler approximations are usually
employed---see \cite[Ch.\,7]{barbero:evangelista:06} or
\cite[Sec.\,8.5]{jakli:saupe:06}.  As an alternative, one could use AC
fields at a frequency such that the force on the ions ($\rhof\bfE$)
effectively time-averages to zero, while the director (which couples
to $E^2$) effectively responds to the field associated with the RMS
voltage.  In a similar vein, the gravitational potential $\rhom gh$ is
not usually encountered but is included for illustration.  All of the
other terms discussed above are common.  The simplest prototype of
\eqref{eqn:Wgeneral} that embodies liquid-crystal distortional
elasticity and a coupled electric field is
\begin{equation*}
  W(\nhat,\!\nabla\nhat,\!\nabla U) =
  \frac12 K | \nabla\nhat |^2 -
  \frac12 \epstensor(\nhat) \nabla U \cdot \nabla U ,
\end{equation*}
with the dielectric tensor $\epstensor$ having the form
\eqref{eqn:eps-tensor}.  In most of what follows, we simply work with
$W$ in the generic form $W(x,\nhat,\!\nabla\nhat,U,\!\nabla U)$.

\subsection{Surface anchoring energy}

\label{sec:surface_anchoring_energy}

The surface anchoring energy $\Ws$ can take a variety of forms, as
there are a number of different techniques used to try to control the
orientational properties of a liquid-crystal material at an interface
to a substrate---see for example
\cite{barbero:evangelista:06,guo:zheng:palffy:19,%
  sluckin:95,sluckin:poniewierski:86,sonin:95}.  A reasonable model is
to assume that $\Ws$ depends on the director, the normal to the
surface, and a distinguished direction tangent to the surface at the
point.  Thus we assume
\begin{equation*}
  \Ws = \Ws(x,\nhat;\nuhat,\tauhat) ,
\end{equation*}
where $\nuhat$ is the unit normal to the surface at the point $x$ and
$\tauhat$ is a unit-vector tangent to the surface at that point.  We
note that in circumstances in which curvature effects of the anchoring
surface are significant, that would warrant the inclusion of
additional terms in $\Ws$ involving surface gradients of $\nhat$
and/or $\nuhat$ and/or $\tauhat$---see \cite{faetti:virga:97} and
references contained therein.

The anchoring potential should conform to the same symmetries as the
free-energy density: evenness,
\begin{equation*}
  \Ws(x,-\nhat;\nuhat,\tauhat) = \Ws(x,\nhat;\nuhat,\tauhat) ,
\end{equation*}
and either invariance with respect to the full orthogonal group (case
of ``nematic symmetry''),
\begin{equation*}
  \Ws(x,\bfQ\nhat;\bfQ\nuhat,\bfQ\tauhat) = \Ws(x,\nhat;\nuhat,\tauhat) ,
  \quad \forall \, \bfQ \in O(\calV) ,
\end{equation*}
or invariance with respect to only proper orthogonal transformations
(case of ``cholesteric symmetry''),
\begin{equation*}
  \Ws(x,\bfR\nhat;\bfR\nuhat,\bfR\tauhat) = \Ws(x,\nhat;\nuhat,\tauhat) ,
  \quad \forall \, \bfR \in SO(\calV) .
\end{equation*}
Representation theory tells us that in the former case, $\Ws$ must be
of the form
\begin{equation*}
  \Ws(x,\nhat;\nuhat,\tauhat) =
  \tildeWs(x,\nhat\cdot\nuhat,\nhat\cdot\tauhat) ,
\end{equation*}
for some function $\tildeWs$ that satisfies
\begin{equation*}
  \tildeWs(\Arg_1,-\Arg_2,-\Arg_3) = \tildeWs(\Arg_1,\Arg_2,\Arg_3) ,
\end{equation*}
while in the latter case, we must have
\begin{gather*}
  \Ws(x,\nhat;\nuhat,\tauhat) =
  \tildeWs(x,\nhat\cdot\nuhat,\nhat\cdot\tauhat,\nhat\cdot\tautwo) , \quad
  \tautwo := \nuhat \times \tauhat \\
  \tildeWs(\Arg_1,-\Arg_2,-\Arg_3,-\Arg_4) =
  \tildeWs(\Arg_1,\Arg_2,\Arg_3,\Arg_4) .
\end{gather*}
See \cite[Sec.\,11]{truesdell:noll:04}.

A simple example of such an energy (which obeys the full nematic
symmetry) is
\begin{equation}\label{eqn:Ws-example}
  \begin{aligned}
    \Ws &= -\frac12 W_0 (\nhat\cdot\nhat_0)^2 , \quad
           \nhat_0 = a \, \nuhat + b \, \tauhat , \quad a^2 + b^2 = 1 \\
        &= -\frac12 W_0 \bigl[
           a^2 (\nhat\cdot\nuhat)^2 +
           2ab (\nhat\cdot\nuhat) (\nhat\cdot\tauhat) +
           b^2 (\nhat\cdot\tauhat)^2 \bigr] .
  \end{aligned}
\end{equation}
Here the positive constant $W_0$ is the ``anchoring strength'' and
$\nhat_0$ the ``easy axis.''  The potential simply encourages the
director to align with $\pm\nhat_0$ and penalizes it if it doesn't.
Among other possibilities, such an energy allows one to model a weak
``pre-tilt'' boundary condition, in which the preferred orientation
$\nhat_0$ is tilted some amount in a certain direction from the
normal.  The $x$ dependence that we envision is through the anchoring
coefficients.  While parameters such as $W_0$ above are almost always
taken to be constants, it is conceivable that if the anchoring
strength were known to vary appreciably across the surface, then one
may want to model $W_0=W_0(x)$.

For the situation in which the in-plane vector $\tauhat$ is associated
with the ``rubbing direction'' of grooves rubbed into a polymer
alignment layer, the anchoring energy should be indifferent to
$\tauhat$ versus $-\tauhat$, as discussed in
\cite{guo:zheng:palffy:19}.  In that case, we would require the
additional symmetry
\begin{equation*}
  \Ws ( x, \nhat ; \nuhat, - \tauhat ) =
  \Ws ( x, \nhat ; \nuhat, \tauhat ) ,
\end{equation*}
which would necessitate that the nematic function satisfy both
\begin{gather*}
  \tildeWs(\Arg_1,-\Arg_2,\Arg_3) = \tildeWs(\Arg_1,\Arg_2,\Arg_3) \\
  \tildeWs(\Arg_1,\Arg_2,-\Arg_3) = \tildeWs(\Arg_1,\Arg_2,\Arg_3)
\end{gather*}
and that the cholesteric function obey
\begin{gather*}
  \tildeWs(\Arg_1,-\Arg_2,\Arg_3,\Arg_4) =
  \tildeWs(\Arg_1,\Arg_2,\Arg_3,\Arg_4) \\
  \tildeWs(\Arg_1,\Arg_2,-\Arg_3,-\Arg_4) =
  \tildeWs(\Arg_1,\Arg_2,\Arg_3,\Arg_4) .
\end{gather*}

The difference between the types of anchoring energies allowed in the
case of nematic symmetry versus cholesteric symmetry (for the models
invariant with respect to $\pm\tauhat$) can be illustrated by example.
Consider the case of weak homeotropic anchoring, with easy axis
$\nhat_0=\nuhat$.  For the common situation of polynomial expressions
of at most quadratic order, the set of admissible terms comprises
\begin{equation*}
  (\nhat\cdot\nuhat)^2 , ~~
  (\nhat\cdot\tauhat)^2 , ~~
  (\nhat\cdot\tautwo)^2 , ~~
  (\nhat\cdot\tauhat) (\nhat\cdot\tautwo) ,
\end{equation*}
with only two of the first three terms being independent, by virtue of
the relation
\begin{equation*}
  (\nhat\cdot\nuhat)^2 + (\nhat\cdot\tauhat)^2 +
  (\nhat\cdot\tautwo)^2 = 1 .
\end{equation*}
The first three admissible terms satisfy the nematic symmetry
assumptions (since $(\nhat\cdot\tautwo)^2 = 1 - (\nhat\cdot\nuhat)^2 -
(\nhat\cdot\tauhat)^2$), while the fourth term does not.  Thus one can
construct nematic anchoring energies of the form
\begin{equation*}
  \Ws = - \frac12 W_0 (\nhat\cdot\nuhat)^2 ~~~ \text{or} ~~~
  \Ws = \frac12 W_1 (\nhat\cdot\tauhat)^2 +
        \frac12 W_2 (\nhat\cdot\tautwo)^2 .
\end{equation*}
With $W_0$, $W_1$, and $W_2$ all positive, both energies attain their
minimum values at $\nhat=\pm\nuhat$.  The level sets of the first
energy (projected onto the tangent plane to the surface at the point)
are circles, while those of the second are ellipses with axes aligned
with $\tauhat$ and $\tautwo$.
Such potentials are said to be of ``Rapini-Papoular type'' (see for
example \cite[Sec.\,2.2]{barbero:evangelista:06}).  A slight
generalization of this model, which obeys the cholesteric symmetry but
not the nematic symmetry, is
\begin{equation*}
  \Ws = \frac12 W_1 (\nhat\cdot\tauhat)^2 +
  \frac12 W_2 (\nhat\cdot\tautwo)^2 +
  W_3 (\nhat\cdot\tauhat) (\nhat\cdot\tautwo) .
\end{equation*}
With $W_1, W_2 > 0$ and $W_3^2 < W_1 W_2$, this potential still has
its minimum at $\nhat=\pm\nuhat$, but the $W_3$ term penalizes the
twist differently in different directions (clockwise versus
counterclockwise) and causes the axes of the elliptical level sets to
be rotated relative to $\tauhat$ and $\tautwo$.  More aspects of the
physical significance of such a model are discussed in
\cite{guo:zheng:palffy:19}.

The analogous case of weak planar anchoring can be expressed
\begin{equation*}
  \Ws = - \frac12 W_1 (\nhat\cdot\tauhat)^2 -
  \frac12 W_2 (\nhat\cdot\tautwo)^2 +
  W_3 (\nhat\cdot\tauhat) (\nhat\cdot\tautwo) , \quad
  W_1, W_2 \ge 0 , \quad W_3^2 \le W_1 W_2 ,
\end{equation*}
with $W_1$ and $W_2$ not both zero.  The easy axis would be determined
by the weights:
$W_1 > W_2$ and $W_3=0 ~ \Rightarrow ~ \nhat_{\text{min}} = \pm \tauhat$,
$W_2 > W_1$ and $W_3=0 ~ \Rightarrow ~ \nhat_{\text{min}} = \pm \tautwo$.
With $W_3 \not= 0$, the easy axis would lie in the plane
$\operatorname{span}\{\tauhat,\tautwo\}$ but would not be aligned with
either of those directions.


The various representations for $\Ws$ have consequences concerning
quantities such as $\partial\Ws/\partial\nhat$,
$\partial\Ws/\partial\nuhat$, and $\partial\Ws/\partial\tauhat$, but
we shall not require them in what follows and therefore omit those
details.  In our development, we can assume the more general
cholesteric symmetry (with or without the assumption of invariance
with respect to $\pm\tauhat$), viewing the corresponding nematic
energy as a special case.  For the most part, we simply work with the
generic expression $\Ws(x,\nhat;\nuhat,\tauhat)$, which is capable of
modeling most of the cases of interest.

\section{Orientational Equilibrium and the Issue of Variational Compatibility}

\label{sec:Variational_compatibility}

The issue of variational compatibility concerns the following.  The
Euler-Lagrange equations associated with the functional
\eqref{eqn:calF}, when combined with appropriate boundary conditions
and the constraint $|\nhat|=1$, completely determine the equilibrium
director field $\nhat$ and electric potential field $U$.  These
solutions, however, must also satisfy the static limit of the
Ericksen-Leslie equations for liquid crystal flows.  In this section,
we frame this question and present Ericksen's suggested approach to
resolving it.

\subsection{Euler-Lagrange equations}

The equilibrium Euler-Lagrange equations associated with a functional
of the form \eqref{eqn:calF} follow from the variations
$\delta_\nhat\calF = 0$ (subject to $|\nhat|=1$) and $\delta_U\!\calF
= 0$, from which one obtains
\begin{subequations}
\begin{gather}
  \label{eqn:E-L}
  \Div \Bigl( \dWdgn \Bigr) -
  \dWdn + \lambda \nhat = \bfzero , \quad
  \Div \Bigl( \frac{\partial W}{\partial\nabla U} \Bigr) -
  \frac{\partial W}{\partial U} = 0 , \quad \text{in } \Omega \\
  \label{eqn:naturalBC}
  \Bigl( \dWdgn \Bigr) \nuhat -
  \frac{\partial\Ws}{\partial\nhat} + \mu \nhat = \bfzero , \quad
  \text{on } \Gamma_2 .
\end{gather}
\end{subequations}
Here \eqref{eqn:naturalBC} is the natural boundary condition
associated with the weak anchoring potential, and $\lambda$ and $\mu$
are Lagrange-multiplier fields associated with the pointwise
unit-length constraint on $\nhat$.
Equation $(\text{\ref{eqn:E-L}})_2$ can be seen to yield the
appropriate electrostatics equation: using \eqref{eqn:Wgeneral} and
\eqref{eqn:WE},
\begin{equation*}
  \frac{\partial W}{\partial\nabla U} =
  - \epstensor \nabla U + \Pf = \epstensor \bfE + \Pf = \bfD , \quad
  \frac{\partial W}{\partial U} = \rhof ,
\end{equation*}
from which follows
\begin{equation*}
  \Div \Bigl( \frac{\partial W}{\partial\nabla U} \Bigr) -
  \frac{\partial W}{\partial U} = 0 ~~ \Rightarrow ~~
  \Div \bfD = \rhof .
\end{equation*}
We note that in the thermodynamic formalism, $\bfD$ and $-\bfE$ are
conjugate variables \cite[Sec.\,10]{landau:lifshitz:pitaevskii:93}; so
the relationship
$\partial W / \partial \nabla U = - \partial W / \partial \bfE =
\bfD$
is expected (and in fact is built into \eqref{eqn:WEgeneral}).

The Euler-Lagrange equations \eqref{eqn:E-L} plus
\eqref{eqn:naturalBC} and the essential boundary conditions completely
determine equilibrium director fields $\nhat$ and electric potential
fields $U$ in a coupled, self-consistent way.  The problem is
nonlinear, however, and so it is possible for there to be more than
one distinct solution pair for a given set of parameters.  In this
case, any solution pair of least free energy provides a globally
stable state of the system.  The issue of ``variational
compatibility'' concerns the consistency of these solutions with the
static limit of the Ericksen-Leslie equations, which form the
hydrodynamic theory of liquid crystal flows at this modeling scale.

\subsection{Static limit of Ericksen-Leslie hydrodynamics}

The hydrostatic limit of the Ericksen-Leslie equations is discussed in
\cite[Sec.\,3.5]{degennes:prost:93},
\cite[Sec.\,3.1]{sonnet:virga:12}, and \cite[Sec.\,2.4]{stewart:04}.
In this limit, the hydrodynamic system reduces to a balance of forces
(associated with conservation of linear momentum) and a balance of
torques (associated with conservation of angular momentum).  In the
classical setting, these limiting equations are usually written
\begin{subequations}
  \begin{gather}
    \Div \bfT + \bfF = \bfzero \label{eqn:lin_mom} \\
    \Div \Bigl( \dWedgn \Bigr) -
    \dWedn + \bfG + \lambda \nhat = \bfzero . \label{eqn:ang_mom}
  \end{gather}
\end{subequations}
Here $\bfT$ is the Cauchy stress tensor (associated with contact
forces)
\begin{equation*}
  \bfT = - p \, \bfI - (\nabla\nhat)^T
  \dWedgn ,
\end{equation*}
$\bfF$ is the external body force (force per unit volume), $\bfG$ is
the external generalized force acting on the director (force per unit
area)---the torque density (torque per unit volume) is
$\nhat \times \bfG$---the Lagrange multiplier field $\lambda$ is
associated with the pointwise constraint $|\nhat|=1$, and $p$ is the
pressure field associated with incompressibility.  The fields $\bfF$
and $\bfG$ are prescribed and depend on the particular system being
studied.  One can readily interpret \eqref{eqn:lin_mom} as a force
balance.  The actual torque balance takes a form different from
\eqref{eqn:ang_mom}; we shall encounter it later in our development.
To show that \eqref{eqn:ang_mom} is equivalent to the torque balance
requires some effort but is well documented---see the reference cited
above.

For a free-energy density $W$ of the form \eqref{eqn:Wgeneral}, the
director equilibrium equation $(\text{\ref{eqn:E-L}})_1$ is consistent
with \eqref{eqn:ang_mom} if we identify $\bfG$ as
\begin{equation}\label{eqn:bfG}
  \bfG =
  \Div \Bigl[ \frac{\partial(W-\We)}{\partial\nabla\nhat} \Bigr] -
  \frac{\partial(W-\We)}{\partial\nhat} =
  \Div \Bigl( \frac{\partial\WE}{\partial\nabla\nhat} \Bigr) -
  \frac{\partial\WE}{\partial\nhat} -
  \frac{\partial\WH}{\partial\nhat} .
\end{equation}
In the absence of flexoelectric terms, we would have
$\partial\WE / \partial\nabla\nhat = \bfzero$ and a reduced expression
for $\partial\WE/\partial\nhat$, and $\bfG$ would take the familiar
form
\begin{equation}\label{eqn:GEH}
  \bfG = \eps_0 \epsa ( \bfE \cdot \nhat ) \bfE +
  \mu_0 \Delta\chi ( \bfH \cdot \nhat ) \bfH ,
\end{equation}
with $\nhat\times\bfG$ comprising the dielectric and magnetic torques.
The full expression for $\bfG$ (with flexoelectric terms included) is
less intuitive.
The flexoelectric contributions are complicated: distortion of the
director field $\nhat$ induces the flexoelectric polarization $\Pf$,
as given in \eqref{eqn:Pf}, which the system strives to align with
$\bfE$ in order to minimize the term $-\Pf\cdot\bfE$ in $W$.  It is
also the case that an imposed electric field can induce distortion of
$\nhat$ so as to minimize $- \Pf \cdot \bfE$ as well (the so-called
``inverse flexoelectric effect'').
In the literature, a fair amount of attention has been given to
various aspects of flexoelectricity in nematics.  Some discussion and
references can be found in \cite[Sec.\,3.3.2]{degennes:prost:93},
\cite[Sec.\,8.7]{jakli:saupe:06}, and \cite[Ch.\,4]{lagerwall:99}.
Most of these issues are not important for our development.


It is not clear that with $\bfG$ as in \eqref{eqn:bfG} with
flexoelectric terms included that \eqref{eqn:ang_mom} would still
imply an appropriate balance of torques.  An argument would be
required, as the system now contains terms and couplings that were not
present in the original analysis relating \eqref{eqn:ang_mom} to
torque balance.  We shall see later in our development that working
from a free energy that contains the potentials for all fields of
external origin leads automatically to the satisfaction of the
appropriate force and torque balances, thereby guaranteeing
variational compatibility.  This is one of the attractive features of
our approach.

\subsection{Ericksen's compatibility potential}

\label{sec:Ericksen_compatability_potential}

It is conceivable that for certain choices of $\bfF$ in
\eqref{eqn:lin_mom} and $\bfG$ in \eqref{eqn:ang_mom}, the variational
equilibrium fields do not provide a consistent system for the
determination of the pressure field $p$.  This is referred to as the
issue of ``variational compatibility.''  It was addressed by Ericksen
\cite{ericksen:62} and is recounted in
\cite[Sec.\,3.1.4]{sonnet:virga:12} and \cite[Sec.\,2.4]{stewart:04}.
A sufficient condition for variational compatibility is that there
exist a function $\Psi(x,\nhat)$ (the ``compatibility potential'')
that gives rise to both $\bfF$ and $\bfG$ via
\begin{equation*}
  \bfF = \frac{\partial\Psi}{\partial x} , \quad
  \bfG = \frac{\partial\Psi}{\partial\nhat} .
\end{equation*}
Here the partial derivatives are meant to be computed with $x$ and
$\nhat$ treated as separate independent variables---the $x$ dependence
of $\nhat$ is ignored in computing $\partial\Psi/\partial x$.
If such a potential $\Psi$ exists, then it can be shown that
consistency of \eqref{eqn:lin_mom} and \eqref{eqn:ang_mom} is assured
with $p = p_0 - \We + \Psi$, $p_0 = \text{const}$ (see
\cite{ericksen:62} or \cite[Sec.\,3.1.4]{sonnet:virga:12} or
\cite[Sec.\,2.4]{stewart:04}).  If one simply incorporates the
potential $\Psi$ into the free-energy density,
\begin{equation*}
  W(x,\nhat,\!\nabla\nhat) = \We(\nhat,\!\nabla\nhat) - \Psi(x,\nhat) ,
\end{equation*}
then the balances \eqref{eqn:lin_mom} and \eqref{eqn:ang_mom} result
naturally, as suggested by Ericksen himself later
\cite[Sec.\,II.A.1]{ericksen:76}.

In \cite[Sec.\,3.1.4]{sonnet:virga:12} and
\cite[Sec.\,2.4]{stewart:04}, it is shown that the case of a uniform
external magnetic field fits comfortably into this framework, using
\begin{equation*}
  \Psi = \frac12 \mu_0 \Delta \chi ( \bfH \cdot \nhat )^2 ,
\end{equation*}
as we have already used in \eqref{eqn:Wgeneral}.
A gravitational body force fits comfortably into Ericksen's framework
as well \cite[Sec.\,4.2.4]{stewart:04}, with a compatibility potential
\begin{equation*}
  \Psi = - \rhom g h(x) ,
\end{equation*}
as we have also used.
The case of an external electric field is more complicated and does
not fit as naturally into Ericksen's framework.  The electric field is
coupled to the director field (and to its gradient, if flexoelectric
terms are included) and enters the free energy density as a dependent
variable through $U$ and $\nabla U$.  At least that is the case in the
approach taken here---a possible alternate approach would be to treat
electrostatics as a PDE constraint.  While a magnetic field and a
gravitational field can be regarded as ``external'' fields, an
electric field cannot.  We now describe the extension of these ideas
that serves to handle the general case, in essence following the
suggestion of Ericksen, though in a more general context.

\section{Virtual Work Principle}

\label{sec:VirtualWorkPrinciple}

To determine the quantities of interest to us (body forces and
couples, stress tensors, boundary tractions), we follow the same path
as originally followed by Ericksen \cite{ericksen:61} (as recounted in
\cite[Sec.\,2.4.2]{stewart:04} and
\cite[Secs.\,127--8]{truesdell:noll:04}).  Before involving boundary
conditions and developing general expressions for variations, we
determine the forms of the balance of forces and balance of torques
that pertain to our system.

\subsection{Force balance and torque balance}

\label{sec:force:torque:balance}

The free energy of an arbitrary subregion $V\subset\interior(\Omega)$
is given by
\begin{equation}\label{eqn:calFV}
  \calF[\nhat,U;V] =
  \int_V W(x,\nhat,\!\nabla\nhat,U,\!\nabla U) \, \text{d}V ,
\end{equation}
with $W$ as in \eqref{eqn:Wgeneral}.  We consider small displacements
of the form
\begin{gather}
  x^*\! = \Phi_{\eps}(x) = x + \eps \bfphi(x) + o(\eps) \notag \\
  \nhat^*\!(x^*\!) = \nhat(x) + \eps \bfpsi(x) + o(\eps)
  \label{eqn:displacements} \\
  U^*\!(x^*\!) = U(x) + \eps \Psi(x) + o(\eps) . \notag
\end{gather}
Here we follow the notation of \cite[Sec.\,37]{gelfand:fomin:63}.  The
perturbations $\bfphi$, $\bfpsi$, and $\Psi$ can depend on $\nhat$,
$\nabla\nhat$, $U$, and $\nabla U$, though we do not express this.  In
the notation of \cite{ericksen:61}, we would have
\begin{equation*}
  \delta\bm{x} = \eps\bfphi , \quad
  \Delta\nhat = \eps\bfpsi , \quad
  \Delta U = \eps\Psi .
\end{equation*}
The change in $\calF$ caused by such a displacement is
\begin{equation*}
  \Delta\calF = \calF[\nhat^*\!,U^*\!;V^*] - \calF[\nhat,U;V] , \quad
  V^*\! = \Phi_{\eps}(V) ,
\end{equation*}
with the variation $\delta\!\calF$ the first-order part
\begin{equation*}
  \delta\!\calF = \eps \frac{d}{d\eps}
  \calF[\nhat^*\!,U^*\!;V^*] \bigr|_{\eps=0} \,,
\end{equation*}
which is a linear form in $\bfphi$, $\bfpsi$, and $\Psi$.

The state of the system in $V$ is also affected by influences from
outside of $V$\!, which can include forces and fluxes of various
types.  In equilibrium, all influences must be in balance, in some
appropriate sense.  The Principle of Virtual Work employed in
\cite{ericksen:61} and utilized here embodies this by expressing the
characterization that $(\nhat,U)$ is an equilibrium pair if and only
if
\begin{equation}\label{eqn:virtual_work}
  \delta\!\calF = \eps \! \int_S (
  \bft\cdot\bfphi + \bfs\cdot\bfpsi + r \, \Psi ) \, \text{d}S ,
\end{equation}
for all virtual displacements consistent with the constraints of
incompressibility and $|\nhat|=1$.  Here $S$ is the boundary of $V$\!,
and $\bft$, $\bfs$, and $r$ represent generalized force densities,
which are to be identified.  The physical dimensions of $\bft$,
$\bfs$, and $r$ are force per unit area, couple per unit area, and
charge per unit area, which leads one to anticipate the forms they
will take.  Such a principle would ordinarily also involve work
associated with volume densities, supported in $V$\!, in addition to
the densities supported on $S$.  These are not present in our
principle because all such influences have been built into $W$.  For
\eqref{eqn:displacements} to correspond to a virtual displacement
``consistent with constraints'' means that it must conform to the
constraints to first order.  Incompressibility demands that
displacements preserve volume pointwise (be isochoric), which implies
$\Div\bfphi=0$; while satisfaction of $|\nhat|=1$ to first order
requires $\nhat\cdot\bfpsi=0$ at each point.

The expression \eqref{eqn:virtual_work} can be contrasted with that
originally employed by Ericksen, which in our notation would take the
form
\begin{equation*}
  \delta \! \int_V \We \, \text{d}V =
  \eps \! \int_V ( \bfF\cdot\bfphi + \bfG\cdot\bfpsi ) \, \text{d}V +
  \eps \! \int_S ( \bft\cdot\bfphi + \bfs\cdot\bfpsi ) \, \text{d}S .
\end{equation*}
The main differences are that the work function above involves only
the distortional elasticity of the director field $\We$ and that the
densities $\bfF$ and $\bfG$ are prescribed in the first integral on
the right-hand side---\eqref{eqn:virtual_work} corresponds to
\cite[Eqn.\,(3.14)]{ericksen:62}, in the more general context of a
coupled electric potential field.  The expression above leads to the
classical balance of forces and balance of torques by following the
same path we shall follow below.

Expressions of force balance and torque balance can be deduced from
\eqref{eqn:virtual_work} by considering infinitesimal rigid
displacements, as done in \cite[Sec.\,II.A]{leslie:79},
\cite[Sec.\,2]{leslie:87b}, \cite[Sec.\,3.1.4]{sonnet:virga:12}, and
\cite[Sec.\,2.4, Remark~(i)]{stewart:04}.  We shall write a general
rigid displacement in the form
\begin{equation*}
  x^*\! = o + \bfR ( x - o^*\! ) , \quad \bfR \in SO(\calV) .
\end{equation*}
In this notation, the coordinates of the point $x^*\!$ with respect to
the frame $(o,\ehat_i)$ are the same as the coordinates of $x$ with
respect to $(o^*\!\!,\ehat^*_i\!)$, with $\ehat^*_i\!= \bfR^T\!\!\ehat_i$.
The transformed fields
\begin{equation*}
  U^*\!(x^*\!) = U(x) , \quad \nhat^*\!(x^*\!) = \bfR \nhat(x)
\end{equation*}
satisfy
\begin{equation*}
  \nabla^*\! U^*\!(x^*\!) = \bfR \nabla U(x) , \quad
  \nabla^*\! \nhat^*\!(x^*\!) = \bfR \nabla\nhat(x) \bfR^T \!\! .
\end{equation*}
Such transformations are isochoric and preserve $|\nhat|=1$.  Special
cases include rigid translations ($\bfR=\bfI$) and rigid rotations
($o^*\!\!=o$).

The variation in $\calF$ associated with such transformations can be
determined by expansion of
\begin{align*}
  \Delta\calF &= \calF[\nhat^*\!,U^*\!;V^*] - \calF[\nhat,U;V] \\
  &= \int_{V^*} \!
     W(x^*\!,\nhat^*\!,\nabla^*\!\nhat^*\!,U^*\!,\nabla^*\!U^*) \,
     \text{d}V^* \! -
     \int_V W(x,\nhat,\!\nabla\nhat,U,\!\nabla U) \, \text{d}V \\
  &= \int_V \bigl[
     W(x^*\!(x),\bfR\nhat,\bfR\nabla\nhat\bfR^T\!,U,\bfR\nabla U) -
     W(x,\nhat,\!\nabla\nhat,U,\!\nabla U) \bigr] \, \text{d}V .
\end{align*}
Here we have used the fact that $\text{d}V^* \! = \text{d}V$ for an
isochoric displacement---the volume dilation factor is one everywhere.
The parts of $W$ associated with $\We$ and $\WE$ do not contribute to
this expression, since
\begin{equation*}
  \We(\bfR\nhat,\bfR\nabla\nhat\bfR^T\!) = \We(\nhat,\!\nabla\nhat) , \quad
  \WE(\bfR\nhat,\bfR\nabla\nhat\bfR^T\!\!,\bfR\nabla U) =
  \WE(\nhat,\!\nabla\nhat,\!\nabla U) , 
\end{equation*}
$\forall \, \bfR \in SO(\calV)$, as can be verified directly.
Contributions to $\Delta\calF$ can come only from $\WH$ (by virtue of
the factor $(\bfH\cdot\nhat)^2$) and from the terms with explicit $x$
dependence ($\rhof(x)U$ and $\rhom gh(x)$).

For a rigid translation of the form
\begin{equation*}
  x^*\! = x + \eps\bfphi_0 , \quad
  \nhat^*\!(x^*\!) = \nhat(x) , \quad
  U^*\!(x^*\!) = U(x) ,
\end{equation*}
which corresponds to
\begin{equation*}
  \bfR = \bfI , \quad o - o^*\! = \eps\bfphi_0 , \quad
  \bfphi=\bfphi_0 , ~~ \bfpsi=\bfzero , ~~ \Psi=0 ,
\end{equation*}
we obtain
\begin{equation*}
  \delta\!\calF = \eps \bfphi_0 \cdot \! \mathlarger\int_V
  \dWdx \, \text{d}V .
\end{equation*}
Utilizing this in \eqref{eqn:virtual_work}, along with the
arbitrariness of $\bfphi_0$, we conclude that
\begin{equation*}
  \mathlarger\int_V \dWdx \, \text{d}V =
  \int_S \bft \, \text{d}S .
\end{equation*}
In equilibrium this must hold for all $V$, no matter what form the
stress vector $\bft$ takes.  Writing this in the manner
\begin{equation}\label{eqn:force_balance}
  \int_V \bfF \, \text{d}V + \int_S \bft \, \text{d}S = \bfzero , \quad
  \bfF = - \dWdx ,
\end{equation}
we see that here $- \partial W / \partial x$ plays the role of a body
force, which for $W$ as in \eqref{eqn:Wgeneral} is given by
\begin{equation}\label{eqn:FdWdx}
  \bfF = - \dWdx =
  - U \nabla\!\rhof - \rhom g \nabla h .
\end{equation}
By the definition of the height
function $h$, $\nabla h$ is just a unit vector in the vertical
direction.

An infinitesimal rigid rotation can be produced via
\begin{align*}
  x^*\! &= o + \bfR(\eps) (x-o) , \quad
          \bfR(\eps) = \bfI + \eps \bfW_0 + o(\eps) , \quad
          \bfW_0 \in \Skew(\calV) \\
        &= x + \eps \bfW_0 (x-o) + o(\eps) =
           x + \eps \bfw_0\times\bfr(x) + o(\eps) ,
\end{align*}
where $\bfw_0$ is the axial vector of the skew-symmetric tensor
$\bfW_0$ and $\bfr(x)=x-o$, the position vector of the point $x$, and
\begin{equation*}
  \nhat^*\!(x^*\!) = \bfR(\eps) \nhat(x) =
  \nhat(x) + \eps \, \bfW_0 \nhat(x) + o(\eps) =
  \nhat(x) + \eps \, \bfw_0 \times \nhat(x) + o(\eps) , ~~~
  U^*\!(x^*\!) = U(x) .
\end{equation*}
From this we obtain
\begin{equation*}
  \delta\!\calF =
  \eps \, \bfW_0 \cdot \! \mathlarger\int_V \Bigl(
  \dWdx \otimes \bfr +
  \frac{\partial\WH}{\partial\nhat} \otimes \nhat \Bigr) \text{d}V =
  \eps \, \bfw_0 \cdot \! \mathlarger\int_V \Bigl(
  \bfr \times \dWdx +
  \nhat \times \frac{\partial\WH}{\partial\nhat} \Bigr) \text{d}V .
\end{equation*}
Utilizing this in \eqref{eqn:virtual_work}, together with $\bfphi =
\bfw_0\times\bfr$, $\bfpsi = \bfw_0\times\nhat$, and the arbitrariness
of $\bfw_0$, we obtain
\begin{equation*}
  \mathlarger\int_V \Bigl( \bfr\times\dWdx +
  \nhat\times\frac{\partial\WH}{\partial\nhat} \Bigr) \text{d}V =
  \int_S ( \bfr\times\bft + \nhat\times\bfs ) \, \text{d}S .
\end{equation*}
As before, this can be written in a form identical to the classical
case:
\begin{equation}\label{eqn:torque_balance}
  \int_V ( \bfr\times\bfF + \nhat\times\bfG ) \, \text{d}V +
  \int_S ( \bfr\times\bft + \nhat\times\bfs ) \, \text{d}S =
  \bfzero , \quad \bfF = - \dWdx , \quad
  \bfG = - \frac{\partial\WH}{\partial\nhat} ,
\end{equation}
with $- \partial W / \partial x$ playing the role of the body force
(as before) and $- \partial\WH / \partial\nhat$ playing the role of
the generalized force acting on $\nhat$.  With $\WH$ as in
\eqref{eqn:WH}, $\bfG$ takes the form
\begin{equation*}
  \bfG = - \frac{\partial\WH}{\partial\nhat} =
  \mu_0 \Delta \chi ( \bfH \cdot \nhat ) \bfH ,
\end{equation*}
which is consistent with \cite[Sec.\,3.1.4]{sonnet:virga:12} and
\cite[Secs.\,2.4.2\,Remark\,(ii),\,4.2]{stewart:04}.  The body couple and
couple stress vector are usually denoted
\begin{equation*}
  \bfK := \nhat \times \bfG, \quad \bfl := \nhat \times \bfs .
\end{equation*}
See \cite[Sec.\,II.A]{leslie:79} or \cite[Sec.\,2]{leslie:87b} or
\cite[Sec.\,2.4]{stewart:04}.  Using the expressions for the magnetic
induction $\bfB$ and magnetization $\bfM$ in \eqref{eqn:BM}, which are
valid for a material in a uniaxial nematic liquid crystal phase, we
see that
\begin{equation*}
  \bfK = \bfB \times \bfH = \mu_0 \bfM \times \bfH ,
\end{equation*}
the ``magnetic torque'' density
\cite[Sec.\,2.4.2\,Remark\,(ii)]{stewart:04}.  The expressions for
$\bfF$ and $\bfG$ in \eqref{eqn:torque_balance} are exactly what one
would obtain from the approach of
Sec.\,\ref{sec:Ericksen_compatability_potential} after merging the
appropriate compatibility potentials into $W$.  The electric-field
contributions (from $\WE$) enter via a different route, as we shall
see.

In our context, then, the conservation laws for linear momentum (force
balance) and angular momentum (torque balance) are given in integral
form by \eqref{eqn:force_balance} and \eqref{eqn:torque_balance}.  It
will be seen below that $\bft$ and $\bfl$ can be expressed in terms of
appropriate tensors:
\begin{equation*}
  \bft = \bfT \nuhat , \quad \bfl = \bfL \nuhat .
\end{equation*}
Expressions for the stress tensor $\bfT$ and couple stress tensor
$\bfL$ will be deduced below.  Using these in
\eqref{eqn:force_balance} and \eqref{eqn:torque_balance}, combined
with the arbitrariness of $V$ and the Divergence Theorem, one can
express these balances in point form, precisely as in the classical
case (\cite[Sec.\,3.1.3]{sonnet:virga:12},
\cite[Sec.\,2.4]{stewart:04}):
\begin{equation}\label{eqn:balances}
  \bfF + \Div \bfT = \bfzero , \quad
  \eps_{ijk} T_{kj} \ehat_i + \bfK + \Div\bfL = \bfzero .
\end{equation}
Here $\eps_{ijk}$ is the Ricci alternator,
$\{ \ehat_1, \ehat_2, \ehat_3 \}$ is a fixed orthonormal basis for
$\calV\!$, and summation over repeated indices is assumed.  The
condition $(\ref{eqn:balances})_1$ leads to the elimination of the
$\bfr$-dependent parts of \eqref{eqn:torque_balance}, leaving a
balance of couples in $(\ref{eqn:balances})_2$ that is independent of
the origin of the coordinate system.  The equations
\eqref{eqn:balances} correspond to the static limit of Cauchy's first
and second laws of motion for a polar material, as found in
\cite[Eqns.\,(205.2), (205.10)]{truesdell:toupin:60} (where the latter
is expressed in terms of the skew tensors associated with these axial
vectors).  The first term in the torque balance is the torque density
(per unit volume) associated with the skew part of $\bfT$.  It is
related to the axial vector of $\myskew(\bfT)$ via
\begin{equation*}
  \axial(\myskew(\bfT)) = \frac12 \eps_{ijk} T_{kj} \ehat_i ,
\end{equation*}
as observed in \cite[Secs.\,2.1.5,\,3.1.3]{sonnet:virga:12}.  In the
absence of any body couple and couple stress ($\bfK=\bfzero$,
$\bfL=\bfzero$), the torque balance would simply dictate that $\bfT$
be symmetric, as is the case in nonpolar materials.

\subsection{Variations and constraints}

To study more general displacements, we use a variational formula of
\cite[Sec.\,37]{gelfand:fomin:63} for an integral functional of a
collection of scalar fields of the form
\begin{equation}\label{eqn:Jfnl}
  J[u_1,\ldots,u_m] = \! \int_V
  F(x,u_1,\ldots,u_m,\!\nabla u_1,\ldots,\!\nabla u_m) \, \text{d}V .
\end{equation}
In the notation of that book, the variations take the form
\begin{equation}\label{eqn:PhiPsi}
  \begin{split}
    x^*\! &= \Phi(x,u,\!\nabla u;\eps) =
    x + \eps \bfphi(x,u,\!\nabla u) + o(\eps) \\
    u^*_i\!(x^*\!) &= \Psi_i(x,u,\!\nabla u;\eps) =
    u_i(x) + \eps \psi_i(x,u,\!\nabla u) + o(\eps) ,
  \end{split}
\end{equation}
where
\begin{equation*}
  u = (u_1,\ldots,u_m) , \quad \nabla u = (\nabla u_1,\ldots,\!\nabla u_m) .
\end{equation*}
Such variations involve both the independent and dependent variables
and are sometimes referred to as ``combined'' or ``generalized''
variations.  In \cite[Sec.\,2.4.1]{stewart:04}, they are termed
``non-contemporaneous variations.''  The expression for $\delta J$
(defined in the same way we have used in the previous section) can be
written
\begin{equation}\label{eqn:deltaJ}
  \begin{split}
    \delta J = {} & \eps \! \mathlarger\int_V \Bigl(
      \frac{\partial F}{\partial x}\cdot\bfphi + \Tc\cdot\nabla\bfphi +
      \frac{\partial F}{\partial u_i}\psi_i +
      \frac{\partial F}{\partial\nabla u_i}\cdot\nabla u_i \Bigr) \text{d}V \\
    = {} & \eps \! \mathlarger\int_V \Bigl\{ \Bigl(
      \frac{\partial F}{\partial x} - \Div\Tc \Bigr) \cdot \bfphi +
      \Bigl[ \frac{\partial F}{\partial u_i} -
      \Div \Bigl( \frac{\partial F}{\partial\nabla u_i} \Bigr) \Bigr]
      \psi_i \Bigr\} \, \text{d}V \\
    & {} + \eps \! \mathlarger\int_S \Bigl[ \Tc\nuhat\cdot\bfphi +
      \Bigl( \frac{\partial F}{\partial\nabla u_i}\cdot\nuhat \Bigr)
      \psi_i \Bigr] \text{d}S ,
  \end{split}
\end{equation}
where
\begin{equation}\label{eqn:Tc}
  \Tc = F \, \bfI -
  \nabla u_1 \otimes \frac{\partial F}{\partial\nabla u_1} - \cdots -
  \nabla u_m \otimes \frac{\partial F}{\partial\nabla u_m} .
\end{equation}
In all formulas, summation over repeated indices is implied.

Though written in a different way, the expression above for $\delta J$
is equivalent to \cite[Eqn.\,(105)]{gelfand:fomin:63} and
\cite[Eqn.\,(2.125)]{stewart:04}.  In \cite[Sec.\,2.d]{toupin:60}, the
tensor $\Tc$ is referred to as the ``canonical stress tensor,'' a term
used in field theories in relativistic physics.  In one space
dimension, such expressions are related to canonical variables and
first integrals \cite[Sec.\,17]{gelfand:fomin:63}.  If, for example,
$F$ were the strain-energy density of a hyperelastic material, then
such a relation ($\Tc=\partial F/\partial\nabla\bfphi$) would
correspond to the relationship between the Piola-Kirchoff stress (or
nominal stress) and the deformation gradient in that setting
\cite[Sec.\,28]{gurtin:81}, \cite[Sec.\,4.3.1]{ogden:97}.  For our
free-energy functional \eqref{eqn:calFV}, with displacements denoted
as in \eqref{eqn:displacements}, the general variational formula above
assumes the form
\begin{align}
  \delta\!\calF = {} & \eps \! \mathlarger\int_V \Bigl(
    \dWdx\cdot\bfphi + \Tc\cdot\nabla\bfphi +
    \dWdn\cdot\bfpsi + \dWdgn\cdot\nabla\bfpsi +
    \frac{\partial W}{\partial U}\Psi +
    \frac{\partial W}{\partial\nabla U} \cdot \nabla\Psi \Bigr) \text{d}V
    \notag \\
  = {} & \eps \! \mathlarger\int_V \Bigl\{ \Bigl(
    \dWdx - \Div\Tc \Bigr) \cdot \bfphi +
    \Bigl[ \dWdn -
    \Div \Bigl( \dWdgn \Bigr) \Bigr]
    \cdot \bfpsi + \Bigl[ \frac{\partial W}{\partial U} -
    \Div \Bigl( \frac{\partial W}{\partial\nabla U} \Bigr) \Bigr]
    \Psi \Bigr\} \, \text{d}V \notag \\
  & {} + \eps \! \mathlarger\int_S \Bigl[ \Tc\nuhat\cdot\bfphi +
    \Bigl( \dWdgn \Bigr) \nuhat \cdot \bfpsi +
    \Bigl( \frac{\partial W}{\partial\nabla U}\cdot\nuhat \Bigr)
    \Psi \Bigr] \text{d}S , \label{eqn:unconstrained_delta_calF}
\end{align}
with
\begin{equation*}
  \Tc = W \bfI - (\nabla\nhat)^T \!
  \dWdgn -
  \nabla U \otimes \frac{\partial W}{\partial\nabla U} .
\end{equation*}
This simply corresponds to \eqref{eqn:deltaJ} and \eqref{eqn:Tc} with
$m=4$ and the role of $u_1$, $u_2$, $u_3$, and $u_4$ played by $n_1$,
$n_2$, $n_3$, and $U$ (where $n_i$ are the Cartesian components of
$\nhat$).

In order to be used in our virtual-work principle, the displacements
must be consistent with the constraints of incompressibility and
$|\nhat|=1$, which imply $\Div\bfphi=0$ and $\bfpsi\cdot\nhat=0$
\cite[Sec.\,2.1.2]{sonnet:virga:12}, \cite[Sec.\,2.4.2]{stewart:04}.
These constraints can be implemented in various ways.  We impose the
incompressibility constraint in the traditional way, by introducing
the term $-p\Div\bfphi$ into the variation, with $p$ the Lagrange
multiplier pressure field.  The unit-length constraint on the director
can be enforced via Lagrange multipliers as well, but here we choose
instead to impose it by writing $\bfpsi=\bfchi\times\nhat$, with
$\bfchi$ an arbitrary rotation vector.  With these changes, the
constrained variation $\delta\!\calF$ takes the form
\begin{align}
  \delta\!\calF = {} & \eps \! \mathlarger\int_V \Bigl\{ \Bigl(
    \dWdx - \Div\bfT \Bigr) \cdot \bfphi +
    \nhat \times \Bigl[ \dWdn -
    \Div \Bigl( \dWdgn \Bigr) \Bigr]
    \cdot \bfchi + \Bigl[ \frac{\partial W}{\partial U} -
    \Div \Bigl( \frac{\partial W}{\partial\nabla U} \Bigr) \Bigr]
    \Psi \Bigr\} \, \text{d}V \notag \\
  & {} + \eps \! \mathlarger\int_S \Bigl[ \bfT\nuhat\cdot\bfphi +
    \nhat \times \Bigl( \dWdgn \Bigr)
    \nuhat \cdot \bfchi +
    \Bigl( \frac{\partial W}{\partial\nabla U}\cdot\nuhat \Bigr)
    \Psi \Bigr] \text{d}S , \label{eqn:constrained_delta_calF}
\end{align}
with
\begin{equation*}
  \bfT = \Tc - p \, \bfI = ( W - p ) \bfI -
  (\nabla\nhat)^T \! \dWdgn -
  \nabla U \otimes \frac{\partial W}{\partial\nabla U} .
\end{equation*}
At this stage, the pressure field is undetermined.  Using the general
form for the constrained variation $\delta\!\calF$ above in the
virtual-work principle \eqref{eqn:virtual_work}, along with the
arbitrariness of $\bfphi$, $\bfchi$, $\Psi$, and the subregion $V$, we
obtain the following conditions (to be satisfied in equilibrium):
\begin{subequations}
\begin{gather}
  \dWdx - \Div\bfT = \bfzero , \quad
  \nhat \times \Bigl[ \dWdn -
  \Div \Bigl( \dWdgn \Bigr) \Bigr] =
  \bfzero , \quad
  \frac{\partial W}{\partial U} - \Div \Bigl(
  \frac{\partial W}{\partial\nabla U} \Bigr) = 0
  \label{eqn:field_equations} \\
  \bft = \bfT \nuhat , \quad
  \nhat \times \bfs = \nhat \times \Bigl(
  \dWdgn \Bigr) \nuhat , \quad
  r = \frac{\partial W}{\partial\nabla U} \cdot \nuhat .
  \label{eqn:fluxes}
\end{gather}
\end{subequations}
The field equations \eqref{eqn:field_equations} recover the local form
of the force balance $(\ref{eqn:balances})_1$ along with the
Euler-Lagrange equations for the equilibrium director field and
electric potential \eqref{eqn:E-L}, while \eqref{eqn:fluxes} gives the
formulas for the generalized force densities and fluxes associated
with the influences outside of a fluid element.  We note that
\eqref{eqn:field_equations} gives the ``strong form'' or ``point
form'' of the field equations, and we have tacitly assumed sufficient
regularity of the underlying fields to justify the various
integrations by parts that we have employed.

One can show that when evaluated on an equilibrium pair $(\nhat,U)$,
we necessarily have
\begin{equation*}
  \Div \Tc = \dWdx .
\end{equation*}
This can be verified by direct calculation, using the Euler-Lagrange
equations satisfied by $\nhat$ and $U$ together with the relation
$(\nabla\nhat)^T \! \nhat = \bfzero$ (which is a consequence of
$|\nhat|=1$).  It can also be deduced from the necessary vanishing of
the inner variation $\delta_x\calF$, that is, the unconstrained
variation \eqref{eqn:unconstrained_delta_calF} with $\bfpsi=\bfzero$,
$\Psi=0$, and $\text{supp}(\bfphi) \subset \text{int}(V)$, which
corresponds to a local parallel displacement of $\nhat$ and $U$---see
\cite[Sec.\,2.5]{ball:17}.  As a consequence of this, we obtain
\begin{equation*}
  \dWdx - \Div\bfT = \bfzero
  ~~ \Rightarrow ~~ \nabla p = \bfzero
  ~~ \Rightarrow ~~ p = p_0 = \text{const} ,
\end{equation*}
which is consistent with the fact that in a model that assumes
incompressibility, the pressure can only be determined to within an
arbitrary constant hydrostatic pressure
\cite[Sec.\,30]{truesdell:noll:04}.  This then provides us with the
final form for our stress tensor:
\begin{equation}\label{eqn:Tfinal}
  \bfT = ( W - p_0 ) \bfI - (\nabla\nhat)^T \!
  \dWdgn + \bfE \otimes \bfD .
\end{equation}
Here we have used $\bfE = - \nabla U$ and
$\bfD = \partial W / \partial \nabla U$ to simplify the last term.
This is the total stress tensor containing all mechanical and
electromagnetic influences, not just contact forces.  It is perhaps
more appropriate to refer to it as the total momentum-flux tensor.
With this definition of $\bfT$, we readily see that any equilibrium
pair $(\nhat,U)$ necessarily satisfies the force balance
$(\ref{eqn:balances})_1$.  The form of the couple stress tensor
follows from $(\text{\ref{eqn:fluxes}})_2$:
\begin{equation}\label{eqn:couple_stress_tensor}
  \bfl = \nhat \times \bfs = \nhat \times \Bigl(
  \dWdgn \Bigr) \nuhat =
  \bfL \nuhat ~~ \Rightarrow ~~
  L_{ij} = \eps_{ikl} n_k \frac{\partial W}{\,\partial n_{l,j}} ,
\end{equation}
as in the classical case \cite[Eqn.\,(3.73)]{sonnet:virga:12},
\cite[Eqn.\,(2.165)]{stewart:04}.  The identification of $\bfT$ and
$\bfL$ completes the picture foreshadowed at the end of
Sec.\,\ref{sec:force:torque:balance}.  The term
\begin{equation*}
  r = \frac{\partial W}{\partial\nabla U} \cdot \nuhat =
  \bfD \cdot \nuhat
\end{equation*}
corresponds to a flux of electric displacement, as in the integral
form of the Gauss Law of electrostatics ($\int_S \bfD\cdot\nuhat \,
\text{d}S = Q$, with $Q$ the total charge contained in $V$).

\subsection{Satisfaction of force balance and torque balance}

In our development, everything follows from the free energy and
associated virtual-work principle.  This includes the force balance
and torque balance \eqref{eqn:balances}, the equilibrium field
equations \eqref{eqn:field_equations}, and the expressions for the
stress tensor \eqref{eqn:Tfinal} and couple stress tensor
\eqref{eqn:couple_stress_tensor}.  It stands to reason that an
equilibrium pair $(\nhat,U)$ (coupled solutions of the Euler-Lagrange
equations \eqref{eqn:E-L}) should necessarily satisfy the force
balance and torque balance, guaranteeing consistency with the
hydrostatic limit of the Ericksen-Leslie equations.  Satisfaction of
the force balance is clear, as we have already observed.
Satisfaction of the torque balance is less clear.  The traditional
approach to verify this is to use an identity due to Ericksen, which
follows from the invariance of the distortional elasticity $\We$ with
respect to rigid rotations.  This was done first in the absence of
external fields in \cite{ericksen:61} and later generalized to include
gravitational and magnetic fields in \cite{ericksen:62}.  Textbook
accounts can be found in \cite[Sec.\,3.1.3]{sonnet:virga:12} and
\cite[Sec.\,2.4 Remark~(i)]{stewart:04}.  That approach could be
followed here, deriving and employing a modified version of
``Ericksen's Identity.''  We choose instead a different approach.

In a simpler variational setting, such balances could be obtained as
direct consequences of Noether's Theorem.  Broadly stated, this
theorem shows that continuous symmetries of a system imply
conservation laws for that system.  These ideas hold a prominent place
in variational mechanics, thermodynamics, relativistic physics, and
elsewhere.  Classical discussions can be found in
\cite[Ch.\,21]{callen:85}, \cite[Secs.\,20,\,37.5]{gelfand:fomin:63},
and \cite[Ch.\,XI Sec.\,20, App.\,II]{lanczos:70}.  A more modern
point of view is presented in \cite[Ch.\,7]{mansfield:10}, for
example.  Applications (in general terms) include symmetries with
respect to time translation (which can imply conservation of energy),
space translation (linear momentum), rotation (angular momentum), and
the gauge symmetry of the Maxwell equations (which implies
conservation of charge).  In continuum mechanics, precursors of these
ideas are found in early theories of elasticity---see
\cite[Sec.\,98]{truesdell:noll:04}, where a version of these ideas is
referred to as the Cosserat-Toupin Fundamental Equivalence Theorem for
hyperelastic materials.

The simplest instance of Noether's Theorem relevant to our interests
can be stated as follows.  If a functional of the form
\eqref{eqn:Jfnl} is invariant under transformations of the form
\eqref{eqn:PhiPsi}---in the sense that
$J[u_1^*,\ldots,u_m^*;V^*] = J[u_1,\ldots,u_m;V]$, for all $\eps$
sufficiently small---and this holds for arbitrary subdomains $V$, then
\begin{equation*}
  \Div \Bigl( \Tc^T \! \bfphi +
  \psi_i \frac{\partial F}{\partial\nabla u_i} \Bigr) = 0
\end{equation*}
when evaluated on equilibrium fields $u_1,\ldots,u_m$.  Here $\Tc$ is
the canonical stress tensor \eqref{eqn:Tc}.  The result simply follows
from the expression for $\delta J$ in \eqref{eqn:deltaJ}: invariance
implies $\delta J = 0$, and equilibrium implies
\begin{equation*}
  \frac{\partial F}{\partial x} - \Div\Tc = \bfzero
  ~~~ \text{and} ~~~
  \frac{\partial F}{\partial u_i} -
  \Div \Bigl( \frac{\partial F}{\partial\nabla u_i} \Bigr) = 0 ,
  ~~ i = 1,\ldots,m ,
\end{equation*}
leaving
\begin{equation*}
  0 = \mathlarger\int_S \Bigl[ \Tc\nuhat\cdot\bfphi +
  \Bigl( \frac{\partial F}{\partial\nabla u_i} \cdot \nuhat \Bigr)
  \psi_i \Bigr] \text{d}S = \mathlarger\int_S \Bigl( \Tc^T \! \bfphi +
  \psi_i \frac{\partial F}{\partial\nabla u_i} \Bigr) \cdot \nuhat \,
  \text{d}S ,
\end{equation*}
with the conclusion following via the Divergence Theorem and the
arbitrariness of $V$.  See \cite[Sec.\,37.5]{gelfand:fomin:63}.  An
equivalent form of the conclusion is
\begin{equation}\label{eqn:Noether_concl}
  \Div\Tc\cdot\bfphi + \Tc\cdot\nabla\bfphi +
  \Div \Bigl( \frac{\partial F}{\partial\nabla u_i} \Bigr) \psi_i +
  \frac{\partial F}{\partial\nabla u_i} \cdot \nabla\psi_i = 0 .
\end{equation}
Note that in the absence of invariance ($\delta J \not= 0$), the
conclusion of Noether's Theorem for this case becomes
\begin{equation*}
  \delta J = \eps \! \mathlarger\int_V \Bigl[ \Div\Tc\cdot\bfphi +
  \Tc\cdot\nabla\bfphi +
  \Div \Bigl( \frac{\partial F}{\partial\nabla u_i} \Bigr) \psi_i +
  \frac{\partial F}{\partial\nabla u_i} \cdot \nabla \psi_i \Bigr]
  \text{d}V ,
\end{equation*}
for all subdomains $V$ and $\eps$ sufficiently small.

A functional of the form
\begin{equation}\label{eqn:Jfnl_nox}
  J[u_1,\ldots,u_m] = \int_V
  F(u_1,\ldots,u_m,\nabla u_1,\ldots,\nabla u_m) \, \text{d}V ,
\end{equation}
where $F$ satisfies
\begin{equation*}
  F(u_1,\ldots,u_m,\bfR\nabla u_1,\ldots,\bfR\nabla u_m ) =
  F(u_1,\ldots,u_m,\nabla u_1,\ldots,\nabla u_m) ,
  ~~ \forall \bfR \in SO(\calV) ,
\end{equation*}
is invariant under rigid translations and rigid rotations.  A rigid
translation can be expressed
\begin{equation*}
  x^*\! = x + \eps \bfphi_0 , ~~ u_i^*\!(x^*\!) = u_i(x) ,
\end{equation*}
in the notation of \eqref{eqn:PhiPsi}, for which
\begin{equation*}
  \bfphi = \bfphi_0 , \quad \nabla\bfphi = \bfzero , \quad
  \psi_i = 0 , \quad \nabla\psi_i = \bfzero .
\end{equation*}
Using these in \eqref{eqn:Noether_concl}, combined with the
arbitrariness of $\bfphi_0$, gives
\begin{equation*}
  \Div \Tc = \bfzero ,
\end{equation*}
which can be interpreted as a balance of forces (or conservation of
linear momentum).  In one space dimension, such a relation simply
corresponds to a first integral.  If $F$ were to contain explicit $x$
dependence, as in \eqref{eqn:Jfnl}, then the functional would no
longer necessarily be invariant under translation, and this relation
would become
\begin{equation*}
  \Div \Tc = \frac{\partial F}{\partial x} .
\end{equation*}
An infinitesimal rigid rotation can be expressed
\begin{equation*}
  x^*\! = x + \eps \bfW_0 \bfr(x) , ~~~
  u_i^*\!(x^*\!) = u_i(x) , \quad \text{where }
  \bfW_0 \in \Skew(\calV) \text{ and }
  \bfr(x) = x - o ,
\end{equation*}
for which
\begin{equation*}
  \bfphi = \bfW_0 \bfr , \quad \nabla\bfphi = \bfW_0 , \quad
  \psi_i = 0 , \quad \nabla\psi_i = \bfzero .
\end{equation*}
Using these in \eqref{eqn:Noether_concl}, gives
\begin{equation*}
  \Div\Tc \cdot \bfW_0\bfr + \Tc\cdot\bfW_0 = 0 .
\end{equation*}
Using $\Div\Tc=\bfzero$ and the arbitrariness of $\bfW_0$ leads to
the conclusion that
\begin{equation*}
  \Tc \in \text{Symm}(\calV) ,
\end{equation*}
which can be viewed as a balance of torques (or conservation of
angular momentum).  Thus for a frame-indifferent equilibrium model of
the form \eqref{eqn:Jfnl_nox}, we necessarily have
\begin{equation*}
  \Div \Tc = \bfzero ~~ \text{and} ~~ \Tc \in \text{Symm}(\calV)
\end{equation*}
when evaluated on fields satisfying the associated strong-form
Euler-Lagrange equations.  Similar examples, most in the context of
relativistic physics, can be found in
\cite[Sec.\,38]{gelfand:fomin:63}.

Our liquid-crystal system is not so simple: it is neither translation
invariant nor rotation invariant, and it is subject to constraints
(incompressibility and $|\nhat|=1$).  Nevertheless, the same path can
be followed to show that conservation of angular momentum (balance of
torques) is satisfied by equilibrium fields---linear-momentum
conservation (balance of forces) is clear.  For $\calF[\nhat,U;V]$ in
the form \eqref{eqn:calFV} and transformations of the form
\eqref{eqn:displacements}, the variation $\delta\!\calF$ is given by
\eqref{eqn:unconstrained_delta_calF}.  When evaluated on an
equilibrium pair $(\nhat,U)$, this can be put in the form of the
conclusion of Noether's Theorem:
\begin{equation}\label{eqn:deltaF-a-la-Noether}
  \begin{split}
    \delta\!\calF = {} & \eps \! \mathlarger\int_V \Bigl\{
    \Div\Tc\cdot\bfphi + \Tc\cdot\nabla\bfphi + \Bigl[
    \Div \Bigl( \dWdgn \Bigr) +
    \lambda \nhat \Bigr] \cdot \bfpsi \\
    & {} + \dWdgn \cdot \nabla\bfpsi +
    \Div \Bigl( \frac{\partial W}{\partial\nabla U} \Bigr) \Psi +
    \frac{\partial W}{\partial\nabla U} \cdot \nabla \Psi \Bigr\}
    \, \text{d}V ,
  \end{split}
\end{equation}
for all subdomains $V$ and $\eps$ sufficiently small.  For an
infinitesimal rigid rotation of the form
\begin{equation*}
  x^* \! = x + \eps \bfW_0 \bfr(x) , \quad
  \nhat^*\!(x^*\!) = \nhat(x) + \eps \bfW_0 \nhat(x) , \quad
  U^*\!(x^*\!) = U(x) , \quad \bfW_0 \in \Skew(\calV) ,
\end{equation*}
we have already obtained in Sec.\,\ref{sec:force:torque:balance} that
\begin{equation*}
  \delta\!\calF =
  \eps \, \bfW_0 \cdot \!\! \mathlarger\int_V \Bigl(
  \dWdx \otimes \bfr +
  \frac{\partial\WH}{\partial\nhat} \otimes \nhat \Bigr) \text{d}V .
\end{equation*}
For such a transformation, we have
\begin{equation*}
  \bfphi = \bfW_0\bfr , ~~ \nabla\bfphi = \bfW_0 , ~~
  \bfpsi = \bfW_0\nhat , ~~ \nabla\bfpsi = \bfW_0\nabla\nhat , ~~
  \Psi = 0 , ~~ \nabla\Psi = \bfzero ,
\end{equation*}
which when substituted into \eqref{eqn:deltaF-a-la-Noether} gives
\begin{align*}
  \delta\!\calF &= \eps \! \mathlarger\int_V \Bigl\{ \Div\Tc\cdot\bfW_0\bfr +
    \Tc\cdot\bfW_0 + \Bigl[ \Div \Bigl(
    \dWdgn \Bigr) + \lambda\nhat \Bigr]
    \cdot \bfW_0\nhat + \dWdgn \cdot
    \bfW_0\nabla\nhat \Bigr\} \, \text{d}V \\
  &= \eps\,\bfW_0 \cdot \!\! \mathlarger\int_V \Bigl[ \Div\Tc\otimes\bfr + \Tc +
    \Div \Bigl( \dWdgn \Bigr) \otimes \nhat +
    \dWdgn (\nabla\nhat)^T \! \Bigr]
    \text{d}V .
\end{align*}
Equating these two expressions for $\delta\!\calF$ and using the fact
that $\Div\Tc = \partial W / \partial x$, combined with the
arbitrariness of $\eps$, $V$, and $\bfW_0$, gives
\begin{equation*}
  \Tc - \frac{\partial\WH}{\partial\nhat} \otimes \nhat +
  \dWdgn (\nabla\nhat)^T \! +
  \Div \Bigl( \dWdgn \Bigr) \otimes \nhat
  \in \text{Symm}(\calV) ,
\end{equation*}
and the above remains true with $\Tc$ replaced by $\bfT$ (since they
differ by the symmetric tensor $p_0 \bfI$).  Expressing in terms of
Cartesian components and equating the skew part of the expression
above to zero, we find this to be equivalent to
\begin{equation*}
  \eps_{ijk} T_{kj} - \eps_{ijk} \frac{\partial\WH}{\partial n_k} n_j +
  \frac{\partial}{\partial x_l} \Bigl( \eps_{ijk}
  \frac{\partial W}{\partial n_{k,l}} n_j \Bigr) = 0 ,
\end{equation*}
which is precisely $(\ref{eqn:balances})_2$.  Thus an equilibrium pair
$(\nhat,U)$ necessarily satisfies balance of torques, in addition to
balance of forces, and therefore is consistent with the hydrostatic
limit of the Ericksen-Leslie equations (properly formulated with a
coupled electric field).

The total stress tensor $\bfT$ in \eqref{eqn:Tfinal} contains both
mechanical and electromagnetic influences.  The contribution to the
torque density $\eps_{ijk}T_{kj}\ehat_i$ from the term
$\bfE\otimes\bfD$ in $\bfT$,
\begin{equation*}
  \eps_{ijk} ( \bfE \otimes \bfD )_{kj} =
  \eps_{ijk} E_k D_j = ( \bfD \times \bfE )_i \, ,
\end{equation*}
admits an easy interpretation.  The expression $\bfD\times\bfE$ is
referred to as the ``dielectric torque''
\cite[Sec.\,8.4]{jakli:saupe:06}.  As a consequence of the relation
$\bfD = \eps_0 \bfE + \bmP$, we obtain
\begin{equation*}
  \bfD \times \bfE = \bmP \times \bfE ,
\end{equation*}
and we see that this terms gives the couple per unit volume exerted by
the electric field on the polarization, generalizing the formula for
the torque on a point dipole in an external field
$\bfp_0 \times \Eext(x_0)$ \cite[Sec.\,3.9]{stratton:41}.  In the case
of a uniaxial liquid crystal with no flexoelectricity taken into
account ($\bfD = \epstensor \bfE$,
$\epstensor = \eps_0 [ \eperp\bfI + \Delta\eps ( \nhat\otimes\nhat ) ]$),
this becomes the familiar expression
\begin{equation*}
  \bfD \times \bfE =
  \eps_0 \Delta\eps ( \bfE\cdot\nhat ) ( \nhat\times\bfE ) ,
\end{equation*}
which strives to rotate $\nhat$ into alignment parallel to $\pm\bfE$
(if $\Delta\eps>0$), perpendicular to $\bfE$ (if $\Delta\eps<0$).  We
note that for an isotropic linear dielectric ($\bfD = \eps \bfE$,
$\eps$ a scalar field), we would have $\bfD \times \bfE = \bfzero$.

\subsection{Boundary conditions}

In order to determine the conditions that hold on the segment
$\Gamma_2$ of the boundary of $\Omega$, we require the variation of
the functional associated with the anchoring energy.  Let us separate
the total free energy \eqref{eqn:calF} into volume and surface parts:
\begin{equation*}
  \calF[\nhat,U] = \calFv[\nhat,U;\Omega] + \calFs[\nhat;\Gamma_2] ,
\end{equation*}
where
\begin{equation*}
  \calFv[\nhat,U;\Omega] = \!
  \int_{\Omega} W(x,\nhat,\!\nabla\nhat,U,\!\nabla U) \, \text{d}V , \quad
  \calFs[\nhat;\Gamma_2] = \!
  \int_{\Gamma_2} \! \Ws(x,\nhat;\nuhat,\tauhat) \, \text{d}S .
\end{equation*}
Using the same perturbations as in \eqref{eqn:displacements}, the
variation $\delta\!\calFs$ can be determined by expansion of
\begin{equation*}
  \Delta \calFs = \calFs[\nhat^*\!;\Gamma^*_2] -
  \calFs[\nhat;\Gamma_2] , \quad \Gamma^*_2\!\! = \Phi_{\eps}(\Gamma_2) .
\end{equation*}
This is now a true ``domain variation,'' as we now allow deformation
of the boundary.  In order to expand with respect to $\eps$, we
require the transformation properties of $\nuhat$, $\tauhat$, and
$\text{d}S$, in addition to those for $x$ and $\nhat$ in
\eqref{eqn:displacements}.  The additional formulas that we need are
given by
\begin{gather*}
  \nuhat^*\! = \nuhat - \eps \, \bfP(\nuhat) (\nabla\bfphi)^T \! \nuhat
  + o(\eps) , ~~
  \tauhat^*\! = \tauhat + \eps \, \bfP(\tauhat) (\nabla\bfphi) \tauhat
  + o(\eps) , \\
  \text{d}S^*\! = \bigl[ 1 + \eps \, \bfP(\nuhat) \cdot \nabla\bfphi
  + o(\eps) \bigr] \text{d}S .
\end{gather*}
Here $\bfP(\nuhat)$ and $\bfP(\tauhat)$ denote the projections
transverse to $\nuhat$ and $\tauhat$,
\begin{equation*}
  \bfP(\nuhat) = \bfI - \nuhat \otimes \nuhat , \quad
  \bfP(\tauhat) = \bfI - \tauhat \otimes \tauhat ,
\end{equation*}
and $\nabla\bfphi$ denotes limiting values from the interior of
$\Omega$.  Formulas similar to those for $\nuhat^*\!$ and
$\text{d}S^*\!$ above can be found in \cite[Sec.\,6]{gurtin:81} and
\cite[Sec.\,5.2]{virga:94}.  The formula for $\tauhat^*\!$ follows
from
\begin{equation*}
  \tauhat^*\! =
  \frac{(\nabla\Phi_{\eps})\tauhat}{|(\nabla\Phi_{\eps})\tauhat|} , \quad
  \nabla\Phi_{\eps} = \bfI + \eps \nabla\bfphi + o(\eps) .
\end{equation*}
In terms of these, we obtain
\begin{align*}
  \Delta \calFs = {} & \! \int_{\Gamma_2^*\!} \!
  \Ws(x^*\!,\nhat^*\!;\nuhat^*\!,\tauhat^*\!) \, \text{d}S^*\! - \!
  \int_{\Gamma_2} \! \Ws(x,\nhat;\nuhat,\tauhat) \, \text{d}S \\
  = {} & \! \int_{\Gamma_2} \bigl[ \Ws(x^*\!(x),\nhat^*\!;\nuhat^*\!,\tauhat^*\!)
         - \Ws(x,\nhat;\nuhat,\tauhat) \bigr] \text{d}S \\
  & {} + \eps \! \int_{\Gamma_2}
  \! \Ws(x,\nhat;\nuhat,\tauhat) [ \bfP(\nuhat)\cdot\nabla\bfphi ]
  \, \text{d}S + o(\eps) \\
  = {} & \eps \! \mathlarger\int_{\Gamma_2} \Bigl[
  \frac{\partial\Ws}{\partial\xS} \cdot \bfphi +
  \frac{\partial\Ws}{\partial\nhat} \cdot \bfpsi -
  \frac{\partial\Ws}{\partial\nuhat} \cdot \bfP(\nuhat)
         (\nabla\bfphi)^T\! \nuhat \\
  & {} + \frac{\partial\Ws}{\partial\tauhat} \cdot \bfP(\tauhat)
  (\nabla\bfphi) \tauhat + \Ws \, \bfP(\nuhat)\cdot\nabla\bfphi \Bigr]
  \text{d}S + o(\eps) ,
\end{align*}
which simplifies to
\begin{subequations}
\begin{equation}\label{eqn:deltaFs}
  \delta\!\calFs = \eps \! \mathlarger\int_{\Gamma_2} \Bigl(
  \frac{\partial\Ws}{\partial\xS} \cdot \bfphi +
  \Ts \cdot \nablaS \bfphi +
  \frac{\partial\Ws}{\partial\nhat} \cdot \bfpsi \Bigr) \text{d}S ,
\end{equation}
with
\begin{equation}\label{eqn:Ts}
  \Ts = \Ws \, \bfP(\nuhat) -
  \nuhat \otimes \bfP(\nuhat) \frac{\partial\Ws}{\partial\nuhat} +
  \bfP(\tauhat) \frac{\partial\Ws}{\partial\tauhat} \otimes \tauhat .
\end{equation}
\end{subequations}
Here $\partial\Ws/\partial\xS$ and $\nablaS\bfphi$ denote surface
gradients.  For the small amount of surface calculus that we require,
we rely on the formulations in \cite{gurtin:88,gurtin:murdoch:75},
\cite[Sec.\,5.2.3]{sonnet:virga:12}, and
\cite[Secs.\,2.3.6,\,5.2]{virga:94}.  While the approaches in these
references differ somewhat in their details, they agree in the aspects
that we require.  Related developments relying more on tools of
differential geometry and invariant calculus can be found in
\cite[Chs.\,XIV-XV]{mcconnell:57},
\cite[App.\,A]{slattery:sagis:oh:07}, and
\cite[Ch.\,XII]{weatherburn:61}.  Considerations of surface anchoring
energies depending on only $\nhat$ and $\nuhat$ can be found in
\cite{barratt:jenkins:73,jenkins:barratt:74} and
\cite[Ch.\,5]{virga:94}.

Thus for a scalar field $f$ defined on a surface $S$, the surface
gradient $\nablaS f$ is defined as the tangential vector field
characterized by
\begin{equation*}
  \frac{d}{dt} f(x(t)) \bigr|_{t=0} = \nablaS f(x_0) \cdot \dot{x}(0) ,
\end{equation*}
for all paths $x(t) \subset S$ through $x_0 = x(0)$.  For any
extension $\tilde{f}$ of $f$ to a neighborhood of the surface, one can
show that
\begin{equation*}
  \nablaS f = \bfP(\nuhat) \nabla \tilde{f} .
\end{equation*}
The notation $\partial\Ws/\partial\xS$ above refers to such a vector
field.  Thus for a simple anchoring energy of the form considered in
Sec.\,\ref{sec:surface_anchoring_energy}, we would have
\begin{equation*}
  \Ws = - \frac12 W_0(x) ( \nhat\cdot\nhat_0 )^2
  ~~ \Rightarrow ~~
  \frac{\partial\Ws}{\partial\xS} =
  - \frac12 ( \nhat\cdot\nhat_0 )^2 \, \nablaS W_0 ,
\end{equation*}
which implies a tangential force in the direction of decreasing values
of $W_0$ (zero, if $W_0$ is constant).  For a vector field $\bfv$ on
$S$, the surface gradient $\nablaS\bfv$ is defined in a similar way:
\begin{equation*}
  \frac{d}{dt} \bfv(x(t)) \bigr|_{t=0} = \nablaS \bfv (x_0) \, \dot{x}(0) .
\end{equation*}
Thus at each point $x \in S$, $\nablaS\bfv(x)$ is a linear
transformation that maps vectors tangent to $S$ at $x$ into vectors in
$\calV$.  In terms of an extension $\tilde{\bfv}$ of $\bfv$, one can
take
\begin{equation*}
  \nablaS \bfv = ( \nabla \tilde{\bfv} ) \bfP(\nuhat) ,
\end{equation*}
which produces the correct result when applied to a tangent vector and
is extended to annihilate any component of a vector in the normal
direction.  Thus in \eqref{eqn:deltaFs}, we simply take
\begin{equation*}
  \nablaS \bfphi = ( \nabla \bfphi ) \bfP(\nuhat) ,
\end{equation*}
again with $\nabla\bfphi$ denoting limiting values from
$\text{int}(\Omega)$.

To move the derivatives off $\nablaS\bfphi$ in \eqref{eqn:deltaFs}, we
make use of the identity
\begin{equation*}
  \divS ( \bfL^T \!\! \bfv ) = \bfL \cdot \nablaS\bfv + \divS\bfL \cdot \bfv
\end{equation*}
and the Surface Divergence Theorem for a tangential vector field $\bft$,
\begin{equation*}
  \int_S \divS\bft \,\, \text{d}S =
  \int_{\partial S} \bft \cdot \nutwo \, \text{d}s .
\end{equation*}
The identity can be found in
\cite[Eqn.\,$(\mathrm{A}14)_1$]{gurtin:88} and
\cite[Eqn.\,(5.42)]{virga:94}; while the Surface Divergence Theorem
can be found in \cite[App.\,A]{gurtin:88},
\cite[Sec.\,2]{gurtin:murdoch:75}, and \cite[Sec.\,2.3.6]{virga:94}.
Here $\nutwo$ denotes the binormal (unit vector pointing outward from
$S$ on the boundary $\partial S$), and the surface divergences of a
vector field $\bfv$ and tensor field $\bfL$ are defined
\begin{equation*}
  \divS \bfv = \tr \bigl( \nablaS \bfv \bigr) , \quad
  \divS \bigl( \bfL^T \!\! \bfa \bigr) =
  \divS \bfL \cdot \bfa , ~ \forall \bfa \in \calV ,
\end{equation*}
as in \cite[Sec.\,2]{gurtin:murdoch:75} and
\cite[Sec.\,2.3.6]{virga:94}.  In our setting, then,
\begin{equation*}
  \divS \bfphi = \bfP(\nuhat) \cdot \nabla\bfphi .
\end{equation*}
The surface stress tensor $\Ts$ satisfies $\Ts\nuhat=\bfzero$, and so
$\Ts^T\!\bfv$ is tangential for any vector field $\bfv$ on $S$
($\Ts^T\!\bfv\cdot\nuhat = \bfv\cdot\Ts\nuhat = 0$).  The identity and
integral formula above combine to give
\begin{equation*}
  \int_{\Gamma_2} \! \Ts \cdot \nablaS \bfphi \, \text{d}S = \!
  \int_{\partial\Gamma_2} \!\! \Ts \nutwo \cdot \bfphi \, \text{d}s - \!
  \int_{\Gamma_2} \divS \! \Ts \cdot \bfphi \, \text{d} S ,
\end{equation*}
and we obtain the final expression for $\delta\!\calFs$:
\begin{equation}\label{eqn:deltaFs_final}
  \delta\!\calFs = \eps \! \mathlarger\int_{\Gamma_2} \Bigl[
  \Bigl( \frac{\partial\Ws}{\partial\xS} - \divS \Ts \Bigr)
  \cdot \bfphi + \frac{\partial\Ws}{\partial\nhat} \cdot \bfpsi \Bigr]
  \text{d}S +
  \eps \! \int_{\partial\Gamma_2} \! \Ts \nutwo \cdot \bfphi \, \text{d}s .
\end{equation}

Combining \eqref{eqn:constrained_delta_calF} (for the case $V=\Omega$)
and \eqref{eqn:deltaFs_final}, we have
\begin{align*}
  \delta\!\calF = {} & \eps \! \mathlarger\int_{\partial\Omega} \Bigl[
  \bfT \nuhat \cdot \bfphi + \nhat \times \Bigl(
  \dWdgn \Bigr) \nuhat \cdot \bfchi +
  \Bigl( \frac{\partial W}{\partial\nabla U}\cdot\nuhat \Bigr) \Psi
  \Bigr] \text{d}S \\
  & + \eps \! \mathlarger\int_{\Gamma_2} \Bigl[ \Bigl(
  \frac{\partial\Ws}{\partial\xS} - \divS \Ts \Bigr) \cdot \bfphi +
  \nhat \times \frac{\partial\Ws}{\partial\nhat} \cdot \bfchi \Bigr]
  \text{d}S + \eps \! \int_{\partial\Gamma_2} \!
  \Ts\nuhat_2 \cdot \bfphi \, \text{d}s ,
\end{align*}
where $\bfT$ is as in \eqref{eqn:Tfinal}, $\Ts$ is as in
\eqref{eqn:Ts}, and $\delta\!\calF$ above is assumed to be evaluated
on an equilibrium pair $(\nhat,U)$ (which is the reason the volume
terms in \eqref{eqn:constrained_delta_calF} are not present).  The
virtual-work principle \eqref{eqn:virtual_work} now takes the form
\begin{equation*}
  \delta\!\calF = \eps \! \int_{\partial\Omega} ( \bft\cdot\bfphi +
  \bfl\cdot\bfchi + r\,\Psi ) \, \text{d}S +
  \eps \! \int_{\partial\Gamma_2} \! \bff\cdot\bfphi \,\, \text{d}s .
\end{equation*}
Balancing expressions above and using the arbitrariness of $\bfphi$,
$\bfchi$, and $\Psi$, along with previously identified essential
boundary conditions (see the Fig.\,\ref{fig:domain} caption) and
natural boundary condition \eqref{eqn:naturalBC}, we can state the
full set of conditions that hold on each of the boundary segments.

On $\Gamma_1$, both $\nhat$ and $U$ satisfy Dirichlet boundary
conditions ($\nhat=\nhat_0$ and $U=-V/2$), and we have the relations
\begin{equation*}
  \bft = \bfT \nuhat , \quad \bfl = \nhat \times \Bigl(
  \dWdgn \Bigr) \nuhat = \bfL \nuhat , \quad
  r = \frac{\partial W}{\partial\nabla U} \cdot \nuhat =
  \bfD \cdot \nuhat = - \sigmaf .
\end{equation*}
Here $\bft$ is the traction from the substrate, $\bfl$ the couple
stress exerted by the interface on the director field adjacent to it
(with $\bfL$ as given in \eqref{eqn:couple_stress_tensor}), and
$\sigmaf$ the density of free charge at a point on the surface of the
electrode.  The minus sign in front of $\sigmaf$ above is due to the
fact that $\nuhat$ is outward from $\Omega$, not outward from the
electrode.  One can understand the connection between $r$, $\sigmaf$,
$\Psi$, and work as follows.  The quantity $\eps\Psi$ corresponds to
an infinitesimal change $\delta U$ in the electric potential $U$ on
$\Gamma_1$, i.e., an infinitesimal change in the voltage on the lower
electrode.  The associated work done by the variable voltage source to
effect this change can be deduced from the relations for a
parallel-plate capacitor:
\begin{equation*}
  Q = C \, V , \quad W = \frac12 \, Q \, V = \frac12 \, C \, V^2 \! ,
\end{equation*}
where $Q$ is the total charge on the positively charged capacitor
plate ($-Q$ on the opposite plate), $C$ is the capacitance, $V$ is the
potential difference, and $W$ is the work (electrostatic energy) to
charge the capacitor in an incremental, reversible way (see
\cite[Sec.\,6.6]{reitz:milford:67}).  The incremental work associated
with an incremental change in voltage is thus
\begin{equation*}
  \delta W = C \, V \delta V = Q \, \delta V ,
\end{equation*}
which is a special case of
\cite[Eqn.\,(2.6)]{landau:lifshitz:pitaevskii:93}.  The coupling
between $r=-\sigmaf$ and $\eps\Psi=\delta U$ is reflective of this in
a pointwise sense on $\Gamma_1$ (one periodic cell of the lower
electrode plate).

On $\Gamma_2$, $U$ satisfies $U=V/2$, while $\nhat$ satisfies the
natural boundary condition \eqref{eqn:naturalBC},
and we have the relations
\begin{equation*}
  \bft = \bfT \nuhat + \frac{\partial\Ws}{\partial\xS} - \divS \Ts , ~~
  \bfl = \bfL \nuhat + \nhat \times \frac{\partial\Ws}{\partial\nhat} =
  \nhat \times \Bigl[ \Bigl( \dWdgn \Bigr)
  \nuhat + \frac{\partial\Ws}{\partial\nhat} \Bigr] = \bfzero , ~~
  \bff = \Ts \nuhat_2 ,
\end{equation*}
and $r=-\sigmaf$ (as on $\Gamma_1$).  In contrast to the relation on
$\Gamma_1$, the stress vector is not necessarily continuous across
$\Gamma_2$: the limiting values are $\bfT\nuhat$ from the interior
versus $\bft$ above from the exterior.  It suffers instead a jump
($\partial\Ws/\partial\xS - \divS\Ts$) related to the anchoring energy
supported on $\Gamma_2$.  The behavior is analogous to that of a
surface tension, which can lead to a difference in pressure on
opposite sides of an interface.  The vector $\partial\Ws/\partial\xS$
is tangential, while $\divS\Ts$ can have both tangential and normal
components.  The couple stress vector as well can be discontinuous,
with limit $\bfL\nuhat$ from inside and $\bfl=\bfzero$ from outside
(the vanishing of $\bfl$ a consequence of the natural boundary
condition satisfied by $\nhat$ on $\Gamma_2$).  Thus no torque is
transmitted to the substrate by the couple stress; it is instead
absorbed by the anchoring energy.  The quantity $\bff$ is the force
per unit length exerted by the substrate on the boundary of
$\Gamma_2$.  The couple balance on $\Gamma_2$ found here is consistent
with \cite{jenkins:barratt:74}, as conveyed in
\cite[Eqn.\,(2.13)]{barratt:jenkins:73}; while the traction balance
here contains additional terms not present in those earlier papers
(which considered the interface between a liquid crystal and an
isotropic fluid), by virtue of the more general form of anchoring
energy employed here.  In the section that follows, an illustration is
given of the forms that these boundary conditions and jumps take for a
simple example anchoring energy.

On $\Gamma_3$, we have periodic conditions on both $\nhat$ and $U$ (on
opposing sides).  In terms of the periodic extensions of the fields,
these conditions along with those related to the fluxes can be
expressed as continuity conditions:
\begin{equation*}
  \llbracket \nhat \rrbracket = \bfzero , \quad
  \llbracket U \rrbracket = 0 , \quad
  \llbracket \bfT \nuhat \rrbracket = \bfzero , \quad
  \llbracket \bfL \nuhat \rrbracket = \bfzero , \quad
  \llbracket \bfD \cdot \nuhat \rrbracket = 0 ,
\end{equation*}
where $\llbracket \cdot \rrbracket$ denotes the difference between the
limit from the exterior of $\Omega$ and the limit from the interior of
$\Omega$ at a point on $\Gamma_3$.

\subsection{Weak-anchoring boundary condition example}

For the sake of definiteness, we illustrate the boundary conditions
and jump conditions on $\Gamma_2$ for the prototypical anchoring
energy in \eqref{eqn:Ws-example},
\begin{align*}
  \Ws &= - \frac12 W_0 (\nhat\cdot\nhat_0)^2 , \quad
        \nhat_0 = a \nuhat + b \tauhat , \quad a^2 + b^2 = 1 \\
      &= - \frac12 W_0 \bigl[ a^2 (\nhat\cdot\nuhat)^2 +
        2 a b (\nhat\cdot\nuhat) (\nhat\cdot\tauhat) +
        b^2 (\nhat\cdot\tauhat)^2 \bigr] ,
\end{align*}
with $W_0$ a positive constant.  This potential conforms to the full
nematic symmetry (indifferent with respect to $O(\calV)$).  Special
cases include weak homeotropic anchoring ($a=\pm1$, $b=0$) and weak
planar anchoring ($a=0$, $b=\pm1$); the general case with both $a$ and
$b$ nonzero corresponds to a weak pre-tilt boundary condition.  For a
planar boundary $\Gamma_2$ (as in our model geometry in
Fig.\,\ref{fig:domain}), with a Cartesian frame $\ehat_1$, $\ehat_2$,
$\ehat_3$ aligned with $\tauhat$, $\tautwo$, $\nuhat$, the Cartesian
components of the surface stress tensor $\Ts$ in \eqref{eqn:Ts}
associated with $\Ws$ above are given by
\begin{equation*}
  \bigl[ \Ts \bigr] = - W_0 (\nhat\cdot\nhat_0) \!\!
  \begin{bmatrix}
    \frac12(\nhat\cdot\nhat_0) & 0 & 0 \\
    b(\nhat\cdot\tautwo) & \frac12(\nhat\cdot\nhat_0) & 0 \\
    \nhat\cdot(b\nuhat-a\tauhat) & -a(\nhat\cdot\tautwo) & 0
  \end{bmatrix} \!\! .
\end{equation*}

The extrema of $\Ws$ are at $\nhat=\pm\nhat_0$ (minima) and
$\nhat\perp\nhat_0$ (maxima), and at any point on $\Gamma_2$ at which
$\nhat$ takes on these values, $\Ts$ assumes simple forms:
\begin{equation*}
  \nhat = \pm \nhat_0 ~ \Rightarrow ~ \Ts = - \frac12 W_0
  \bigl( \ehat_1\otimes\ehat_1 + \ehat_2\otimes\ehat_2 \bigr) , \quad
  \nhat \perp \nhat_0 ~ \Rightarrow ~ \Ts = \bfzero .
\end{equation*}
It follows that
\begin{equation*}
  \nhat = \pm \nhat_0 ~ \Rightarrow ~ \bff = \Ts \nuhat_2 =
  - \frac12 W_0 \nuhat_2 ,
\end{equation*}
that is, at any point on $\partial\Gamma_2$ at which the director is
aligned with the easy axis, the external force per unit length on
$\partial\Gamma_2$ must be inward and normal.  For the model geometry
we assume in Fig.\,\ref{fig:domain}, these correspond to equal and
opposite forces exerted by adjacent periodic cells on each other.

Since $W_0$ is assumed constant, we have $\partial\Ws/\partial\xS =
\bfzero$, and the boundary condition on the stress vector above
becomes
\begin{equation*}
  \bft = \bfT \nuhat - \divS \Ts ,
\end{equation*}
with
\begin{equation*}
  \divS \Ts = \bigl( T_{11,1} + T_{12,2} \bigr) \ehat_1 +
  \bigl( T_{21,1} + T_{22,2} \bigr) \ehat_2 +
  \bigl( T_{31,1} + T_{32,2} \bigr) \ehat_3 . 
\end{equation*}
The formulas for these components are not especially illuminating.  We
give them here, for the record, excluding a common factor of $-W_0$
($T_{ij,k} = - W_0 \, \Ttilde_{ij,k}$):
\begin{gather*}
  \Ttilde_{11,1} = (\nhat\cdot\nhat_0)
  \Bigl(\dndx\cdot\nhat_0\Bigr) , \quad \Ttilde_{12,2} = 0 , \\
  \Ttilde_{21,1} = b \Bigl[ (\nhat\cdot\nhat_0)
  \Bigl(\dndx\cdot\tautwo\Bigr) + (\nhat\cdot\tautwo)
  \Bigl(\dndx\cdot\nhat_0\Bigr) \Bigr] , \quad
  \Ttilde_{22,2} = (\nhat\cdot\nhat_0)
  \Bigl(\dndxx\cdot\nhat_0\Bigr) , \\
  \Ttilde_{31,1} = (\nhat\cdot\nhat_0) \Bigl[
  \dndx\cdot(b\nuhat-a\tauhat) \Bigr] +
  \bigl[\nhat\cdot(b\nuhat-a\tauhat)\bigr]
  \Bigl(\dndx\cdot\nhat_0\Bigr) , \\
  \Ttilde_{32,2} = -a \Bigl[ (\nhat\cdot\nhat_0)
  \Bigl(\dndxx\cdot\tautwo\Bigr) +
  (\nhat\cdot\tautwo) \Bigl(\dndxx\cdot\nhat_0\Bigr) \Bigr] .
\end{gather*}
Some simplifications occur when special cases (such as $\nhat =
\nhat_0 = \nuhat$, $\nhat \perp \nhat_0 = \nuhat$, etc.)\ are 
considered.  We conclude that in general $\divS\Ts$ has nontrivial
planar and normal components and that at any point on $\Gamma_2$ at
which $\partial\nhat/\partial x_1 = \partial\nhat/\partial x_2 =
\bfzero$, we have $\divS\Ts=\bfzero$.  In particular, for
one-dimensional problems (in which fields depend only on $x_3$), the
terms from $\Ws$ play no role ($\partial\Ws/\partial\xS = \divS\Ts =
\bfzero$), and the stress vector suffers no jump across $\Gamma_2$:
$\bft = \bfT \nuhat$.

The contribution from our example $\Ws$ to the boundary condition on
the couple stress vector,
\begin{equation*}
  \bfl = \bfL \nuhat + \nhat \times
  \frac{\partial\Ws}{\partial\nhat} = \bfzero ,
\end{equation*}
is readily obtained from
\begin{equation*}
  \Ws = - \frac12 W_0 (\nhat\cdot\nhat_0)^2 ~~ \Rightarrow ~~
  \nhat \times \frac{\partial\Ws}{\partial\nhat} =
  - W_0 (\nhat\cdot\nhat_0) (\nhat\times\nhat_0)
\end{equation*}
and admits the following interpretation.  Here $\nuhat$ is the outward
normal from $\Omega$ and $-\nuhat$ the outward normal from the
substrate.  Thus $-\bfL\nuhat$ is the torque (per unit area)
transmitted to $\nhat$ on $\Gamma_2$ from the interior of $\Omega$,
and the boundary condition gives a torque balance on $\Gamma_2$ that
reads
\begin{equation*}
  \bfzero = - \bfL \nuhat -
  \nhat \times \frac{\partial\Ws}{\partial\nhat} =
  - \bfL \nuhat + W_0 (\nhat\cdot\nhat_0) (\nhat\times\nhat_0) ,
\end{equation*}
with the last term giving the restoring torque from the anchoring
potential striving to bring $\nhat$ into alignment with $\pm\nhat_0$.
If the anchoring on $\Gamma_2$ were infinitely strong (the director on
$\Gamma_2$ fixed to the boundary), then the torque from the interior
would be transmitted to the substrate.  With weak anchoring, the
torque is instead absorbed by the anchoring potential.

\section{Interpretation}

\label{sec:Interpretation}

We wish to put our findings in the context of other established
results.  We ignore, for the moment, issues related to weak anchoring
and summarize our results as follows.  For a free-energy functional of
a director field $\nhat$ and electric potential $U$ of the form
\begin{equation*}
  \calF[\nhat,U] = \!
  \int_\Omega \! W(x,\nhat,\nabla\nhat,U,\nabla U) \, \text{d}V ,
\end{equation*}
the coupled Euler-Lagrange equations are given by
\begin{equation*}
  \Div \Bigl( \dWdgn \Bigr) -
  \dWdn + \lambda\nhat = \bfzero , \quad
  \Div \Bigl( \frac{\partial W}{\partial\nabla U} \Bigr) -
  \frac{\partial W}{\partial U} = 0 ,
\end{equation*}
the latter equation being equivalent to
\begin{equation*}
  \Div\bfD = \rhof , ~~ \text{with} ~
  \bfD = \eps_0 \bfE + \bmP , ~
  \bfE = - \nabla U
\end{equation*}
when $W$ is constructed as in Sec.\,\ref{sec:free-energy-density}
using \eqref{eqn:WEgeneral}.  An equilibrium pair $(\nhat,U)$
(solutions of the strong-form Euler-Lagrange equations above)
necessarily satisfies an appropriate balance of forces
\begin{subequations}\label{eqn:summary}
\begin{equation}
  \Div \bfT + \bfF = \bfzero , \quad
  \bfF = - \dWdx
\end{equation}
and balance of torques
\begin{equation}\label{eqn:summary-torque-balance}
  2 \bfw + \bfK + \Div \bfL = \bfzero ,
\end{equation}
where
\begin{equation}\label{eqn:summary-TwKGL}
\begin{gathered}
  \bfT = ( W - p_0 ) \bfI -
  ( \nabla\nhat )^T \! \dWdgn +
  \bfE \otimes \bfD , \quad
  \bfw = \axial ( \myskew(\bfT) ) =
  \frac12 \eps_{ijk} T_{kj} \ehat_i \\
  \bfK = \nhat\times\bfG , \quad
  \bfG = - \frac{\partial\widetilde{W}}{\partial\nhat} , \quad
  \bfL = \eps_{ikl} n_k \frac{\partial W}{\partial n_{l,j}}
  \ehat_i \otimes \ehat_j .
\end{gathered}
\end{equation}
\end{subequations}
Here $\widetilde{W}$ denotes the collection of all terms included in
$W$ that depend on $\nhat$ and are \emph{not} invariant under rigid
rotations, i.e., $\widetilde{W}(\ldots,\bfR\nhat,\ldots) \not=
\widetilde{W}(\ldots,\nhat,\ldots)$ for some $\bfR\in SO(\calV)$.  For
$W$ as in \eqref{eqn:Wgeneral}, we have $\widetilde{W}=\WH$, and the
electric-field contributions to the torque balance come from
$\myskew(\bfT)$ (and $\bfL$, if flexoelectric terms are included).
These balances assure us that variational equilibrium
($\delta\!\calF=0$) would be consistent with the hydrostatic limit of
the Ericksen-Leslie equations, properly formulated with a coupled
electric field.  We first compare these results with the
``compatibility potential'' approach discussed in
Sec.\,\ref{sec:Variational_compatibility} above.

\subsection{Contrast with compatibility potential}

As discussed in Sec.\,\ref{sec:Ericksen_compatability_potential}, the
original ideas of Ericksen are quite sufficient to establish
consistency between variational equilibrium equations and
Ericksen-Leslie hydrostatics for the cases of gravitational fields
and/or magnetic fields.  In that setting, the director equilibrium
equation is written
\begin{subequations}\label{eqn:summary_one}
\begin{equation}
  \Div \Bigl( \dWedgn \Bigr) -
  \dWedn + \bfG_1 + \lambda\nhat
  = \bfzero , \quad \bfG_1 = \frac{\partial\Psi}{\partial\nhat} ,
\end{equation}
and the force and torque balances
\begin{equation}
  \Div\bfT_1 + \bfF_1 = \bfzero , \quad
  \bfF_1 = \frac{\partial\Psi}{\partial x} , \quad
  2 \bfw_1 + \bfK_1 + \Div\bfL_1 = \bfzero ,
\end{equation}
where
\begin{equation}
\begin{gathered}
  \bfT_1 = - p \bfI -
  (\nabla\nhat)^T \! \dWedgn , \quad
  p = p_0 - \We + \Psi , \quad
  \bfw_1 = \axial ( \myskew(\bfT_1) ) \\
  \bfK_1 = \nhat\times\bfG_1 , \quad
  \bfL_1 = \eps_{ikl} n_k \frac{\partial\We}{\partial n_{l,j}}
  \ehat_i \otimes \ehat_j .
\end{gathered}
\end{equation}
\end{subequations}
In the case of a combined magnetic field and gravitational field, the
compatibility potential can be written
\begin{equation*}
  \Psi = \frac12 \mu_0 \Delta\chi ( \bfH\cdot\nhat )^2 - \rhom g h =
  - \WH - \rhom g h ,
\end{equation*}
and the two approaches can be seen to be equivalent by identifying $W
= \We - \Psi$ and dropping any equations or terms involving $\bfE$ or
$\bfD$ or $U$ (which aren't present in the model in this case).

For the case of an electric field without flexoelectric polarization
taken into account, we would have
\begin{equation*}
  W = \We - \Psi , \quad \Psi = \frac12\bfD\cdot\bfE - \rhof U , \quad
  \bfD = \epstensor(\nhat) \bfE =
  \eps_0 \bigl[ \eperp \bfE + \Delta\eps (\bfE\cdot\nhat) \nhat \bigr] .
\end{equation*}
Here one could proceed with the compatibility-potential approach by
analogy with the magnetic-field case, treating $U$ and $\bfE$ as given
fields, as originally hinted at by the language in \cite{ericksen:62}
and as suggested in \cite[Sec.\,3.1.4]{sonnet:virga:12},
\cite[Secs.\,2.5,\,4.2]{stewart:04}, and
\cite[Sec.\,4.1.1]{virga:94}---though \cite[Sec.\,4.2]{stewart:04}
expresses that proceeding in this way ``may be subject to more
scrutiny.''  The results one would obtain are consistent to a good
degree with those produced by the coupled-electric-field approach we
have taken here, though there are some issues that arise.  The fields
in \eqref{eqn:summary} versus \eqref{eqn:summary_one} would be related
as follows:
\begin{gather*}
  \bfT = (W - p_0) \, \bfI -
  (\nabla\nhat)^T \! \dWdgn +
  \bfE \otimes \bfD = \bfT_1 + \bfE \otimes \bfD \\
  \bfF = - \dWdx = - U \nabla\rhof , \quad
  \bfF_1 = \frac{\partial\Psi}{\partial x} =
  (\nabla\bfE)\bfD + \rhof\bfE - U \nabla\rhof \\
  \bfw = \axial(\myskew(\bfT)) =
  \axial(\myskew(\bfT_1)) +
  \axial(\myskew(\bfE\otimes\bfD)) =
  \bfw_1 + \frac12 \bfD\times\bfE \\
  \bfG = - \frac{\partial\widetilde{W}}{\partial\nhat} = \bfzero , \quad
  \bfG_1 = \frac{\partial\Psi}{\partial\nhat} =
  \eps_0 \Delta\eps (\bfE\cdot\nhat) \bfE \\
  \bfK = \nhat\times\bfG = \bfzero , \quad
  \bfK_1 = \nhat\times\bfG_1 =
  \eps_0 \Delta\eps (\bfE\cdot\nhat) \nhat\times\bfE = \bfD\times\bfE \\
  \bfL = \eps_{ikl} n_k \frac{\partial W}{\partial n_{l,j}}
  \ehat_i \otimes \ehat_j =
  \eps_{ikl} n_k \frac{\partial\We}{\partial n_{l,j}}
  \ehat_i \otimes \ehat_j = \bfL_1 .
\end{gather*}
Using these, one sees that
\begin{align*}
  \Div \bfT + \bfF &=
  \Div \bfT_1 + \Div(\bfE\otimes\bfD) - U\nabla\rhof \\
  &= \Div \bfT_1 + (\Div\bfD)\bfE + (\nabla\bfE)\bfD - U\nabla\rhof \\
  &= \Div \bfT_1 + \rhof\bfE + (\nabla\bfE)\bfD - U\nabla\rhof \\
  &= \Div \bfT_1 + \bfF_1
\end{align*}
and
\begin{equation*}
  2 \bfw + \bfK + \Div \bfL =
  2 \bfw_1 + \bfD \times \bfE + \bfzero + \Div \bfL_1 =
  2 \bfw_1 + \bfK_1 + \Div \bfL_1 .
\end{equation*}
Thus the force systems $\bfF$, $\bfT$, $\bfK$, $\bfL$ and $\bfF_1$,
$\bfT_1$, $\bfK_1$, $\bfL_1$ are equivalent, in the sense that
\begin{equation*}
  \int_V \bfF \, \text{d}V + \! \int_S \bfT\nuhat \, \text{d}S = \!
  \int_V \bfF_1 \, \text{d}V + \! \int_S \bfT_1\nuhat \, \text{d}S
\end{equation*}
and
\begin{equation*}
  \int_V ( \bfr\times\bfF +\bfK ) \, \text{d}V + \!
  \int_S ( \bfr\times\bfT\nuhat + \bfL\nuhat ) \, \text{d}S = \!
  \int_V ( \bfr\times\bfF_1 +\bfK_1 ) \, \text{d}V + \!
  \int_S ( \bfr\times\bfT_1\nuhat + \bfL_1\nuhat ) \, \text{d}S ,
\end{equation*}
for any volume element $V$ (with boundary $S$).  That is, both systems
produce the same net force and the same net torque on all subregions.
These relationships reflect the arbitrariness of how one chooses to
define the ``body force,'' ``body couple,'' etc.\ (commented upon in
many places).

The main issue that we see with this approach to dealing with electric
fields concerns the definition of $\bfT$ versus $\bfT_1$ and the
implications for determining proper boundary tractions.  In our
development, $\bfT$ emerges as a total momentum flux tensor, and so
the total force per unit area exerted on one of the boundary
electrodes is given by
\begin{equation*}
  \bfT\nuhat = ( \bfT_1 + \bfE\otimes\bfD ) \nuhat =
  \bfT_1\nuhat + (\bfD\cdot\nuhat) \bfE =
  \bfT_1\nuhat + \sigmaf\bfE ,
\end{equation*}
where $\nuhat$ is the outward normal from the electrode and $\sigmaf$
the surface charge density at a point.  This last piece,
$\sigmaf\bfE$, is necessary---without it, the electric-field force
density on the electrode (from $\bfT_1$) is $- \frac12 \sigmaf \bfE$
instead of the correct value $\frac12 \sigmaf \bfE$.  Thus
$\bfT_1\nuhat$ does not give the correct expression for the stress
vector on an electrode, the reason for this being that the term
$\bfE\otimes\bfD$ in $\bfT$ has been put in $\bfF_1$ instead (as
$\Div(\bfE\otimes\bfD)$).  Other issues that one might raise with this
approach are that one must append electrostatics ($\Div\bfD=\rhof$) as
a side condition, instead of having it embedded naturally in the field
equations, and the point that the expression $(\nabla\bfE)\bfD$ in
$\bfF_1$ does not seem to admit a natural interpretation as a body
force.  One can make a case for the following as a candidate for the
total/net electrostatic body force density (in a linear dielectric in
equilibrium):
\begin{equation*}
  \bfF_2 = \Div \Bigl[ \bfE\otimes\bfD -
  \frac12 ( \bfD\cdot\bfE ) \bfI \Bigr] =
  \rhof\bfE + (\nabla\bfE)\bmP - \frac12 \nabla ( \bmP\cdot\bfE ) .
\end{equation*}
This expression is discussed more in what follows.  It has the
dependencies $\bfF_2 = \bfF_2(x,\nhat,\nabla\nhat)$, however, and so
it can't be expressed in the form $\bfF_2 = \partial\Psi/\partial x$,
for any $\Psi = \Psi(x,\nhat)$.  The term that prevents $\bfF_2$ from
being such a potential field is the last term,
$-\frac12\nabla(\bmP\cdot\bfE)$, which arises from a pressure gradient
and would be handled separately in the compatibility potential
framework.  The case of electric fields \emph{with} flexoelectricity
taken into account does not fit in the compatibility-potential
approach at all, and so it is not possible to make a comparison.

\subsection{Electric field without flexoelectricity}

It is worthwhile to examine more closely and contrast the cases of
an electric field without flexoelectric effects taken into account
versus an electric field with flexoelectricity, and we do this now.
As we have already observed, the former case is the more common and
falls under the umbrella of a linear dielectric medium; while the
latter case represents a nonlinear medium and also introduces a
coupling between the electric-field variables and the gradient of the
director field.

\subsubsection{No free-charge distribution}

In the absence of any free-charge distribution (the most common case),
the free-energy density would be given by
\begin{equation}\label{eqn:WDE}
  W = \We - \frac12 \bfD \cdot\bfE , \quad
  \bfD = \epstensor(\nhat) \bfE , \quad
  \bfE = - \nabla U ,
\end{equation}
where the dielectric tensor $\epstensor$ is as in \eqref{eqn:eps-tensor}.
The field equations \eqref{eqn:field_equations} would take the form
\begin{equation*}
  \Div \bfT = \bfzero , \quad
  \nhat \times \Bigl[ \Div \Bigl(
  \dWedgn \Bigr) -
  \dWedn \Bigr] +
  \bfD \times \bfE = \bfzero , \quad
  \Div \bfD = 0 ,
\end{equation*}
with the total stress tensor $\bfT$ given by
\begin{equation}\label{eqn:TeTM}
  \bfT = - p_0 \bfI + \Te + \TM , \quad
  \Te := \We \, \bfI - (\nabla\nhat)^T \!
  \dWedgn , \quad
  \TM := \bfE \otimes \bfD - \frac12 ( \bfD \cdot \bfE ) \, \bfI .
\end{equation}
Here again $\bfD\times\bfE$ is the dielectric torque, and we have
decomposed $\bfT$ into the part arising from distortional elasticity
and the part arising from the electric field.  The tensor $\TM$ is
familiar and is known as the ``Minkowski stress tensor''
\cite[Sec.\,6.9]{jackson:75}.

The effective total force per unit volume due to the electric field is
given by
\begin{equation*}
  \Div \TM = ( \Div \bfD ) \bfE + ( \nabla \bfE ) \bfD -
  \frac12 \nabla ( \bfD \cdot \bfE ) =
  ( \nabla \bfE ) \bmP - \frac12 \nabla ( \bmP \cdot \bfE ) .
\end{equation*}
Here we have used $\Div\bfD=0$ and $\bfD=\eps_0\bfE+\bmP$ to simplify
above.  The term $(\nabla\bfE)\bmP$ is familiar and is known as the
``Kelvin force'' \cite[Sec.\,13.1]{degroot:69}, \cite[Ch.\,II,
Secs.\,5.f,\,8.a]{degroot:suttorp:72}.  It is the analogue for a
polarization distribution of the formula for the force on a dipole
$\bfp_0$ at a point $x_0$ in an external field (to which the dipole
does not contribute): $\nabla \Eext(x_0)\bfp_0$
\cite[Sec.\,3.9]{stratton:41}.  The term
$- \frac12 \nabla (\bmP\cdot\bfE)$, which is also found in
\cite[Eqn.\,(347)]{degroot:suttorp:72}, is less familiar and admits
various interpretations.  The expression $- \frac12 \bmP \cdot \bfE$
gives the electrostatic energy density of a polarization in an
electric field of its own generation, to which it is linearly coupled
\cite[Sec.\,4.7]{jackson:75}, \cite[Sec.\,2.10]{stratton:41}.  This
derives from the expression for the energy of a point dipole in an
external field: $- \bfp_0 \cdot \Eext(x_0)$
\cite[Sec.\,2-8]{reitz:milford:67}, \cite[Sec.\,3.9]{stratton:41}.
Thus the term $- \frac12 \nabla ( \bmP \cdot \bfE )$ can be viewed as
a pressure-related force per unit volume (the negative gradient of a
pressure contribution $\frac12 \bmP \cdot \bfE$) expressing the desire
of the material to expand in regions of large polarization density, so
as to reduce the local concentration of polarization.  The
distinctions among the various pressure contributions (``Kelvin
pressure,'' ``Helmholtz pressure,'' etc.) is discussed in
\cite[Ch.\,II, Sec.\,8.a]{degroot:suttorp:72}.  The polarization is
not expected to be uniform, and so one would expect some form of
contribution to the force from $\nabla\bmP$.  The term
$-\frac12\nabla(\bmP\cdot\bfE)$ provides that.  We note that terms
involving gradients of $\bfE$ versus gradients of $\bmP$ are related
by various identities, such as
\begin{equation*}
  \int_V \bigl[ ( \Div\bmP ) \bfE + ( \nabla \bfE ) \bmP \bigr]
  \text{d}V = \! \int_S ( \bmP \cdot \nuhat ) \bfE \,\, \text{d}S ,
\end{equation*}
which is valid for any regular subdomain $V$ with boundary $S$.  Using
this, one can obtain
\begin{equation}\label{eqn:rhoP_sigmaP}
  \mathlarger \int_V \Bigl[ (\nabla\bfE)\bmP -
  \frac12\nabla(\bmP\cdot\bfE) \Bigr] \text{d}V = \!
  \int_V \rho_\bmP \bfE \, \text{d}V + \!
  \mathlarger \int_S \Bigl[ \sigma_\bmP \bfE -
  \frac12(\bmP\cdot\bfE)\nuhat \Bigr] \text{d}S ,
\end{equation}
where $\rho_\bmP$ and $\sigma_\bmP$ are the effective volume and
surface charge densities associated with $\bmP$
\begin{equation*}
  \rho_\bmP := - \Div \bmP , \quad
  \sigma_\bmP := \bmP \cdot \nuhat .
\end{equation*}

\subsubsection{Self-field correction}

\label{sec:self-field-correction}

There are other analogies and interpretations to attach to the term
$-\frac12\nabla(\bmP\cdot\bfE)$, and one of them concerns the notions
of ``self energies'' and ``self fields.''  Continuous distributions of
charge (volume densities and surface densities) are idealizations and
embody ``self energies,'' as is illustrated in
\cite[Sec.\,1.11]{jackson:75}.  For example, the electrostatic energy
of a discrete collection of charges $q_1,\ldots,q_n$ located at points
$p_1,\ldots,p_n$ (in the electric field of their own generation) can
be written
\begin{equation*}
  W = \frac12 \sum_i q_i U_i(p_i) , \quad
  U_i(x) = \frac1{4\pi\eps_0} \sum_{j\not=i} \frac{q_j}{|x-p_j|} ,
\end{equation*}
while the analogous formula for a volume distribution $\rho$ is
\begin{equation*}
  W = \frac12 \int_V \rho U \, \text{d}V , \quad
  U(x) = \frac1{4\pi\eps_0} \mathlarger \int_V
  \frac{\rho(x')}{|x-x'|} \, \text{d}V' .
\end{equation*}
The electrostatics of discrete systems is filled with such summations,
which exclude indices associated with divergent ``self energies,''
while potential theory is filled with improper integrals similar to
that above.  It makes no sense to exclude $x=x'$ from such an integral
expression, and the singularity usually causes no difficulty.  There
are, however, situations in which complications can arise, such as in
the case of the ``Kelvin cavity conditions''
\cite[Sec.\,6.10]{jeffreys:jeffreys:56},
\cite[Sec.\,3.28]{stratton:41}, \cite[Sec.\,6]{toupin:56} (in which
one tries to compute the electric field inside a polarized material
via the negative gradient of an integral similar to that above and
finds that the value of the limit characterizing the improper integral
depends on the shape of the shrinking cavity).  Another area in which
trouble can arise concerns the force on a surface charge distribution,
which we now recount.

Coulomb's Law states that the force on a point charge $q_0$ at a point
$x_0$ in an external electric field $\Eext$ (to which $q_0$ does not
contribute) is given by $q_0\Eext(x_0)$.  On the basis of this, one
would expect that the force density on a volume charge distribution
$\rho$ would be given by $\rho\bfE$ and the force density on a surface
charge distribution $\sigma$ would be $\sigma\bfE$.  The first formula
is correct; while the second is not.  The true force per unit area on
$\sigma$ is $\frac12\sigma\bfE$, and one can find multiple derivations
and explanations of this---the force on an electrode in a
parallel-plate capacitor is a special case.  In
\cite[Sec.\,5]{landau:lifshitz:pitaevskii:93}, the force on $\sigma$
is computed using the Maxwell stress tensor; while in
\cite[Sec.\,1.11]{jackson:75}, it is computed by what amounts to a
virtual-work argument.  In introductory books on electromagnetism,
this discrepancy ($\sigma\bfE$ versus $\frac12\sigma\bfE$) is often
explained in terms of ``self-field corrections.''  This is the point
of view taken in \cite[Sec.\,4-10]{reitz:milford:67}, where it is
argued that since electric charges do not feel that part of the
electric field to which they contribute, then the force on a surface
charge distribution should be given by
\begin{equation*}
  \bfF_\sigma = \int_S \sigma \bfE' \text{d}S , \quad
  \bfE' = \bfE - \bfE_{\text{self}} ,
\end{equation*}
where $\bfE$ is the total electric field at a point and
$\bfE_{\text{self}}$ is that part of $\bfE$ due to $\sigma$ on
$\text{d}S$ itself.  A limit argument (which is sketched in
\cite[Sec.\,4-10]{reitz:milford:67} and can be made rigorous using
results from potential theory \cite[Ch.\,VI, Sec.\,5]{kellogg:29})
gives $\bfE_{\text{self}} = \frac12 \bfE$, and the result follows.

The reason that the force on $\sigma$ requires adjustment while the
force on $\rho$ does not can be explained in terms of scaling---here we
use simple examples to illustrate results that are valid in more
generality.  The electric potentials of a volume charge density in a
ball of radius $R$ and a surface charge density on a sphere of the
same radius are given by
\begin{equation*}
  U_\rho(x) = \frac1{4\pi\eps_0} \mathlarger \int_{B_R} \!
  \frac{\rho(x')}{|x-x'|} \, \text{d}V' , \quad
  U_\sigma(x) = \frac1{4\pi\eps_0} \mathlarger \int_{S_R} \!
  \frac{\sigma(x')}{|x-x'|} \, \text{d}S' ,
\end{equation*}
with associated electric fields $\bfE=-\nabla U$, from which one can
establish that
\begin{equation*}
  U_\rho = O(R^2) , \quad \bfE_\rho = O(R) , \quad
  U_\sigma = O(R) , \quad \bfE_\sigma = O(1) , \quad
  \text{as } R \rightarrow 0 .
\end{equation*}
Thus the self field associated with $\rho$ is of vanishing importance,
while that of $\sigma$ persists in the infinitesimal limit.  The
analogous expressions for a polarization density $\bmP$ resemble those
for $\sigma$, as we now illustrate---we note in addition that $\bmP$
and $\sigma$ have the same physical dimensions (charge per unit area),
whereas $\rho$ has the dimensions of charge per unit volume.

The force on a point dipole $\bfp_0$ at a point $x_0$ in an external
field $\Eext$ (to which $\bfp_0$ does not contribute) is given by
$\nabla\Eext(x_0)\bfp_0$ (as we have already indicated above).  On the
basis of this, one would guess that the force per unit volume on a
polarization (sometimes called the ``ponderomotive force'') would be
given by $(\nabla\bfE)\bmP$ (the Kelvin force), and this has been
proposed in various places, including
\cite[Sec.\,3.5.B]{eringen:maugin:90a} and
\cite[Sec.\,4.1]{tiersten:90}.  However, the electric potential of a
prescribed polarization $\bmP$ in a ball of radius $R$ (as used for
illustration with $\rho$ and $\sigma$ above) is
\begin{equation*}
  U_\bmP(x) = \frac1{4\pi\eps_0} \mathlarger \int_{B_R} \!\!
  \frac{\bmP(x')\cdot(x-x')}{|x-x'|^3} \, \text{d}V' ,
\end{equation*}
from which we deduce that
\begin{equation*}
  U_\bmP = O(R) , \quad \bfE_\bmP = O(1) , \quad
  \text{as } R \rightarrow 0 .
\end{equation*}
Thus there is a non-negligible self field, and one would expect that a
self-field correction would be needed to produce the proper force
density on $\bmP$ (as in the case of a surface charge distribution).
While the situation does not seem to be as amenable to direct
calculation as the case of a surface charge density, the term
$-\frac12\nabla(\bmP\cdot\bfE)$ is at least a plausible candidate for
this correction: $-\frac12\bmP\cdot\bfE$ is the self energy of $\bmP$
and the negative gradient $\frac12\nabla(\bmP\cdot\bfE)$ the
associated self field (which must be subtracted as a ``correction'').
This point of view is reinforced by \eqref{eqn:rhoP_sigmaP}.  In fact,
if $\bfE$ were parallel to $\nuhat$ on $S$, then the expression in
\eqref{eqn:rhoP_sigmaP} would simplify exactly as in the case of the
force on a surface charge distribution on a conductor:
\begin{equation*}
  \bfE = E \nuhat ~~ \Rightarrow ~~
  \sigma_\bmP \bfE - \frac12 ( \bmP\cdot\bfE ) \, \nuhat =
  \sigma_\bmP \bfE - \frac12 \sigma_\bmP \bfE =
  \frac12 \sigma_\bmP \bfE .
\end{equation*}
This is just for the sake of illustration; there is no reason to
expect $\bfE$ to be parallel to $\nuhat$ in general---it would only do
so at points at which $S$ coincided with or osculated to the
equipotential surface through the point.  As a final note concerning
\eqref{eqn:rhoP_sigmaP}, we observe that the expression in the surface
integral can be represented
\begin{equation*}
  \sigma_\bmP \bfE - \frac12 ( \bmP \cdot \bfE ) \, \nuhat =
  \TP \nuhat , \quad
  \TP := \bfE \otimes \bmP - \frac12 ( \bmP \cdot \bfE ) \, \bfI ,
\end{equation*}
where $\TP$ is the part of the Minkowski tensor that involves the
polarization $\bmP$.  This is analogous to the way that the force on a
surface charge density on a conductor in vacuo can be calculated using
the Maxwell stress tensor, as mentioned earlier.

In any event, one doesn't expect the Kelvin force $(\nabla\bfE)\bmP$
(with $\bfE$ the total macroscopic electric field at a point) to
provide the net force per unit volume on a distribution of
polarization.  The point is emphasized as well in
\cite[Sec.\,16]{landau:lifshitz:pitaevskii:93}, where the following
formula is given for the net force on a dielectric solid inserted into
an electric field (for which the sources remain fixed):
\begin{equation*}
  \bfF = \int_\Omega ( \nabla \Eext ) \bmP \, \text{d}V .
\end{equation*}
The authors there hasten to point out (in a footnote) that
$(\nabla\Eext)\bmP$ does \emph{not} give the force density at a point.
Other influences are present; however, these cancel out when the
integration is done over the entire body.  Thus it seems that even the
most literal extrapolation from $\nabla\Eext(x_0)\bfp_0$ fails to give
an acceptable force per unit volume on $\bmP$.

\subsubsection{Free-charge distribution included}

\label{sec:free-charge-included}

If a free-charge distribution is included, then the expressions for
$\bfD$ and $\bfE$ above in \eqref{eqn:WDE} do not change, but $W$
would now be given by
\begin{equation*}
  W = \We + \rhof U - \frac12 \bfD\cdot\bfE .
\end{equation*}
The field equations would then take the form
\begin{equation*}
  \Div \bfT = U \nabla \rhof , \quad
  \nhat \times \Bigl[ \Div \Bigl(
  \dWedgn \Bigr) -
  \dWedn \Bigr] +
  \bfD \times \bfE = \bfzero , \quad
  \Div \bfD = \rhof ,
\end{equation*}
with the total stress tensor $\bfT$ given by
\begin{equation*}
  \bfT = ( \rhof U - p_0 ) \, \bfI + \Te + \TM ,
\end{equation*}
with $\Te$ and $\TM$ as before in \eqref{eqn:TeTM}.  The net force per
unit volume due to electrostatics (in equilibrium) would now given by
\begin{equation}\label{eqn:divTE}
  \begin{aligned}
    \nabla ( \rhof U ) + \Div \TM &=
    \rhof \nabla U + U \nabla \rhof +
    ( \Div \bfD ) \bfE + ( \nabla \bfE ) \bfD -
    \frac12 \nabla ( \bfD \cdot \bfE ) \\
    &=  - \rhof \bfE + U \nabla \rhof + \rhof \bfE +
    ( \nabla \bfE ) \bmP - \frac12 \nabla ( \bmP \cdot \bfE ) \\
    &= U \nabla \rhof + ( \nabla \bfE ) \bmP -
    \frac12 \nabla ( \bmP \cdot \bfE ) .
  \end{aligned}
\end{equation}
The Coulomb force density $\rhof\bfE$ is balanced by part of the
gradient of the $\rhof U$ term.  The other part of that
pressure-related term ($U \nabla \rhof$) is balanced by an identical
term on the right-hand side of the force balance
$\Div\bfT = U \nabla\rhof$ above, similar to the way in which the
gravitational force $- \rhom g \nabla h$ in \eqref{eqn:FdWdx} would be
balanced by an appropriate pressure gradient (as would be the case
with any body force entering via $- \partial W / \partial x$).  The
last two terms give an expression for the force per unit volume on the
polarization, and these terms must be balanced by the stresses due to
distortional elasticity, as the force balance has now been reduced to
\begin{equation}\label{eqn:divTe_vs_divTM}
  \Div \Te + ( \nabla\bfE ) \bmP - \frac12 \nabla ( \bmP \cdot \bfE )
  = \bfzero .
\end{equation}

The term $-U\nabla\rhof$ represents a force that arises due to the
artificial way that the free-charge density enters our model, in which
we treat $\rhof$ as a prescribed (fixed) scalar field.  As we have
indicated earlier, the more realistic situation in the setting of
liquid crystals is for free charges to arise from mobile ionic
impurities.  In that case, the distribution $\rhof$ would influence
and be influenced by the electric field and potential, which in turn
influence and are influenced by the director field, and $\rhof$, $U$,
and $\nhat$ would need to be computed in a coupled, self-consistent
way.  Poisson-Boltzmann theory would need to be employed
\cite[Ch.\,7]{barbero:evangelista:06},
\cite[Sec.\,8.5]{jakli:saupe:06}---we do not pursue this here.  While
the force $-U\nabla\rhof$ is thus somewhat artificial in our setting,
the pressure gradient associated with $\rhof U$ properly balances both
this force and the Coulomb force on the free charges in equilibrium,
as seen above.

The expression $\rhof\bfE + (\nabla\bfE)\bmP$ can be found in several
places in the continuum-mechanics literature as a proposed form for
the body force due to an electric field in a dielectric fluid or
solid.  In \cite[Sec.\,4.1]{tiersten:90}, for example, a derivation of
it is suggested based upon the formulas for the force on a point
charge and that on a point dipole in an external electric field.  In
\cite[Sec.\,3.5.B]{eringen:maugin:90a}, on the other hand, the authors
put forward a derivation based on statistical averaging of the
microscopic Maxwell-Lorentz equations, along the lines of
\cite{degroot:suttorp:72}.  The development in
\cite[Sec.\,3.5.B]{eringen:maugin:90a} (once magnetic terms are
removed) produces three contributions to the force density: the
Coulomb force $\rhof\bfE$, a force on polarization of the form
$(\nabla\bfE)\bmP$, and the negative gradient of a pressure, which is
left to be modeled constitutively.  For the constitutive theory of
electromagnetic fluids later developed
\cite[Sec.\,5.12]{eringen:maugin:90a}, the authors incorporate
$-\frac12\bmP\cdot\bfE$ into the free energy ($\frac12\bmP\cdot\bfE$
into the pressure), effectively reproducing the third, fourth, and
fifth terms in the middle step in \eqref{eqn:divTE} above.  A somewhat
similar statistical-averaging approach can be found in \cite{eu:86}.

The authors of \cite{eringen:maugin:90a,eringen:maugin:90b}
acknowledge the arbitrariness of what contributions are put in the
``body force'' versus the stress tensor
\cite[Sec.\,3.10]{eringen:maugin:90a}, and while they choose to take a
body force of the form $\rhof\bfE + (\nabla\bfE)\bmP$ (in the
electrostatic setting), when they derive jump conditions at material
interfaces and mechanical tractions on boundaries, they move these
terms into associated tensors---see
\cite[Secs.\,3.10,\,3.14]{eringen:maugin:90a}.  The relation
\begin{equation*}
  \Div\TM = \rhof\bfE + (\nabla\bfE)\bmP -
  \frac12\nabla(\bmP\cdot\bfE) ,
\end{equation*}
with the Minkowski tensor $\TM$ as given in \eqref{eqn:TeTM},
can in fact be derived as an identity directly from the macroscopic
Maxwell equations of electrostatics, using only $\curl\bfE=\bfzero$,
$\bfD=\eps_0\bfE+\bmP$, and $\Div\bfD=\rhof$ (with no assumptions
about how $\bmP$ depends on $\bfE$)
\cite{bocker:16,eu:86,eu:oppenheim:86}.  By rearranging terms,
however, other similar-looking identities can be obtained.  The
arbitrariness of various such expressions and decompositions
(including those involving electromagnetic versus mechanical forces)
is emphasized in \cite{bocker:16}, where the author encourages working
with \emph{total} momentum-flux tensors (as is urged by others as
well).

\subsubsection{Representations and tensors in the literature}

The same expressions can be written in different (but equivalent) ways
and can be found in different forms in the literature.  The
relations $\curl \bfE = \bfzero$ and $\bfD = \eps_0 \bfE + \bmP$ imply
\begin{equation*}
  ( \nabla \bfE )^T \!\! = \nabla \bfE ~~~ \text{and} ~~~
  ( \nabla \bfE ) \bfD - \frac12 \nabla ( \bfD \cdot \bfE ) =
  ( \nabla \bfE ) \bmP - \frac12 \nabla ( \bmP \cdot \bfE ) ,
\end{equation*}
which in the notation of \cite{eu:86,eu:oppenheim:86} is written
$\frac12 [\bfD,\bfE] = \frac12 [\bmP,\bfE]$, with
$[\mathbf{A},\mathbf{B}] := (\nabla\mathbf{B})^T \! \mathbf{A} -
(\nabla\mathbf{A})^T \mathbf{B}$.  When combined with the assumption
of a linear dielectric ($\bfD = \epstensor \bfE$) with $\epstensor^T
\!\! = \epstensor$, these become
\begin{equation*}
  ( \nabla \bfE ) \bfD - \frac12 \nabla ( \bfD \cdot \bfE ) =
  - \frac12 \Eps_{ij,k} E_i E_j \ehat_k ,
\end{equation*}
which is the form of the ponderomotive force found in
\cite{bocker:16,eu:oppenheim:86}.  In a perfect linear dielectric, the
dielectric tensor $\epstensor$ is taken to be real and symmetric
\cite[Sec.\,13]{landau:lifshitz:pitaevskii:93},
\cite[Sec.\,15-3]{reitz:milford:67}.  If the linear medium is
isotropic (the dielectric tensor scalar), then the above expression
simplifies to $- \frac12 E^2 \nabla \eps$, which forms part of the
Helmholtz ponderomotive force as found in
\cite[Sec.\,13.1]{degroot:69}, \cite[Ch.\,II,
Sec.\,8.a]{degroot:suttorp:72},
\cite[Sec.\,15]{landau:lifshitz:pitaevskii:93},
\cite[Sec.\,2.22]{stratton:41}.  Whether isotropic or not, the force
density vanishes if the medium is homogeneous.  For the case of a
uniaxial nematic liquid crystal (absent flexoelectric effects), we
would have a linear, anisotropic dielectric medium
\begin{equation*}
  \bfD = \epstensor \bfE , \quad
  \epstensor = \eps_0 \bigl[ \eperp \bfI +
  \epsa ( \nhat\otimes\nhat ) \bigr] ,
\end{equation*}
as in \eqref{eqn:eps-tensor}, and the polarization force above would
simplify to
\begin{equation*}
    ( \nabla \bfE ) \bfD - \frac12 \nabla ( \bfD \cdot \bfE ) =
    - \eps_0 \epsa ( \bfE\cdot\nhat ) ( \nabla\nhat )^T \!\! \bfE .
\end{equation*}
As is the case with the general expressions, this force density
depends on the gradient of the dielectric tensor, here through the
term $\nabla\nhat$, vanishing in regions of uniform orientation
($\nhat=\text{const}$).  This force density also vanishes at any
points where $\bfE$ is either parallel to $\nhat$ (because of
$(\nabla\nhat)^T\!\nhat=\bfzero$) or perpendicular to $\nhat$ (because
of $\bfE\cdot\nhat=\bfzero$).  We note that this expression is
invariant with respect to both $\pm\nhat$ and $\pm\bfE$, as it should
be.

The form of the tensor $\bfT$ that we have derived in
\eqref{eqn:Tfinal} is consistent with results in the physics
literature, including those found in \cite{degroot:69,jackson:75,%
  landau:lifshitz:pitaevskii:93,stratton:41}, the closest related
results being those found in
\cite[Sec.\,15]{landau:lifshitz:pitaevskii:93}.  There a dielectric
fluid is modeled as a single species of fixed total mass in a fixed
volume, with a thermodynamic state that depends on mass density
$\rhom$, temperature $T$\!, and electric field $\bfE$ at each point.
The authors employ what amounts to a virtual-work argument, involving
an isothermal deformation at constant electric potential, and deduce
the following expression for a general stress tensor (in the notation
of that book):
\begin{equation*}
  \sigma_{ik} = E_i D_k + \Bigl[ \widetilde{F} - \rhom \Bigl(
  \frac{\partial\widetilde{F}}{\,\partial\rhom} \Bigr) \Bigr]
  \delta_{ik} .
\end{equation*}
Here $\widetilde{F} = \widetilde{F}(\rhom,T,\bfE)$ is the Helmholtz
free energy per unit volume, and we have transcribed from expressions
written in terms of Gaussian units in the book.  No additional
assumptions about the nature of $\widetilde{F}$ have been made at this
point.  For an incompressible material (and isochoric deformation),
the term involving $\partial \widetilde{F} / \partial \rhom$ would not
arise, and the expression would read
\begin{equation}\label{eqn:sigma}
  \bfsigma = \bfE \otimes \bfD + \widetilde{F} \, \bfI ,
\end{equation}
in our notation.

For the case of a linear isotropic dielectric ($\bfD = \eps \bfE$,
$\eps$ a scalar field), $\widetilde{F}$ would be given by
\begin{equation*}
  \widetilde{F} = F_0(\rhom,T) - \frac12 \eps E^2 ,
\end{equation*}
where $F_0$ is the free energy density when no electric field is
present.  In this case, the stress tensor would take the form
\begin{equation*}
  \sigma_{ik} = - P_0 \delta_{ik} - \frac12 \Bigl[ \eps -
  \rhom \Bigl( \frac{\partial\eps}{\,\partial\rhom} \Bigr) \Bigr]
  E^2 \delta_{ik} + \eps E_i E_k ,
\end{equation*}
where
\begin{equation*}
  P_0 = P_0(\rhom,T) =
  - \frac{\partial\,\,}{\,\,\partial\rho_{\text{m}}^{-1}} \Bigl(
  \frac{F_0}{\,\rhom} \Bigr) ,
\end{equation*}
the thermodynamic pressure in the absence of an electric field
(characterized as the negative of the derivative of the free energy
per unit mass with respect to the specific volume $1/\rhom$
\cite[Ch.\,II, Sec.\,8.a]{degroot:suttorp:72}).
Discarding the term $\partial\eps/\partial\rhom$ (which would not be
present in a model of an incompressible material) and generalizing to
an anisotropic linear dielectric ($\bfD = \epstensor \bfE$, $\epstensor$ a
tensor field), the expression can be written
\begin{equation*}
  \bfsigma = ( \WE - P_0 ) \bfI + \bfE \otimes \bfD , \quad
  \WE = - \frac12 \bfD \cdot \bfE , \quad
  \bfD = \epstensor \bfE .
\end{equation*}
This is equivalent to \eqref{eqn:Tfinal} when $W$ does not depend on
$\nabla\nhat$ and we make the identification $P_0 = p_0 + \WE - W$.
The thermodynamic pressure in the absence of an electric field, $P_0$,
contains the constant hydrostatic pressure $p_0$ and all contributions
from $W$ apart from $\WE$.  Thus except for the contributions due to
distortional elasticity and flexoelectric polarization (both of which
involve $\nabla\nhat$), \eqref{eqn:Tfinal} agrees with
\cite[Eqn.\,(15.9)]{landau:lifshitz:pitaevskii:93} under the
assumption of incompressibility.  Similar expressions can be found in
\cite[Sec.\,13]{degroot:69} and \cite[Secs.\,2.21,\,2.23]{stratton:41},
both of which are referenced in the brief discussion in
\cite[Sec.\,6.9]{jackson:75}.

The Minkowski tensor $\TM$ in \eqref{eqn:TeTM}, which arises in a
natural way in our development (in the case of a linear dielectric
medium), is one of several tensors found in the electromagnetics
literature.  Two other prominent ones are the Einstein-Laub tensor
\begin{equation*}
  \TEL = \bfE \otimes \bfD - \frac12 \eps_0 E^2 \bfI
\end{equation*}
and the Hertz-Abraham tensor, given by the symmetric part of $\TEL$
(see \cite{bocker:16,eu:86,eu:oppenheim:86} or
\cite[Sec.\,3.6]{eringen:maugin:90a}).  Here we are restricting
attention to the case of electrostatics---these tensors all have
magnetic-field contributions in their general forms.  All of these
tensors also reduce to the classical Maxwell stress tensor
$\eps_0 \bigl[ \bfE\otimes\bfE - \frac12 E^2 \bfI \bigr]$ in the
absence of matter.  The Minkowski and Einstein-Laub tensors are
closely related:
\begin{equation*}
  \bfD = \eps_0 \bfE + \bmP ~~ \Rightarrow ~~
  \TM = \TEL - \frac12 ( \bmP\cdot\bfE ) \bfI ,
\end{equation*}
so that
\begin{equation*}
  \Div \TEL = \rhof \bfE + ( \nabla\bfE ) \bmP ~~ \text{vs} ~~
  \Div \TM = \rhof \bfE + ( \nabla\bfE ) \bmP -
  \frac12 \nabla ( \bmP\cdot\bfE ) .
\end{equation*}
The Einstein-Laub tensor appears in several places in the
continuum-mechanics literature on electromagnetics, including
\cite[Sec.\,3.6]{eringen:maugin:90a} and
\cite[Sec.\,4.2]{tiersten:90}, because $\rhof\bfE+(\nabla\bfE)\bmP$ is
the adopted expression for the electrostatic body force in those
theories (with $\frac12\bmP\cdot\bfE$ relegated to a pressure
contribution).  The Hertz-Abraham tensor is found in developments that
insist on the stress tensors being symmetric.

\subsection{Electric field with flexoelectricity}

The introduction of flexoelectric terms into our model changes some
aspects.  As previously noted, the material can no longer be viewed as
a linear dielectric medium, and the flexoelectric terms also introduce
couplings between $\bfE$ and $\nabla\nhat$, in addition to the
couplings between $\bfE$ and $\nhat$ already present in the
non-flexoelectric case.  It is somewhat surprising that this has such
a small impact on our general development: the basic forms of the
equilibrium equations \eqref{eqn:E-L}, force and torque balance
\eqref{eqn:balances}, stress tensor $\bfT$ \eqref{eqn:Tfinal}, and
couple stress tensor $\bfL$ \eqref{eqn:couple_stress_tensor} all read
the same whether flexoelectric terms are present or not.  There are,
however, some new contributions made by the flexoelectric terms, and
we explore them in this section.  It is difficult to find in the
literature discussions of forces and stresses in liquid-crystal
materials subjected to electric fields with flexoelectric effects
included.  An exception is \cite{eringen:79}, which treats macroscopic
models of liquid crystals as a special case of a ``micropolar''
continuum mechanics theory.

We set aside the contributions of a free-charge density $\rhof$,
which would play out as already seen in
Sec.\,\ref{sec:free-charge-included}, and consider a free-energy
density of the form
\begin{gather*}
  W = \We(\nhat,\nabla\nhat) + \WE(\nhat,\nabla\nhat,\nabla U) , ~~
  \WE =- \frac12 \epstensor(\nhat) \bfE \cdot \bfE -
  \Pf(\nhat,\nabla\nhat) \cdot \bfE , ~~ \bfE = - \nabla U \\
  \epstensor = \eps_0 \bigl[ \eperp \bfI +
  \Delta\eps ( \nhat\otimes\nhat ) \bigr] , ~~
  \Pf = \es (\Div\nhat) \nhat + \eb (\nhat\times\curl\nhat) , ~~
  \bfD = \epstensor(\nhat) \bfE + \Pf(\nhat,\nabla\nhat) .
\end{gather*}
The main elements of interest to us (equilibrium Euler-Lagrange
equations, stress tensor and force balance, couple stress tensor and
torque balance) all follow from this.  The equilibrium equation for
the electric potential has the general form
\begin{equation*}
  \Div \Bigl( \frac{\partial W}{\partial\nabla U} \Bigr) -
  \frac{\partial W}{\partial U} = 0 .
\end{equation*}
For $W$ as taken above, we have
\begin{equation*}
  \frac{\partial W}{\partial\nabla U} = \epstensor \bfE + \Pf = \bfD , ~~
  \frac{\partial W}{\partial U} = 0 ~~\: \Rightarrow ~~
  \Div \bfD = \bfzero ~~\: \text{or} ~~ \Div(\epstensor\bfE) = - \Div\Pf ,
\end{equation*}
with $-\Div\Pf$ the effective charge density associated with the
flexoelectric polarization $\Pf$.  If we had included a free-charge
distribution in the model, then the above would have become
\begin{equation*}
  \Div \bfD = \rhof ~~\: \text{or} ~~ \Div(\epstensor\bfE) = \rhof - \Div\Pf .
\end{equation*}

The equilibrium equation for the director can be written
\begin{equation*}
  \Div \Bigl( \dWdgn \Bigr) -
  \dWdn + \lambda \nhat = 0 .
\end{equation*}
For our present model, these terms take the forms
\begin{equation*}
  \dWdn =
  \dWedn - \frac12
  \frac{\partial\,}{\partial\nhat}
  \bigl[ \epstensor(\nhat) \bfE \cdot \bfE \bigr] -
  \frac{\partial\,}{\partial\nhat}
  \bigl[ \Pf(\nhat,\nabla\nhat) \cdot \bfE \bigr] ,
\end{equation*}
with
\begin{equation*}
  \frac{\partial\,}{\partial\nhat}
  \bigl[ \epstensor(\nhat) \bfE \cdot \bfE \bigr] =
  2 \eps_0 \Delta\eps ( \bfE\cdot\nhat ) \bfE , ~~
  \frac{\partial\,}{\partial\nhat}
  \bigl[ \Pf(\nhat,\nabla\nhat) \cdot \bfE \bigr] =
  \es (\Div\nhat) \bfE + \eb ( \curl\nhat\times\bfE ) ,
\end{equation*}
and
\begin{equation*}
  \dWdgn =
  \dWedgn -
  \frac{\partial\,\,}{\partial\nabla\nhat}
  \bigl[ \Pf(\nhat,\nabla\nhat) \cdot \bfE \bigr] ,
\end{equation*}
with
\begin{equation}\label{eqn:dPfdgn}
  \frac{\partial\,\,}{\partial\nabla\nhat}
  \bigl[ \Pf(\nhat,\nabla\nhat) \cdot \bfE \bigr] =
  \es (\nhat\cdot\bfE) \bfI +
  \eb ( \nhat\otimes\bfE - \bfE\otimes\nhat ) .
\end{equation}
Using these, we can write the Euler-Lagrange equation for the director
in the form
\begin{equation*}
  \Div \Bigl( \dWedgn \Bigr) - \dWedn + \bfG' + \lambda\nhat = \bfzero ,
\end{equation*}
with
\begin{equation*}
  \bfG' = \eps_0 \Delta\eps (\bfE\cdot\nhat) \bfE +
  (\es+\eb) \bigl[ (\Div\nhat) \bfE - (\nabla\nhat)^T\!\!\bfE \bigr] +
  (\eb-\es) (\nabla\bfE) \nhat .
\end{equation*}
Here we have deliberately segregated the terms arising from
distortional elasticity in an attempt to write the equation in a form
resembling the classical form \eqref{eqn:ang_mom}.  We see that the
part associated with the induced polarization ($\epstensor\bfE$)
contributes to the generalized force $\bfG'$ in the usual way (the
first term of $\bfG'$ above coinciding with the first term of $\bfG$
in \eqref{eqn:GEH}), while the flexoelectric polarization contributes
the second and third terms of $\bfG'$ above.  The associated couple
density on the director field is $\nhat\times\bfG'$, the first term of
which ($\eps_0 \Delta\eps (\bfE\cdot\nhat) (\nhat\times\bfE)$) is the
familiar dielectric torque.  This term is easy to interpret, as it
simply strives to rotate $\nhat$ into alignment parallel to the
direction of $\pm\bfE$ (for $\Delta\eps>0$), perpendicular to $\bfE$
(for $\Delta\eps<0$).  The second and third terms of
$\nhat\times\bfG'$ are more difficult to interpret, best left for the
analysis of specific systems.  We note however the important
distinction that the dielectric torque
$\eps_0 \Delta\eps (\bfE\cdot\nhat) (\nhat\times\bfE)$ is invariant
with respect to $\bfE$ versus $-\bfE$, while the torque associated
with the flexoelectric polarization reverses when the direction of
$\bfE$ is reversed (a characteristic of flexoelectric effects).  We
point out that $\bfG'$ here is not the same as $\bfG$ in
\eqref{eqn:torque_balance} and \eqref{eqn:summary-TwKGL}.  The vector
field $\bfG'$ here is what one obtains if one writes the director
equilibrium equation for this case in the classical form
\eqref{eqn:ang_mom} and \eqref{eqn:bfG}; whereas $\bfG$ in
\eqref{eqn:torque_balance} and \eqref{eqn:summary-TwKGL} gives the
body couple $\bfK=\nhat\times\bfG$ as it naturally emerges in the
balance of torques in our development $(\ref{eqn:balances})_2$ and
\eqref{eqn:summary-torque-balance} (with the couples from the electric
field coming from $\myskew(\bfT)$ and $\bfL$).

The force balance has the general form
\begin{equation*}
  \Div \bfT - \dWdx = \bfzero , \quad
  \bfT = (W-p_0) \bfI - (\nabla\nhat)^T \! \dWdgn + \bfE\otimes\bfD .
\end{equation*}
For the case of interest in this section, the total stress tensor can
be written
\begin{equation*}
  \bfT = - p_0 \bfI + \Te - (\nabla\nhat)^T \! \dWEdgn +
  \WE \bfI + \bfE\otimes\bfD ,
\end{equation*}
with $\Te$ as in \eqref{eqn:TeTM}.  The last two terms above contain
contributions of Minkowski type plus contributions from the
flexoelectric polarization $\Pf$:
\begin{equation*}
  \WE \bfI + \bfE \otimes \bfD =
  \bfE \otimes \epstensor \bfE -
  \frac12 ( \epstensor\bfE\cdot\bfE ) \bfI +
  \bfE \otimes \Pf - ( \Pf\cdot\bfE ) \bfI .
\end{equation*}
The expression $\WE\bfI + \bfE\otimes\bfD$ is along the lines of the
general expression \eqref{eqn:sigma} from
\cite[Sec.\,15]{landau:lifshitz:pitaevskii:93}, while the coupling
between $\bfE$ and $\nabla\nhat$ introduces the term $-(\nabla\nhat)^T
\! \partial\WE/\partial\nabla\nhat$, which (using \eqref{eqn:dPfdgn})
takes the form
\begin{equation*}
  - ( \nabla\nhat )^T \! \dWEdgn =
  \es ( \nhat \cdot \bfE ) ( \nabla\nhat )^T \!\! -
  \eb ( \nabla\nhat )^T \!\! \bfE \otimes \nhat .
\end{equation*}
As with the torque density associated with $\Pf$, the above
contribution to the stress is linear in $\bfE$ but properly invariant
with respect to $\nhat$ versus $-\nhat$.

As recalled in \eqref{eqn:summary-torque-balance} and
\eqref{eqn:summary-TwKGL}, the torque balance takes the form
\begin{equation*}
  \eps_{ijk} T_{kj} \ehat_i + \bfK + \Div \bfL = \bfzero ,
\end{equation*}
with $\bfT$ the total stress tensor, $\bfK$ the body couple, and
$\bfL$ the couple stress tensor.  For the case of interest in this
section, we have $\bfK = \bfzero$, and $\bfT$ and $\bfL$ can be
written in the forms
\begin{gather*}
  \bfT = ( \We + \WE - p_0 ) \bfI -
  ( \nabla\nhat )^T \! \dWedgn - ( \nabla\nhat )^T \! \dWEdgn +
  \bfE \otimes \bfD \\
  L_{ij} = \eps_{ikl} n_k \frac{\partial W}{\,\partial n_{l,j}} =
  \eps_{ikl} n_k \frac{\partial\We}{\,\partial n_{l,j}} +
  \eps_{ikl} n_k \frac{\partial\WE}{\,\partial n_{l,j}} .
\end{gather*}
Only the skew part of $\bfT$ contributes to the torque balance, and
the terms involving $\We$ contain the usual contributions from
distortional elasticity.  While $\bfD$ is now of the form
$\bfD = \epstensor \bfE + \Pf$ (as opposed to $D = \epstensor \bfE$,
in the case of a linear dielectric medium), the contribution from
$\bfE \otimes \bfD$ retains the same form:
\begin{equation*}
  \eps_{ijk} ( \bfE \otimes \bfD )_{kj} \ehat_i = \bfD \times \bfE .
\end{equation*}
The newly appearing contributions from $\Pf$ again enter by virtue of
the coupling between $\bfE$ and $\nabla\nhat$, via $\partial\WE
/ \partial\nabla\nhat$.  From \eqref{eqn:dPfdgn} we obtain
\begin{equation*}
  - ( \nabla\nhat )^T \! \dWEdgn =
  \es ( \nhat\cdot\bfE ) ( \nabla\nhat )^T \!\! -
  \eb ( \nabla\nhat )^T \!\! \bfE \otimes \nhat ,
\end{equation*}
which can be used to obtain the flexoelectric contribution to the
torque density from the skew part of $\bfT$:
\begin{equation*}
  \eps_{ijk} \Bigl[ - ( \nabla\nhat )^T \! \dWEdgn \Bigr]_{kj} \ehat_i =
  \es ( \nhat\cdot\bfE ) \eps_{ijk} n_{j,k} \ehat_i -
  \eb \, \nhat \times ( \nabla\nhat )^T \!\! \bfE .
\end{equation*}

The flexoelectric terms also contribute to the couple stress tensor
$\bfL$.  Again using \eqref{eqn:dPfdgn}, we obtain
\begin{equation*}
  \eps_{ikl} n_k \frac{\partial\WE}{\,\partial n_{l,j}}
  \ehat_i \otimes \ehat_j = -
  \es ( \nhat\cdot\bfE ) \bfW(\nhat) +
  \eb ( \nhat\times\bfE ) \otimes \nhat ,
\end{equation*}
where $\bfW(\nhat)$ is the skew tensor associated with the vector
$\nhat$ ($W_{ij} = \eps_{ikj} n_k$).  At a point on a surface element
with normal $\nuhat$, the above contribution to the couple stress
gives
\begin{equation*}
  \eps_{ikl} n_k \frac{\partial\WE}{\,\partial n_{l,j}} \nu_j \ehat_i = -
  \es ( \nhat\cdot\bfE ) ( \nhat\times\nuhat ) +
  \eb ( \nhat\cdot\nuhat ) ( \nhat\times\bfE ) ,
\end{equation*}
which involves an interplay among $\nhat$, $\bfE$, and $\nuhat$.  The
first term strives to rotate $\nhat$ towards $\nuhat$ (if
$- \es (\nhat\cdot\bfE) > 0$), away from $\nuhat$ (if
$- \es (\nhat\cdot\bfE) < 0$); while the second term strives to rotate
$\nhat$ toward $\bfE$ (if $\eb (\nhat\cdot\nuhat) > 0$), away from
$\bfE$ (if $\eb (\nhat\cdot\nuhat) < 0$).  Both effects reverse when
$\bfE$ is reversed, and the flexoelectric coefficients $\es$ and $\eb$
can be positive or negative \cite[Sec.\,4.2]{lagerwall:99}---a
collection of results from experimental measurements of flexoelectric
coefficients can be found in \cite[App.\,A]{buka:eber:13}.  These
various flexoelectric contributions are somewhat difficult to
interpret in general.  They involve both forward and inverse
influences: director distortion affecting electric fields by induced
polarization, and electric fields inducing director distortion (the
so-called ``inverse flexoelectric effect''
\cite[Sec.\,3.3.2]{degennes:prost:93},
\cite[Sec.\,4.3]{lagerwall:99}).

\section{Conclusions}

\label{sec:Conclusions}

We have considered macroscopic models for the equilibrium
orientational properties of a nematic liquid crystal and have included
all the force fields of general interest (gravitational, magnetic,
electric), focusing on the most common case of an electric field
arising from electrodes held at constant potential and taking into
account the coupling between the liquid crystal director field $\nhat$
and the electric potential field $U$.  The starting point for the
analysis has been a free energy that is expressed as an integral
functional of $\nhat$ and $U$ and which includes a surface anchoring
energy on part of the boundary of the region containing the material.
From this (and the associated virtual-work principle), we have
followed the same paths as taken by Ericksen and Leslie to deduce
appropriate expressions for force balance (and stress tensor) and
torque balance (and couple stress tensor) corresponding to hydrostatic
equilibrium.  A main difference here is the inclusion of the electric
potential as a state variable to be determined in a coupled,
self-consistent way together with the director field.

We have shown that the solutions of the coupled equilibrium
Euler-Lagrange equations for $\nhat$ and $U$ necessarily satisfy the
force and torque balance laws associated with hydrostatic equilibrium.
We have done this by exploiting the analogy with ideas related to
Noether's Theorem, which guarantees the satisfaction of appropriate
conservation laws when variational models possess continuous
symmetries (here related to translations and rotations).  When one
starts with a well-formed work function (stored-energy function), one
should be able to deduce the satisfaction of such balances by
variational equilibrium fields.  As already observed in
\cite[Sec.\,10]{toupin:56} (in the context of an elastic dielectric
solid with a stored energy and virtual-work principle): ``If the
energy principle is used, we see that a single scalar function of the
variables of state is sufficient to characterize the mechanical and
electrostatic properties of an elastic dielectric completely.''  This
has been borne out here in a similar setting, though involving a
complex dielectric fluid instead of an elastic dielectric solid.  The
issue of ``variational compatibility,'' which confronts versions of
related models for liquid crystals in which the forms of external
force fields are postulated, here becomes a moot point, as all such
fields arise from potentials that are built into the free energy and
hydrostatic equilibrium conditions result automatically.

We have considered the three types of boundary conditions typically
used in modeling liquid crystal systems: strong anchoring (Dirichlet
boundary conditions), weak anchoring (natural boundary conditions
associated with anchoring potentials), and periodic boundary
conditions.  On the part of the boundary subject to weak anchoring,
the surface anchoring energy supported there acts somewhat like a
surface tension in that it can lead to discontinuities in the stress
vector and couple stress vector, in much the same way that a surface
tension can lead to a jump in pressure across the interface supporting
it.  Here there is, in fact, no couple stress exerted by the liquid
crystal material on a weak-anchoring substrate---the couple is instead
absorbed by the anchoring energy.

We have endeavored to put our results in the context of related
results in the physics and continuum-mechanics literatures.  Our
results are identical to those obtained by Ericksen using the approach
of a ``compatibility potential'' in the cases in which the only force
fields involved are gravitational and/or magnetic.  In a sense, our
development merely provides a generalization of those ideas to cases
involving electric fields, in which the coupling between the director
field and the electric field is taken into account (and which can
involve couplings to $\nabla\nhat$ as well, if flexoelectric effects
are included).  We have examined carefully the similarities and
differences between the coupled-electric-field approach taken here and
the approach of treating the electric field (in the non-flexoelectric
setting) in a ``compatibility potential manner'' analogous to the way
that a prescribed non-homogeneous magnetic force field could be
treated.  This latter approach, though advocated in certain places,
appears to us to have some difficulties, which we have discussed.  In
general, these various force fields are distinguished by the facts
that a gravitational field is a true external field (independent of
the state of the material at a point), a magnetic field can be treated
as an external field (by virtue of the weakness of the dependence on
the material state), while an electric field should not be treated as
an external field (since its dependence on the state of the system is
non-negligible).

A main result here is the derivation of the stress tensor
\begin{equation*}
  \bfT = ( W - p_0 ) \bfI - (\nabla\nhat)^T \! \dWdgn + \bfE \otimes \bfD ,
\end{equation*}
found in \eqref{eqn:Tfinal} and \eqref{eqn:summary-TwKGL}.  This
tensor embodies the total stress (total momentum flux) associated with
mechanical and electromagnetic influences in our setting, that is, in
the setting of a macroscopic model for the equilibrium orientational
properties of a material in a nematic liquid crystal phase that is
incompressible, inhomogeneous, anisotropic, subject to gravitational
and/or magnetic and/or electric force fields, and allows for
flexoelectric effects as well as a prescribed distribution of free
electric charge.  This expression for $\bfT$ agrees with all of the
expressions we have been able to find in the literature for all of the
special cases for which we have found results.  In the absence of an
electric field, this tensor coincides with the classical results of
Ericksen for the Cauchy stress in a nematic liquid crystal subject to
gravitational and/or magnetic fields, as found in the textbook
treatments in de\,Gennes and Prost \cite{degennes:prost:93}, Sonnet
and Virga \cite{sonnet:virga:12}, and Stewart \cite{stewart:04}.  In
the absence of flexoelectricity, electrostatic parts may be segregated
into
\begin{equation*}
  \WE \bfI + \bfE \otimes \bfD ,
\end{equation*}
where $\WE$ denotes the electrostatic contribution to the free-energy
density, and this generic expression agrees with those found in Landau
and Lifshitz \cite{landau:lifshitz:pitaevskii:93} and Stratton
\cite{stratton:41} when the results in those books are specialized to
incompressible materials.  The term $\bfE\otimes\bfD$ arises no matter
what constitutive model is used for the polarization $\bmP(\bfE)$, so
long as the recipe \eqref{eqn:WEgeneral} is used (together with
$\bfD = \eps_0 \bfE + \bmP(\bfE)$) to construct $\WE$.

In the classical setting of a linear dielectric medium
($\WE = - \frac12 \bfD\cdot\bfE$, $\bfD = \epstensor \bfE$), the
expression above coincides with the Minkowski stress tensor $\TM$ in
\eqref{eqn:TeTM}.  In that case, the net electrostatic force density
($\Div\TM$) consists of the Coulomb force on the free-charge
distribution ($\rhof\bfE$), the Kelvin force on the polarization
($(\nabla\bfE)\bmP$), plus a term from the pressure
($- \frac12 \nabla (\bmP\cdot\bfE)$), which we have attempted to
interpret in various ways (by analogy to the force of an electric
field on a surface charge density on a conductor, and by the related
notion of a ``self-field correction'').  The combined terms
$(\nabla\bfE) \bmP - \frac12 \nabla (\bmP\cdot\bfE)$ agree with
expressions for the ``ponderomotive force'' found in the literature
for linear dielectrics: $-\frac12 \Eps_{ij,k} E_i E_j \ehat_k$ (in
general), $-\frac12 E^2 \nabla\eps$ (if the medium is isotropic).
While it is more common in continuum mechanics to handle pressure
gradients separately from body forces, the combination of $(\nabla
\bfE) \bmP$ with $-\frac12 \nabla (\bmP\cdot\bfE)$ is necessary here
to obtain agreement with these other expressions of ponderomotive
force.  The $O(1)$ nature of the electric self field associated with
the polarization density in a shrinking volume element $| \Delta V |
\rightarrow 0$ (as discussed in Sec.\,\ref{sec:self-field-correction})
must play a role somehow in the net electric force on this density of
dipoles, and the term $-\frac12 \nabla (\bmP\cdot\bfE)$ seems to be
one way to account for it.

We have highlighted the differences that occur when flexoelectric
terms are included.  These take us out of the realm of a linear
dielectric medium and also introduce couplings between $\nabla\nhat$
and $\bfE$.  While the overall framework of our development remains
unchanged---the expressions for the body force, stress, force balance,
body couple, couple stress, and torque balance maintain their same
general forms---there are new contributions introduced by the
flexoelectric terms.  We have derived and presented expressions for
these new contributions and have tried to provide some interpretation
of them.  Concerning the expression for the total stress tensor $\bfT$
above (and in \eqref{eqn:Tfinal}), we note that there are
flexoelectric contributions to the term
$- (\nabla\nhat)^T \! \partial W / \partial \nabla \nhat$, in addition
to contributions to the expressions for $W$ and $\bfD$.  Flexoelectric
effects are complicated and difficult to interpret in general: they
involve couplings among $\nhat$, $\nabla\nhat$, and $\bfE$, and they
have both forward and inverse influences (director distortion inducing
polarization and contributions to electric fields, and electric fields
encouraging director distortion so as to minimize the coupling term
$- \Pf(\nhat,\nabla\nhat) \cdot \bfE$ in the free-energy density).

Any force field that enters our system in the form $\bfF = - \partial
W / \partial x$ is balanced, in equilibrium, by all or part of a
related pressure gradient.  This follows from the force balance
\begin{equation*}
  \Div \bfT - \dWdx = \bfzero
\end{equation*}
and the fact that $-W$ forms part of the thermodynamic pressure $p =
p_0 - W$:
\begin{equation*}
  \Div \bfT = \Div \Bigl[ ( W - p_0 ) \bfI -
  (\nabla\nhat)^T \! \dWdgn + \bfE\otimes\bfD \Bigr] =
  \nabla W - \cdots = \dWdx + \cdots .
\end{equation*}
The typical case here is that of a gravitational field.  Another case
(which we have not discussed in this report) would be the body force
associated with a prescribed inhomogeneous magnetic field
$\bfH = \bfH(x)$ (see \cite[Sec.\,3.1.4]{sonnet:virga:12} or
\cite[Sec.\,2.4.2 Remark~(ii)]{stewart:04}).  The same type of
balance holds true in the ``compatibility potential'' approach with
$\bfF = \partial\Psi / \partial x$ and $p = p_0 + \Psi - W$ (which, as
we have noted earlier, is equivalent to our approach with $-\Psi$
folded into $W$), and this was observed by Ericksen some time ago
\cite[Sec.\,III.A]{ericksen:76}.  In a more general sense, something
similar holds true in the absence of any such force field at all, in
which case the force balance
\begin{equation*}
  \Div \bfT = \bfzero , ~~ \text{with} ~
  \bfT = ( \We - p_0 ) \bfI - ( \nabla\nhat )^T \! \dWedgn
\end{equation*}
gives
\begin{equation*}
  \nabla \We - \Div \Bigl[ (\nabla\nhat)^T \! \dWedgn \Bigr] = \bfzero ,
\end{equation*}
the effective force density associated with the Ericksen stress (the
negative of the term in brackets above) being balanced by the negative
gradient of the pressure from the distortional elastic energy density.
Electric fields seem to behave in a different way, as seen in
Sec.\,\ref{sec:free-charge-included}.  There, even in the absence of
flexoelectric terms and with no free-charge distribution, the fully
simplified force balance \eqref{eqn:divTe_vs_divTM} reads
\begin{equation*}
  \Div \Te + ( \nabla\bfE ) \bmP -
  \frac12 \nabla ( \bmP\cdot\bfE ) = \bfzero ,
\end{equation*}
with $\Te$ as defined in \eqref{eqn:TeTM}, and we see that a
combination of forces from both distortional elasticity and
electrostatics is needed to obtain an equilibrium balance of forces (a
reflection of the coupling between $\nhat$ and $U$).

Somewhat related to the above is the observation that the stress
tensor $\bfT$ and couple stress tensor $\bfL$ contain mechanical and
electrostatic influences commingled.  This is consistent with the
modern point of view in the continuum mechanics of electromagnetic
mechanical systems.  As already suggested by Toupin
\cite[Sec.\,1]{toupin:60} (dealing with elastic dielectric solids):
``Any division of energy, momentum, stress, and energy flux into
electromagnetic and mechanical components is bound to be somewhat
arbitrary, and it is fruitless to attempt an independent theory of
either component.''  This appears to be the contemporary consensus
view.  It also appears that past attempts at such divisions have led
to some of the confusions and controversies in this area
\cite{barnett:10,barnett:loudon:10,bocker:16}.  As also observed in
\cite[Sec.\,4]{toupin:60}, the field equations and boundary conditions
are independent of any chosen decomposition into mechanical and
electromagnetic components.  There are several ways in which work can
be done in a system such as ours---this includes displacing matter,
rotating $\nhat$, rotating $\bmP$, and moving charge on/off the
electrodes---and these are reflected in various ways in the expression
for the free energy.

We have focused on the case in which the electric field comes from
electrodes held at constant potential by a battery or voltage source,
the common case for liquid crystal experiments and devices.  In such a
case, the appropriate pointwise contribution to the free-energy
density can be written $- \int_0^E \! \bfD \cdot \text{d}\bfE$ and
takes the form $-\frac12 \bfD \cdot \bfE$ if the medium is a linear
dielectric.  While the electric potential remains constant on the
electrodes in such a system, the charge densities on the electrode
surfaces will in general change in response to changes in the state of
the system.  The voltage source may have to push charge on or off the
electrodes to maintain the constant potential, and this work is what
gives rise to the negative sign in the electrostatic contribution to
the free-energy density.  In other analyses of electromagnetic
field/matter interactions, one often finds developments involving
electric fields that are assumed to arise from fixed charge sources
(charge distributions that do not change).  For such systems, the
appropriate contribution to the free-energy density would be
$\int_0^D \! \bfE\cdot\text{d}\bfD$, taking the form
$+\frac12 \bfD\cdot\bfE$ for a linear dielectric
\cite[Sec.\,4.7]{jackson:75},
\cite[Sec.\,10]{landau:lifshitz:pitaevskii:93},
\cite[Sec.\,2.8]{stratton:41}.  The contrasting energetics of
``constant charge'' versus ``constant potential'' are discussed in
\cite[Sec.\,4.7]{jackson:75} and
\cite[Secs.\,5,\,10]{landau:lifshitz:pitaevskii:93} in general, and in
\cite[Secs.\,3.6.2,\,3.6.3]{barbero:evangelista:01} and
\cite[Sec.\,10.1]{collings:hird:97} in the context of liquid crystals.
For the equilibrium electrostatic equations, one would have boundary
conditions on $\bfD\cdot\nuhat$ in the case of fixed charge sources
versus boundary conditions on $U$ in the case of fixed potential.

Thus we have followed through a program to determine what results when
one treats the electric potential as a state variable of equal
standing with the director field.  While the overall analysis and
approach have considerable overlap with earlier works, they provide
generalizations and (we hope) remove some of the question marks about
the differences in handling magnetic fields versus electric fields in
settings such as these.  We also hope that our findings can shed some
light on electric field/matter interactions in other settings, at
least in the circumstances of incompressible fluid hydrostatics.
Electromagnetic stress in ponderable media is complicated and has a
long history that includes a number of controversies.  Here, without
taking sides in any of these debates, we have shown what expressions
result from our starting point (of a free-energy density that contains
potentials for all force fields of external origin) and have observed
how these expressions compare with others found in the literature.  To
extend these ideas to liquid crystal hydrodynamics (at this level of
modeling), a reasonable approach would seem to be to treat the
electric field as adjusting instantaneously to motions of the fluid
and director field (by virtue of the large differences in the time
scales for fluid motion and director orientation changes compared to
dielectric relaxation times \cite{shiyanovskii:lavrentovich:10}).
This would result in a PDE-constrained version of the Ericksen-Leslie
equations (with appropriate terms incorporated into the free-energy
density, such as those we have given in
Sec.\,\ref{sec:free-energy-density}).

\begin{acknowledgements}
  The author is grateful to E.~G.~Virga for helpful comments on an
  earlier draft of this report and to P.~Palffy-Muhoray for helpful
  discussions concerning surface anchoring energies.
\end{acknowledgements}

\bibliography{%
  contin_mech,%
  contin_mech_EM,%
  electromagnetics,%
  frank,%
  interfaces_surfaces,%
  landau,%
  lc-books,%
  lc-flow,%
  lc-misc,%
  math_physics,%
  thermodynamics}

\begin{thebibliography}{10}
\providecommand{\url}[1]{{#1}}
\providecommand{\urlprefix}{URL }
\expandafter\ifx\csname urlstyle\endcsname\relax
  \providecommand{\doi}[1]{DOI~\discretionary{}{}{}#1}\else
  \providecommand{\doi}{DOI~\discretionary{}{}{}\begingroup
  \urlstyle{rm}\Url}\fi

\bibitem{ball:17}
Ball, J.M.: Mathematics and liquid crystals.
\newblock Mol. Cryst. Liq. Cryst. \textbf{647}(1), 1--27 (2017)

\bibitem{barbero:evangelista:01}
Barbero, G., Evangelista, L.R.: An Elementary Course on the Continuum Theory
  for Nematic Liquid Crystals.
\newblock World Scientific, Singapore (2001)

\bibitem{barbero:evangelista:06}
Barbero, G., Evangelista, L.R.: Adsorption Phenomena and Anchoring Energy in
  Nematic Liquid Crystals.
\newblock CRC Press, Boca Raton (2006)

\bibitem{barnett:10}
Barnett, S.M.: Resolution of the {A}braham-{M}inkowski dilemma.
\newblock Phys. Rev. Lett. \textbf{104}, 070401 (2010)

\bibitem{barnett:loudon:10}
Barnett, S.M., Loudon, R.: The enigma of optical momentum in a medium.
\newblock Phil. Trans. R. Soc. A \textbf{368}, 927--939 (2010)

\bibitem{barratt:jenkins:73}
Barratt, P.J., Jenkins, J.T.: Interfacial effects in the magnetohydrostatic
  theory of nematic liquid crystals.
\newblock J. Phys. A: Math. Nucl. Gen. \textbf{6}, 756--769 (1973)

\bibitem{bocker:16}
B\"ocker, J.: Some reflections on electromagnetic force in matter.
\newblock In: 2016 International Symposium on Power Electronics, Electrical
  Drives, Automation and Motion (SPEEDAM), pp. 1433--1440. IEEE (2016)

\bibitem{buka:eber:13}
Buka, A., \'{E}ber, N. (eds.): Flexoelectricity in Liquid Crystals: Theory,
  Experiments and Applications.
\newblock Imperial College Press, London (2013)

\bibitem{bustamante:dorfmann:ogden:09a}
Bustamante, R., Dorfmann, A., Ogden, R.W.: Nonlinear electroelastostatics: a
  variational framework.
\newblock Z. Angew. Math. Phys. \textbf{60}, 154--177 (2009)

\bibitem{bustamante:dorfmann:ogden:09b}
Bustamante, R., Dorfmann, A., Ogden, R.W.: On electric body forces and
  {M}axwell stresses in nonlinearly electroelastic solids.
\newblock Int. J. Engr. Sci. \textbf{47}, 1131--1141 (2009)

\bibitem{callen:85}
Callen, H.B.: Thermodynamics and an Introduction to Thermostatistics, 2nd edn.
\newblock John Wiley \& Sons, New York (1985)

\bibitem{collings:hird:97}
Collings, P.J., Hird, M.: Introduction to Liquid Crystals Chemistry and
  Physics.
\newblock Taylor \& Francis, London (1997)

\bibitem{ericksen:61}
Ericksen, J.L.: Conservation laws for liquid crystals.
\newblock Trans. Soc. Rheol. \textbf{5}, 23--34 (1961)

\bibitem{ericksen:62}
Ericksen, J.L.: Hydrostatic theory of liquid crystals.
\newblock Arch. Ration. Mech. Anal. \textbf{9}(1), 371--378 (1962)

\bibitem{ericksen:76}
Ericksen, J.L.: Equilibrium theory of liquid crystals.
\newblock In: G.H. Brown (ed.) Advances in Liquid Crystals, vol.~2, pp.
  233--298. Academic Press, New York (1976)

\bibitem{eringen:79}
Eringen, A.C.: Continuum theory of nematic liquid crystals subject to
  electromagnetic fields.
\newblock J. Math. Phys. \textbf{20}(12), 2671--2681 (1979)

\bibitem{eringen:maugin:90a}
Eringen, A.C., Maugin, G.A.: Electrodynamics of Continua I: Foundations and
  Solid Media.
\newblock Springer-Verlag, New York (1990)

\bibitem{eringen:maugin:90b}
Eringen, A.C., Maugin, G.A.: Electrodynamics of Continua II: Fluids and Complex
  Media.
\newblock Springer-Verlag, New York (1990)

\bibitem{eu:86}
Eu, B.C.: Statistical foundation of the {M}inkowski tensor for ponderable
  media.
\newblock Phys. Rev. A \textbf{33}(6), 4121--4131 (1986)

\bibitem{eu:oppenheim:86}
Eu, B.C., Oppenheim, I.: On the {M}inkowski tensor and thermodynamics of media
  in an electromagnetic field.
\newblock Physica \textbf{136A}, 233--254 (1986)

\bibitem{faetti:virga:97}
Faetti, S., Virga, E.G.: On a curvature surface energy for nematic liquid
  crystals.
\newblock Arch. Rational Mech. Anal. \textbf{140}(1), 31--52 (1997)

\bibitem{frank:58}
Frank, F.C.: On the theory of liquid crystals.
\newblock Discuss. Faraday Soc. \textbf{25}, {19--28} (1958)

\bibitem{gelfand:fomin:63}
Gelfand, I.M., Fomin, S.V.: Calculus of Variations.
\newblock Prentice-Hall, Englewood Cliffs, N.J. (1963).
\newblock Translated and Edited by Richard A. Silverman

\bibitem{degennes:prost:93}
de~Gennes, P.G., Prost, J.: The Physics of Liquid Crystals, 2nd edn.
\newblock Clarendon Press, Oxford (1993)

\bibitem{degroot:69}
de~Groot, S.R.: The Maxwell Equations.
\newblock North Holland, Amsterdam (1969)

\bibitem{degroot:suttorp:72}
de~Groot, S.R., Suttorp, L.G.: Foundations of Electrodynamics.
\newblock North Holland, Amsterdam (1972)

\bibitem{guggenheim:67}
Guggenheim, E.A.: Thermodynamics: An Advanced Treatment for Chemists and
  Physicists, 5th edn.
\newblock North-Holland, Amsterdam (1967)

\bibitem{guo:zheng:palffy:19}
Guo, T., Zheng, X., Palffy-Muhoray, P.: Surface anchoring energy of cholesteric
  liquid crystals.
\newblock Liquid Crystals  (2019).
\newblock \doi{10.1080/02678292.2019.1660425}

\bibitem{gurtin:81}
Gurtin, M.E.: An Introduction to Continuum Mechanics.
\newblock Academic Press, San Diego (1981)

\bibitem{gurtin:88}
Gurtin, M.E.: Multiphase thermomechanics with interfacial structure 1. {H}eat
  conduction and the capillary balance law.
\newblock Arch. Rational Mech. Anal. \textbf{104}(3), 195--221 (1988)

\bibitem{gurtin:murdoch:75}
Gurtin, M.E., Murdoch, A.I.: A continuum theory of elastic material surfaces.
\newblock Arch. Rational Mech. Anal. \textbf{57}(4), 291--323 (1975)

\bibitem{hutter:vandeven:ursescu:06}
Hutter, K., van~de Ven, A.A.F., Ursescu, A.: Electromagnetic Field Matter
  Interactions in Thermoelastic Solids and Viscous Fluids.
\newblock Springer, Berlin Heidelberg (2006)

\bibitem{jackson:75}
Jackson, J.D.: Classical Electrodynamics, 2nd edn.
\newblock John Wiley \& Sons, New York (1975)

\bibitem{jakli:saupe:06}
J\'akli, A., Saupe, A.: One- and Two-Dimensional Fluids: Properties of Smectic,
  Lamellar and Columnar Liquid Crystals.
\newblock CRC Press, Boca Raton (2006)

\bibitem{jeffreys:jeffreys:56}
Jeffreys, S.H., Jeffreys), B.S.L.: Methods of Mathematical Physics, 3rd edn.
\newblock Cambridge University Press, Cambridge (1956)

\bibitem{jenkins:barratt:74}
Jenkins, J.T., Barratt, P.J.: Interfacial effects in the static theory of
  nematic liquid crystals.
\newblock Quart. J. Mech. Appl. Math. \textbf{27}(1), 111--127 (1974)

\bibitem{kellogg:29}
Kellogg, O.D.: Foundations of Potential Theory.
\newblock Springer, Berlin (1929)

\bibitem{lagerwall:99}
Lagerwall, S.T.: Ferroelectric and Antiferroelectric Liquid Crystals.
\newblock Wiley-VCH, Weinheim (1999)

\bibitem{lanczos:70}
Lanczos, C.: The Variational Principles of Mechanics, 4th edn.
\newblock University of Toronto Press, Toronto (1970)

\bibitem{landau:lifshitz:pitaevskii:93}
Landau, L.D., Lifshitz, E.M., Pitaevskii, L.P.: Electrodynamics of Continuous
  Media, 2nd edn.
\newblock Butterworth-Heinemann, Oxford (1993)

\bibitem{leslie:79}
Leslie, F.M.: Theory of flow phenomena in liquid crystals.
\newblock In: G.H. Brown (ed.) Advances in Liquid Crystals, vol.~4, pp. 1--81.
  Academic Press, New York (1979)

\bibitem{leslie:87b}
Leslie, F.M.: Some topics in equilibrium theory of liquid crystals.
\newblock In: J.L. Ericksen, D.~Kinderlehrer (eds.) Theory and Applications of
  Liquid Crystals, \emph{IMA Volumes in Mathematics and Its Applications},
  vol.~5, pp. 211--234. Springer-Verlag, New York (1987)

\bibitem{liu:13}
Liu, L.: On energy formulations of electrostatics for continuum media.
\newblock J. Mech. Phys. Solids \textbf{61}, 968--990 (2013)

\bibitem{mansfield:10}
Mansfield, E.L.: A Practical Guide to the Invariant Calculus.
\newblock Cambridge University Press, Cambridge (2010)

\bibitem{mcconnell:57}
Mc{C}onnell, A.J.: Applications of Tensor Analysis.
\newblock Dover, New York (1957)

\bibitem{mottram:newton:14}
Mottram, N.J., Newton, C.J.P.: Introduction to {Q}-tensor theory.
\newblock arXiv.org e-Print archive  (2014).
\newblock \urlprefix\url{https://arxiv.org/abs/1409.3542}

\bibitem{ogden:97}
Ogden, R.W.: Non-Linear Elastic Deformations.
\newblock Dover, Mineola, New York (1997)

\bibitem{oseen:33}
Oseen, C.W.: The theory of liquid crystals.
\newblock Trans. Faraday Soc. \textbf{29}, {883--889} (1933)

\bibitem{reitz:milford:67}
Reitz, J.R., Milford, F.J.: Foundations of Electromagnetic Theory, 2nd edn.
\newblock Addison-Wesley, Reading, Massachusetts (1967)

\bibitem{shiyanovskii:lavrentovich:10}
Shiyanovskii, S.V., Lavrentovich, O.D.: Dielectric relaxation and memory
  effects in nematic liquid crystals.
\newblock Liquid Crystals \textbf{37}(6-7), 737--745 (2010)

\bibitem{slattery:sagis:oh:07}
Slattery, J.C., Sagis, L., Eun-Suok, O.: Interfacial Transport Phenomena, 2nd
  edn.
\newblock Springer, New York (2007)

\bibitem{sluckin:95}
Sluckin, T.J.: Anchoring transitions at liquid crystal surfaces.
\newblock Physica A \textbf{213}, 105--109 (1995)

\bibitem{sluckin:poniewierski:86}
Sluckin, T.J., Poniewierski, A.: Orientational wetting transitions and related
  phenomena in nematics.
\newblock In: C.A. Croxton (ed.) Fluid Interfacial Phenomena, chap.~5, pp.
  215--253. John Wiley \& Sons, Chichester (1986)

\bibitem{sonin:95}
Sonin, A.A.: The Surface Physics of Liquid Crystals.
\newblock Gordon and Breach, Luxembourg (1995)

\bibitem{sonnet:virga:12}
Sonnet, A.M., Virga, E.G.: Dissipative Ordered Fluids: Theories for Liquid
  Crystals.
\newblock Springer, New York (2012)

\bibitem{stewart:04}
Stewart, I.W.: The Static and Dynamic Continuum Theory of Liquid Crystals.
\newblock Taylor \& Francis, London (2004)

\bibitem{stratton:41}
Stratton, J.A.: Electromagnetic Theory.
\newblock McGraw-Hill, New York and London (1941)

\bibitem{tiersten:90}
Tiersten, H.F.: A Development of the Equations of Electromagnetism in Material
  Continua.
\newblock Springer-Verlag, New York (1990)

\bibitem{toupin:56}
Toupin, R.A.: The elastic dielectric.
\newblock J. Rational Mech. Anal. \textbf{5}(6), 849--915 (1956)

\bibitem{toupin:60}
Toupin, R.A.: Stress tensors in elastic dielectrics.
\newblock Arch. Ration. Mech. Anal. \textbf{5}(1), 440--452 (1960)

\bibitem{truesdell:noll:04}
Truesdell, C., Noll, W.: The Non-Linear Field Theories of Mechanics, 3rd edn.
\newblock Springer, Berlin and Heidelberg (2004)

\bibitem{truesdell:toupin:60}
Truesdell, C., Toupin, R.A.: The {C}lassical {F}ield {T}heories.
\newblock In: S.~Fl\"{u}gge (ed.) Principles of Classical Mechanics and Field
  Theory, \emph{Encyclopedia of Physics}, vol. III/1, pp. 226--858.
  Springer-Verlag, Berlin Heidelberg (1960)

\bibitem{virga:94}
Virga, E.G.: Variational Theories for Liquid Crystals.
\newblock Chapman \& Hall, London (1994)

\bibitem{weatherburn:61}
Weatherburn, C.E.: Differential Geometry of Three Dimensions, vol.~I.
\newblock Cambridge University Press, Cambridge (1961)

\bibitem{zocher:33}
Zocher, H.: The effect of a magnetic field on the nematic state.
\newblock Trans. Faraday Soc. \textbf{29}, {945--957} (1933)

\end{thebibliography}

\end{document}